\documentclass[a4paper,12pt,times,numbered,print,index,custombib,twoside,custommargin]{Classes/PhDThesisPSnPDF}

\input{Preamble/preamble}

\title{Dynamical Systems Analysis of Various Dark Energy Models}
\author{ Nandan Roy}

\dept{\bf Department of Physical Sciences}

\university{\bf Indian Institute Of Science Education and Research Kolkata}
\crest{\centering \includegraphics[width=0.25\textwidth]{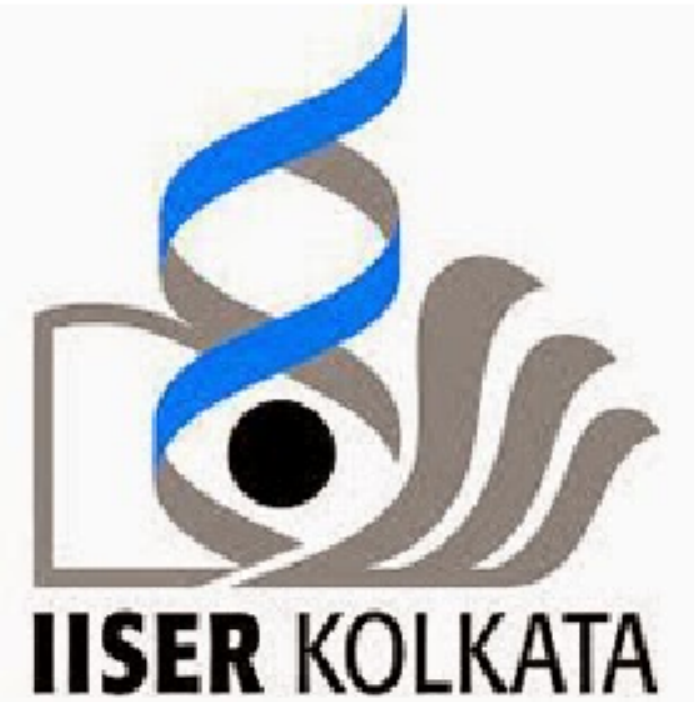}}

\renewcommand{\submissiontext}{\bf Supervisor: Prof. Narayan Banerjee}
\degreetitle{\bf Thesis submitted to IISER Kolkata 
\\
for the Degree of Doctor Of Philosophy in Science}

\subject{LaTeX} 

\ifdefineAbstract
\fi

\begin{document}

\frontmatter
\begin{titlepage}
  \maketitle
\end{titlepage}

% ******************************* Thesis Dedidcation ********************************

\begin{dedication} 

{\Large To

My loving parents 

\&

My teachers.  }

\end{dedication}

\begin{dedication}

\begin{center}

{ \Large \bf \underline{Certificate} }

\end{center}

\begin{flushleft}

{\centering This is to certify that the Ph.D. thesis entitled “ {\bf Dynamical Systems Analysis of Various Dark Energy Models}” submitted by {\bf Nandan Roy } is absolutely based upon his own work under the supervision of  {\bf Prof. Narayan Banerjee} at Indian Institute of Science Education and Research Kolkata (IISER Kolkata) and that neither this thesis nor any part of it has been submitted for either any degree/diploma or any other academic award anywhere before.}

\end{flushleft}

~~

\begin{flushright}
Prof. Narayan Banerjee

Mohanpur, India.
\end{flushright}
\end{dedication}
% ******************************* Thesis Declaration ***************************

\begin{declaration}

This thesis is a presentation of my original research work. Whenever contributions of others are involved, every effort is made to indicate this clearly, with due reference to the literature and acknowledgement of collaborative research and discussions. The work is original and has not been submitted earlier as a whole or in part for a degree or diploma at this or any other Institution or University. This work was done under the guidance of Prof. Narayan Banerjee, at Indian Institute of Science Education and Research Kolkata (IISER Kolkata).

% Author and date will be inserted automatically from thesis.tex \author \degreedate

\end{declaration}

% ************************** Thesis Acknowledgements **************************

\begin{acknowledgements}      

First and foremost I would like to thank my supervisor, Prof. Narayan Banerjee whose guidance, support and infinite patience made this thesis possible. I am also grateful to him for teaching me not only the General Relativity and Cosmology but also the values of life.

My special thanks to Prof. P. K. Panigrahi, Dr. Anandamohan Ghosh and Dr. Golam Mortuza Hossain for their help and valuable suggestions. Thanks also to my co-researchers Arghyada, Vivekda, Abhinav, Barun, Soumitra, Dyuti, Subhrajit, Gopal, Debmalya, Soumya, Ankan, Chiranjib, Subhajit and other members of the Department Of Physical Sciences, IISER Kolkata for their support and help. 

I express my gratitude to my parents and to my uncle Gurudas Roy who have been a constant support and inspiration to me through my whole life. Thanks are due to my in-laws for their love and support. 

Finally, thanks Tapasi for being loving, caring and sharing life with me.

\end{acknowledgements}

  \begin{dedication}
  
  \textbf{\Large Preface}

  \begin{flushleft}
  
  The research work contained in this thesis was carried out at the Indian Institute of Science Education and Research Kolkata, India in the Department of Physical Sciences. Chapter 1 contains an introduction and the other chapters are based on the papers as follows:

  \begin{itemize}
  
  \item Chapter 2 
  
  N. Roy and N. Banerjee, ``Tracking quintessence: a dynamical systems study", Gen. Rel. Grav., {\bf 46}, 1651 (2014).
  
  N. Roy and N. Banerjee, ``Quintessence Scalar Field: A Dynamical Systems Study",  Euro. Phys. J Plus., {\bf 129}, 162 (2014).
  
  \item Chapter 3 
  
  N. Roy and N. Banerjee, ``Stability of chameleon scalar field models", Annals Phys., {\bf 356}, 452, (2015).
  
  \item Chapter 4 
  
 N. Banerjee and N. Roy, ``Phase space analysis of a holographic dark energy model", Gen.Rel.Grav., {\bf 47}, 92 (2015).
  
\end{itemize}    

\end{flushleft}
  
  \end{dedication}
\begin{dedication}

\textbf{\Large List of Papers}

\begin{flushleft}

1.  N. Roy and N. Banerjee, ``Tracking quintessence: a dynamical systems study", Gen. Rel. Grav., {\bf 46}, 1651 (2014).
  
2.  N. Roy and N. Banerjee, ``Quintessence Scalar Field: A Dynamical Systems Study",  Euro. Phys. J Plus., {\bf 129}, 162 (2014).

3.  N. Roy and N. Banerjee, ``Stability of chameleon scalar field models", Annals Phys., {\bf 356}, 452, (2015).

4. N. Banerjee and N. Roy, ``Phase space analysis of a holographic dark energy model", Gen.Rel.Grav., {\bf 47}, 92 (2015).

\end{flushleft}

\end{dedication}

% *********************** Adding TOC and List of Figures ***********************

\tableofcontents

\listoffigures

\listoftables

% \printnomencl[space] space can be set as 2em between symbol and description
%\printnomencl[3em]

\printnomencl

% ******************************** Main Matter ********************************* 
\mainmatter

%*******************************************************************************
%*********************************** First Chapter *****************************
%*******************************************************************************

\chapter{Introduction}  %Title of the First Chapter

\ifpdf
    \graphicspath{{Chapter1/Figs/Raster/}{Chapter1/Figs/PDF/}{Chapter1/Figs/}}
\else
    \graphicspath{{Chapter1/Figs/Vector/}{Chapter1/Figs/}}
\fi

%********************************** %First Section  **************************************
\section{Introduction to Cosmology}
Some very old questions of the human mind: how the universe began? how it works ? what is the ultimate fate of the universe? etc. gave birth to a branch of science, which is called cosmology. Cosmology is the subject which deals with the beginning, evolution and possible ultimate fate of the universe, in brief, cosmology is the study of the universe.

\subsubsection{Cosmological Principle} 
 Copernican principle states that the earth is not the center of the universe and we are not living at a special location of the universe. This principle has been generalized in physical cosmology and named as cosmological principle. Cosmological principle states that the universe is homogeneous and isotropic at any given cosmic time at a sufficiently large scale. Homogeneity means, all points of the universe have same property and isotropy means, there is no preferred direction i,e. the universe looks same in all directions. The universe has anisotropy or inhomogeneity at scales much smaller than the size of the universe ($28 \times 10^{9}$ pc) so as to give room for the structures that we see around us. Observationally, the anisotropy  of the universe is $\frac{\bigtriangleup \rho}{\rho} \simeq 10^{-5}$.

 \subsubsection{Hubble's discovery}
 While observing shift in spectra of nebula and galaxies, Hubble found, spectra of distant objects are redshifted. From this observation, he plotted distance(D) versus velocity($v$) curve for observed objects and arrived at a simple law $ v = H D$. This is the Hubble's law, which simply tells us that distant galaxies are more redshifted and H is the Hubble's parameter. In 1929 he published this result in `` Proceedings of National Academy of Science".

\subsubsection{FRW Cosmology} 
 The most general line element satisfying the cosmological principle is the Friedmann-Robertson-Walker (FRW) metric which is written as \cite{c1, c2, c3, c4}
 \begin{equation} \label{frw}
 ds^2 = - dt^2 + a(t)^2 (\frac{dr^2}{1- k r^2} + r^2(d \theta^2 + \sin^2{\theta} d \phi^2)),
 \end{equation}
where $a(t)$ is scale factor, t is cosmic time and $r, \theta, \phi$ are the spatial coordinates in polar form and $k$ is the curvature constant which describe spatial curvature of the metric.

% If $K=0$, the spatial geometry is flat, if $k = 1$, it is positively curved and if $k = -1$, it is negatively curved. 
 Einstein field equations \cite{c2},
\begin{equation}
G^{\mu} _{\nu} \equiv R^{\mu} _{\nu} - \frac{1}{2} \delta^{\mu}_{\nu} R = 8 \pi G T^{\mu} _{\nu},
\end{equation}
where $G^{\mu} _{\nu}$ is the Einstein tensor, $R^{\mu} _{\nu}$ is Ricci tensor, R is Ricci scalar and $ T^{\mu} _{\nu}$ is the energy momentum tensor, can be written as 
\begin{equation} \label{ch1}
H^2 = (\frac{\dot{a}}{a})^2 = \frac{8 \pi G \rho}{3} - \frac{k}{a^2},
\end{equation} 

\begin{equation} \label{ch2}
\dot{H} = -4 \pi G (\rho + p) + \frac{k}{a^2},
\end{equation}

for an FRW metric. The field equations are written for the universe filled with an ideal fluid. Energy momentum tensor for an ideal fluid is $ T^{\mu} _{\nu} = Diag (-\rho, p, p, p)$, where $\rho$ and $p$ are the energy density and pressure respectively. $H$ is the Hubble parameter, $H = \frac{\dot{a}}{a}$. From Bianchi identity one can write the continuity equation as 
\begin{equation} \label{continuity}
\dot{\rho} + 3H (\rho + p) = 0.
\end{equation}

By rearranging (\ref{ch1}) and (\ref{ch2}) , a $k$ independent equation can be written as 
\begin{equation} \label{ch3}
\frac{\ddot{a}}{a} = - \frac{4 \pi G}{3} (\rho+ 3 p).
\end{equation}
Cosmologists introduced a dimensionless variable, called `deceleration parameter' as $q = - \frac{a \ddot{a}}{\dot{a}^2}$ which is a dimensionless measure of cosmic acceleration. If $q< 0$ , the universe is not only expanding but expanding with an acceleration. If $q > 0$, then the expansion of the universe is decelerating. Now from the recent observations, (which has been briefly discussed in the next section) it is found that the universe is accelerating. 

One can write equation (\ref{ch1}) in a dimensionless form as
\begin{equation} \label{denpara}
\Omega(t) -1 = \frac{k}{(a H)^2},
\end{equation}  
where $\Omega(t) \equiv \frac{\rho(t)}{\rho_c (t)}$, is dimensionless density parameter and $\rho_c (t) = \frac{3 H^2 (t)}{8 \pi G}$, is the critical density.

If ~ $\rho(t) > \rho_c (t), \Omega(t) >1 \Longrightarrow k = + 1$, the universe is closed.

If ~ $\rho(t) < \rho_c (t), \Omega(t) < 1 \Longrightarrow k = - 1$, the universe is open.

If ~ $\rho(t) = \rho_c (t), \Omega(t) = 1 \Longrightarrow k =  0$, the universe is flat.

It has been also confirmed from the observations that the current universe is very close to spatially flat geometry $(\rho(t) \simeq \rho_c (t)) $. This is in fact a result of the early inflation, which washes out the spatial curvature. However, in future evolution, the spatial curvature may play some role. In the present work, the discussion is restricted only to $k=0$, which leads to a great deal of computation simplicity as well.
%Hence, in the rest of our discussion we will consider  a flat universe i,e. $k=0$.

If we consider a spatially flat universe, filled with an ideal barotropic fluid with equation of state $p = (\gamma -1 ) \rho$, where  $\gamma$ is equation of state parameter, the solutions of the Einstein field equations are,

\begin{equation}
H = \frac{2}{3 \gamma (t - t_0)},
\end{equation}
\begin{equation}
a(t) \varpropto (t-t_0)^{\frac{2}{3 \gamma}},
\end{equation}
\begin{equation}
\rho(t) \varpropto a^{-3 \gamma}.
\end{equation}

If $\gamma = 1$, then the universe is dust dominated and if $\gamma = \frac{4}{3}$, the universe is radiation dominated. For these two cases the solutions take the form 

Radiation: $\gamma = \frac{4}{3} , a(t) \varpropto (t - t_0)^{\frac{1}{2}}, \rho \varpropto a^{-4}$,

Dust: $\gamma = 1 , a(t) \varpropto (t - t_0)^{\frac{2}{3}}, \rho \varpropto a^{-3}$.

None of these two solutions correspond to an accelerated expansion($ \frac{\ddot{a}}{a} > 0)$ of the universe. As gravity is attracting, it can not be account for the accelerated expansion of the universe. From equation (\ref{ch3}), for accelerated expansion we need an exotic energy which satisfy the equation of state $\rho + 3 p < 0$, i,e. $\gamma < \frac{2}{3}$ . It essentially tells us that a sufficiently large negative pressure is required to drive the accelerated expansion of the universe. 

\subsubsection{Cosmological redshift and luminosity distance}
Redshift has been used as a popular marker to describe the stage of the evolution, in other words, the age of the universe. As the universe is expanding, the spectrum of light emitted by a stellar object becomes redshifted. The redshift $z$ is defined as  

\begin{equation}
1+z = (\lambda_0 - \lambda)/ \lambda 
\end{equation} 

and for an isotropic universe, it comes out to be $a_o/a$. Here $\lambda$ is the wavelength and $a$ is the scale factor while a subscript zero indicates their observed and the present value respectively.

There are different methods for distance measurement in the expanding universe. The luminosity distance is the most used method to measure the distance. Let us consider an observer at $x=0$ and a source at $x=x_s$. The amount of radiation emitted by the source per unit time per unit area is the absolute luminosity L of the source. The apparent luminosity $l$ is the amount of radiation received by the observer per unit time per unit area. In Minkowski space time one can write the relation between absolute luminosity and apparent luminosity as $l = \frac{L}{4 \pi d^2}$, where d is the distance between the source and the observer.

For an expanding universe it can be generalized to define luminosity distance $d_{L} ^2 = \frac{L}{4 \pi l}$ and one can also obtain the relation between absolute and apparent magnitude as $L = l (1+z)^2$. For an FRW universe the area of the sphere is given by $s = 4 \pi (a_0 f_k (\chi_s))^2$, where\\

$f_k (\chi_s) = \sin \chi,~~ k = +1,$ 

$f_k (\chi_s) =\chi,~~ k = 0,$  

$f_k (\chi_s) = \sinh \chi, ~~ k = -1$.

Using the above relations , we obtain the expression for luminosity  distance in a FRW back ground as 
\begin{equation}
d_l ^2 = a_0 f_k (\chi_s) (1+z).
\end{equation}

\section{Observational Evidences of Accelerated Expansion of the Universe}

The direct observational evidence of the current accelerated expansion of the universe has been found by measuring the luminosity distances of different high red-shift supernovas \cite{c5,c6}.  

To measure the luminosity distances, astronomers use the formula \cite{c7, c8}

\begin{equation}
m - M = 5 \log_{10} (\frac{d_l}{M_{pc}}) + 25.
\end{equation}

Here m and M are the apparent and absolute magnitudes of the source respectively, they are related to the logarithms of apparent luminosity and absolute luminosity respectively. The absolute magnitude(M) is defined as the apparent magnitude of an object if it were placed at a distance 10 parsecs. 
The redshift and magnitude are related by the equation 

\begin{equation}
m = 5 \log z + 1.086 (1 - q_0) z .
\end{equation}

For many years brightest galaxies were the main standard candles for observing the universe. But now type Ia supernova has replaced galaxies and proved itself as an excellent standard candle. Type Ia supernova is observed when a white dwarf exceed the Chandrasekhar limit by accretion of mass from it's companion star of the binary pair and explodes. 

 In 1998 from the observation of type Ia supernova(SNIa), Riess et al \cite{c5} and Perlmutter et al \cite{c6}  indicated that the present universe is undergoing an accelerated expansion. The clue towards this was provided by the magnitude-redshift relation. It was found that for a flat, homogeneous and isotropic universe, about 70 \% of the energy density consists of some kind of mysterious component. This mysterious energy component of the universe which drives the accelerated expansion of the universe is called dark energy. It was also found that matter energy density parameter $\Omega^{(0)}_{m} = 0.28^{+0.09} _{-0.08}$ ($1 \sigma$ statistical). Later, the accelerated expansion of the universe is confirmed by many other observations. Another interesting result was found from these observations that the accelerated expansion of the universe is a recent phenomenon. From high redshift ($z> 1.5$) SNIa data, Riess et al \cite{c9} in 2004 showed that the universe exhibited a transition from a decelerated to an accelerated expansion phase at a redshift of $z \simeq 0.45$ . This result is crucial as indicates that the universe indeed had a decelerated phase which is of utmost importance for the nucleosysthesis and the subsequent formation of structure. Super Nova Legacy Survey \cite{c10} and WMAP \cite{c11} shows excellent agreement and confirm accelerated expansion of the universe. WMAP and SNLS together provide dark energy density parameter $\Omega_{\lambda} = 0.72 \pm 0.04 $.

Apart from SNIa, there are some other cosmological data namely baryon acoustic oscillation, weak lensing, cluster and FRIIb radio galaxies which also indicate accelerated expansion of the universe. Details of this can be found in the references \cite{c12, c13, c14, c15}.

The age of the universe is another interesting evidence for the existence of the dark energy. If we compare age of the universe $(t_0)$ with the age of the oldest stellar populations of the universe $(t_s)$, then the universe must be older than the oldest $(t_0 > t_s)$ stellar population. The lower limit of the age needed to be satisfy $t_{0} > 11 -13 $ Gyr, which has been estimated both from age of the oldest stellar population \cite{c16, c17} and distance to last scattering surface measured by CMB anisotropies \cite{c18}. For a flat FRW universe, in absence of dark energy, one can obtain the age of the universe $t_0 = \frac{2}{3H_0}$. For $H_{0} ^{-1} = 9.776 h^{-1}$ Gyr, $0.64 < h < 0.80$, the age of the universe is $t_0 = 8 - 10$ Gyr. So a flat model without dark energy does not satisfy the lower age bound. But if one consider the same model with dark energy, when $\Omega_m ^{(0)}= 0.3$ and $\Omega_{\lambda} ^{(0)} = 0.7$, $t_0 = 0.964 H_0 ^{-1} = 13.1$ Gyr for $h= 0.72$. This easily satisfy the lower bound $t_0 > 13$ Gyr. Hence the existence of dark energy solves 
the age problem of the universe.

\section{Theoretical models to explain cosmic acceleration}

Theoretical models of cosmology are broadly classified in to two classes:

1) Modified gravity models: 

 In this type of models, the theory of gravity is modified from General Relativity. The interesting feature of these models is that the late time acceleration of the universe can be realized without considering any exotic matter component. 

2) Dark energy models: 

These types of models can also be called as " modified matter models". The energy momentum tensor of the Einstein equations contain an exotic matter component which can generate sufficient negative pressure to drive the accelerated expansion of the universe.

\subsection{Modified gravity models}  
There are many models of modified gravity \cite{m1}. An incomplete list includes $f(R)$ gravity \cite{m2,m3,m4,m5}, scalar-tensor theories \cite{m6,m7,m8,m9,m10,m11}, braneworld models \cite{m12}, Galileon gravity \cite{m13}, Gauss-Bonnet gravity \cite{m14, m15} and so on. 

\subsubsection{$f(R)$ gravity}

The action of $f(R)$ gravity model is 

\begin{equation}
S = \int d^4 x \sqrt{-g} [f(R) + \mathcal{L}_m],
\end{equation}
where $f(R)$ is an arbitrary function of $R$ and $\mathcal{L}_m$ are matter Lagrangian. Depending on the form of $f(R)$, these models can produce an early inflation or a late time acceleration. Models like $f(R) = R^2$ had some success in producing inflationary scenario. Inspired by that $f(R) = \frac{1}{R^n}$ with $n>0$ were proposed to drive a late time acceleration.

By varying the action with respect to the metric, the field equations are obtained as

\begin{equation}
G_{\mu \nu}  = (\frac{\partial f}{\partial R})^{-1} [\frac{1}{2} g_{\mu \nu} (f - \frac{\partial f}{\partial R} R) + \triangledown_{\mu}   \triangledown_{\nu} \frac{\partial f}{\partial R} - \square (\frac{\partial f}{\partial R}) g_{\mu \nu} ] = - 8 \pi G T_{\mu \nu},
\end{equation}
where $T_{\mu \nu} $ is the stress-energy tensor of the matter distribution.

For more details on cosmological dynamics in $f(R)$ see the references \cite{m24,m25,m26,m27,m28}. A phase space analysis of $f(R)$ models is given by Guo and Frolov \cite{m29}. Without considering an explicit form of $f(R)$, one can also constraints $f(R)$ models using observational data \cite{m17}.

Amendola, Polarski and Tsujikawa \cite{m16} showed that all power law $f(R)$ models are cosmologically unacceptable as they disagree with CMB and galaxy red-shift data.  Capozziello and Salzano in \cite{m18}, reviewed the latest results on observation test of $f(R)$ models.

\subsubsection{Scalar tensor theories}

Brans-Dicke theory \cite{m1, m19} is the simplest version of scalar tensor theory. A scalar field is coupled to the Ricci scalar ($R$). The Lagrangian  is written as 

$\mathcal{L} = \frac{\phi R}{2} - \frac{\omega_{BD}}{2 \phi} (\triangledown \phi)^2$,

where $\omega_{BD}$ is Brans-Dicke parameter. In the weak field limit Brans-Dicke theory yields the General Relativity results for $\omega_{BD} \longrightarrow \infty $. However, it was shown that the theory does not reduce to GR in the non linear regime \cite{m30}. With an additional potential $U(\phi)$ and $\omega_{BD} =0$, the generalized Brans-Dicke theory is equivalent to $f(R)$ theory in metric formalism \cite{m20, m21}. By transforming the action of the generalized Brans-Dicke theory with a conformal transformation, one can find that it is equivalent to a coupled quintessence scenario \cite{m22}. Banerjee and Pavon \cite{m23} showed that Brans-Dicke theory by itself can give rise to an accelerated expansion without any exotic matter. However, the drawback is that there is no transition from a decelerated to an accelerated phase.
\subsection{Dark energy models} 

\subsubsection{A. Cosmological constant} 

To achieve a static universe Einstein introduced cosmological constant in his equation in 1917. In 1929, Hubble discovered the accelerated expansion of the universe and Einstein dropped his idea of cosmological constant. But after the discovery of late time acceleration, cosmological constant has been considered as the most popular and simplest possible form of the dark energy.

 %Again discovery of accelerated expansion of the universe gave it a powerful entry in to physical cosmology as a possible driver of accelerated expansion.

Einstein tensor satisfy Bianchi Identity as $\triangledown_{\nu} G^{\mu \nu} = 0$ and the principle of conservation of energy, $\triangledown_{\nu} T^{\mu \nu} = 0$ is satisfied by energy momentum tensor. As the covariant derivatives of the metric is zero ($\nabla_{\alpha} g^{\mu \nu} =0$), the Einstein equations enjoy the freedom of addition of a term $\Lambda g^{\mu \nu}$ in the equations.

\begin{equation} \label{cc}
G^{\mu} _{\nu} \equiv R^{\mu} _{\nu} - \frac{1}{2} \delta^{\mu}_{\nu} R + \Lambda g_{\mu} ^{\nu}  = 8 \pi G T^{\mu} _{\nu}.
\end{equation}

By taking trace of the equation (\ref{cc}), we find $-R + 4 \Lambda = 8 \pi G T$. Using this relation in the equation (\ref{cc})  one can rewrite it as 
\begin{equation}
R_{\mu \nu} - \Lambda g_{\mu \nu} = 8 \pi G (T_{\mu \nu} - \frac{1}{2} T g_{\mu \nu}).
\end{equation}

In a FRW background the modified Einstein field equations are 
\begin{equation}
H^2 = \frac{8 \pi G}{3} \rho - \frac{k}{a^2} + \frac{\Lambda}{3},
\end{equation}
\begin{equation}
\frac{\ddot{a}}{a} = - \frac{4 \pi G}{3} (\rho + 3p) + \frac{\Lambda}{3}.
\end{equation}

It can be clearly seen from the above equations that sufficiently large positive value of cosmological constant ($ \Lambda > 4 \pi G (\rho + 3p)$) can drive the accelerated expansion of the universe. 
  
Though cosmological constant can drive the accelerated expansion of the universe, but it has it's own problem. Observationally $\Lambda$ is of the order of square of present value of Hubble's parameter $H_0$,

$\Lambda \simeq H_0 ^2 = (2.31 h \times 10^{-42} GeV )^2$ .

This corresponds to critical density,

$\rho_{\Lambda} = \frac{\Lambda m_{pl} ^2}{8 \pi} \simeq 10^{- 47} GeV^4$. 

From point of view of particle physics, cosmological constant arises as vacuum energy density  and estimated it to be of the order of $\rho_{vac} \simeq 10^{74} GeV^4$. So estimated value of cosmological constant is $10^{121}$ order larger than the observed value. This discrepancy between theory and observation is known as cosmological constant problem \cite{c19}. 

\subsubsection{B. Scalar field models of dark energy} 

\subsubsection{I. Quintessence}
In order to resolve the problem with a cosmological constant a dark energy with an evolution required attention.
Quintessence is a dynamical alternative to cosmological constant. In quintessence scalar field models of dark energy an ordinary scalar field $\phi$ is minimally coupled to gravity. The scalar field has negative pressure and it slowly rolls down the potential. The relevant action of the quintessence scalar field is written as 

\begin{equation}
S = \int d^4 x (- \frac{1}{2} (\bigtriangledown \phi)^2 - V(\phi)) \sqrt{- g} ,
\end{equation} 

where $(\bigtriangledown \phi)^2 = g^{\mu \nu} \partial_{\mu} \phi \partial_{\nu} \phi$ and $V(\phi)$ is the potential of the quintessence scalar field. The energy momentum tensor for the quintessence scalar field is given by 

\begin{equation}
T_{\mu \nu} = - \frac{2}{\sqrt{- g}} \frac{\delta S}{\delta g^{\mu \nu}} 
            = \partial_{\mu} \phi \partial_{\nu} \phi - g_{\mu \nu} [\frac{1}{2} g^{\alpha \beta} \partial_{\alpha} \phi \partial_{\beta} \phi  + V(\phi)].
\end{equation}

For a flat FRW metric the energy density and pressure of the scalar field are given by \\

\begin{equation}
\rho = - T^0 _0 = \frac{1}{2} \dot{\phi}^2 + V(\phi)
\end{equation}

and

\begin{equation}
p = T^i	 _i = \frac{1}{2} \dot{\phi}^2 - V(\phi)
\end{equation}

respectively. In equation(1.24), no summation is involved.

The Einstein field equations are 

\begin{equation}
H^2 = \frac{8 \pi G}{3} (\frac{1}{2} \dot{\phi}^2 + V(\phi)),
\end{equation}

\begin{equation}
\frac{\ddot{a}}{a} = - \frac{8 \pi G}{3} ( \dot{\phi}^2 - V(\phi)).
\end{equation}

If potential is flat enough so that the scalar field rolls slowly i,e. $V(\phi) > \dot{\phi}^2$, then it can drive the accelerated expansion of the universe. The equation of state of the scalar field is given by 
\begin{equation}
w_{\phi} = \frac{p}{\rho} = \frac{ \dot{\phi}^2 - 2 V(\phi)}{ \dot{\phi}^2 + 2 V(\phi)},
\end{equation}
thus $-1 \leq w_{\phi} \leq 1 $ . By varying the action with respect to $\phi$, one can find equation of motion of the scalar field as
\begin{equation}
\ddot{\phi} + 3 H \dot{\phi} + \frac{dV}{d \phi} = 0.
\end{equation}

Using the field equations and the wave equation the continuity equation is written in the integral form as

\begin{equation} \label{rho1}
\rho = \rho_0 ~ exp(- \int 3 (1 + w_{\phi}) \frac{\dot{a}}{a} dt).
\end{equation}

Here $\rho_0$ is integration constant. Depending on the value of $w_{\phi}$ the evolution of the quintessence scalar field density can be broadly classified in to three classes. When $V(\phi) << \dot{\phi}^2$ ,  $w_{\phi} \simeq 1$ , $\rho \propto a^{-6}$. In the slow roll limit  $V(\phi) >> \dot{\phi}^2$ , $w_{\phi} \simeq -1$ and $\rho = \rho_0$. In the intermediate cases $-1 \leq w_{\phi} \leq 1 $ , $\rho \propto a^{-m}$ and the accelerated expansion can be realized for $0 \leq m < 2$ \cite{c}.

\subsubsection{Slow roll approximation}
The concept of inflation is also based on the slow roll of the scalar field. The slow roll parameters $\epsilon$ and $\eta$ are often used to check the existence of inflationary solution of the model. The slow roll parameters are define as $\epsilon \equiv \frac{m_{pl} ^2}{16 \pi} (\frac{V^{\prime} (\phi)}{V(\phi)})^2$ and $\eta \equiv \frac{m_{pl} ^2}{8 \pi} \frac{V^{\prime \prime} (\phi)}{V (\phi)}$ \cite{c20}. The slow roll inflation occurs when $\epsilon << 1$ and $| \eta | << 1$ are satisfied.

In context of dark energy, due to the existence of barotropic fluid and dark energy these slow roll parameters can not be trusted completely. So, a redefinition of these slow roll parameters is required. By defining slow roll parameters in terms of $H$ and time derivative of $H$ such as $\epsilon = - \frac{\dot{H}}{H^2}$, one can have a very good check of accelerating solutions as it includes both barotropic fluid and dark energy \cite{c}.

\subsubsection{Example: Power law expansion} 
If we are interested in a power law type expansion then $a(t) \propto t^p$. The accelerated expansion occur for $p >1$. We obtain time derivative of Hubble's parameter $\dot{H} = -4 \pi G \dot{\phi}^2$. Then $V(\phi)$ and $\phi$ can be expressed in terms  of $H$ and $\dot{H}$ as $V(\phi) = \frac{3 H^2}{8 \pi G} (1 + \frac{\dot{H}}{3 H^2})$ and $\phi = \int dt (-\frac{\dot{H}}{4 \pi G})^{\frac{1}{2}}$. Hence, the potential corresponds to the power law expansion is given by $V(\phi) = V_0 exp(- \sqrt{\frac{16 \pi}{p}} \frac{\phi}{m_{pl}})$, where $V_0$ is a constant and $\phi \propto \ln t$. Provided that $p > 1$, this potential can be used as dark energy. There is no observation evidences to support a specific potential, so choices of different potential is allowed till now. It is worth to mention here that the first quintessence scalar field model of dark energy was found for the potential $V(\phi) = (\frac{M}{\phi})^{\beta}$, where M is a constant and $\beta$ is a number.

The deficiency of a power law expansion is that it can not account for the transition of the universe from a decelerated expansion phase to an accelerated expansion phase. It rather drives an ever accelerating or ever decelerating expansion, which contradicts observations \cite{c9} as well as theoretical requirement. 

\subsubsection{Tracker models}
A basic problem of the quintessence scalar field model is the coincidence problem \cite{c21}. Observations suggested that at present the energy density of the scalar field $\rho_{\phi}$ and matter energy density are comparable. But we know these two energy densities decay at different rates, so to be comparable today, the initial condition has to be very carefully manipulated. This issue is known as coincidence problem. Zlatev, Wang and Steinhardt \cite{c22} introduced the concept of a tracker field. They showed that the evolution of tracker field is blind to a very wide range of initial conditions and rapidly converge to a common evolution track of $\rho_{\phi} (t)$ and $\rho_m (t)$. Ultimately $\rho_{\phi} (t)$ overtakes the matter density $\rho_m (t)$ and the universe entered into an accelerated expansion phase. It can also be shown that a sufficiently stiff potential, satisfying $\frac{V^{\prime \prime} V}{V^\prime} \geq 1$, shows tracking behaviour \cite{c23}.

\subsubsection{II. K-essence cosmology}
In quintessence scalar field models of dark energy potential energy gives rise to the accelerated expansion of the universe. But it is also possible to have models of dark energy where the kinetic energy drives the acceleration. Originally kinetic energy driven acceleration was introduced to describe inflation of the early universe and this model was named as k-inflation  \cite{c24}. Chiba et al.  \cite{c25} first introduced this idea to describe late time acceleration, it was generalized by Armendariz-Picon et al. \cite{c26, c27} and called it as K-essence models of dark energy. The most general action for the K-essence models is 
\begin{equation}
S= \int d^4 x \sqrt{-g} P(\phi, X).
\end{equation} 
Here $X= - \frac{1}{2} (\bigtriangledown \phi)^2$ and the Lagrangian density $P(\phi, X)$ is in the form of pressure density. Idea of the K-essence is K.E dominated Lagrangian so as $X \longrightarrow 0$ the Lagrangian should also vanish i,e. $P(\phi, X) \longrightarrow 0$. So we can expand the Lagrangian density near $X=0$.
\begin{equation}
P(\phi, X) = K(\phi) X + L(\phi) X^2 + ...
\end{equation}
We have neglected the higher order terms of $X$. If we redefine the field in the form $ \phi_{new} = \int^{\phi_{old}} d\phi \sqrt{\frac{L}{|K|}}$, the Lagrangian transform into 

$P (\phi, X) = f(\phi) (-X + X^2)$,

where $\phi \equiv \phi_{new}, X \equiv X_{new} = (\frac{L}{|K|}) X_{old} $ and $f(\phi) = \frac{K^2 (\phi_{old})}{L (\phi_{old})}$.

In a flat homogeneous and isotropic universe one can write the pressure and energy density of the scalar field in the form 

$p_{\phi} = P(\phi, X) = f(\phi) (-X + X^2),$ 

$\rho_{\phi} = f(\phi) (-X + 3X^2)$.

So from the definition of the equation of state it is given by $w_{\phi} = \frac{p}{\rho} = \frac{1-X}{1-3X}$. The  accelerated expansion occurs when $w_{\phi} < - \frac{1}{3}$ i,e. $X< \frac{2}{3}$ and the equation of state of the cosmological constant can be obtained ($w_{\phi} = -1$) for $X = \frac{1}{2}$. The field equations of the system in flat FRW background is given by 

\begin{equation*}
H^2 = \frac{8 \pi G}{3} f(\phi) [-X + 3 X^2],
\end{equation*}

 \begin{equation*}
 \frac{\ddot{a}}{a} = \frac{8 \pi G}{3} [-2 X + 3 X^2] .
 \end{equation*}

The continuity equation can be written as  $\dot{\rho} + 3 H (\rho + p) = 0$. In a strongly radiation or matter dominated era, when $\rho_m >> \rho_\phi$, the Hubble's parameter evolves as $H = \frac{2}{3 (1 + w_m) (t -t_0)}$. For a constant equation of state of the scalar field i,e. $X =$ constant, $f(\phi) \propto (\phi - \phi_0)^{- \alpha}$, where $\alpha = \frac{2 (1+ w_\phi)}{1 + w_m}$. For $\alpha =2 $ the solution is a scaling solution $w_m = w_\phi$, $f(\phi) \propto (\phi - \phi_0)^{-2}$ and it is the boarder between decelerating and accelerating phase. When $w_{\phi} = -1$, i,e. dark energy dominated epoch, $f(\phi)$ is constant, $X= \frac{1}{2}$ and this corresponds to ghost condensate scenario \cite{c28}.  In order to build a viable model of dark energy $f(\phi)$ has to be fine-tuned to be of order  the present energy density of the universe. More general discussion on K-essence models can be found in the references \citep{c29}.

\subsubsection{III. Tachyon field}
 A tachyon has negative squared mass and speed greater than light (c). The concept of tachyon was already there as particles moving with a speed greater than $c$. A theoretical foundation was later provided by the string theory. The gas which is produced at the time of decay of D-brane, is pressure-less and has finite energy density and resembles classical dust \cite{c30, c31}. Interestingly tachyon has equation of state parameter (EOS) which smoothly varies between -1 to 0 and this leads to choose tachyon as a candidate of dark energy \cite{c32}. Different models shows tachyon with suitable potential can drive the acceleration \cite{c33}. As a consequence of negative squared mass of tachyon, it rests  on the maxima  of the potential and subjected to a very small perturbation it rolls down and it's condensation happen i,e. it gets real mass.
 
 The relevant action for the tachyon model is
 \begin{equation}
 S= - \int d^4 x ~ V(\phi) \sqrt{-det(g_{ab} + \partial_{\alpha} \phi \partial_{\beta} \phi )},
\end{equation}    

$V(\phi)$ is tachyonic potential. In a flat FRW universe the energy density $\rho$ and the pressure $p$ of the tachyon is written as 
\begin{equation*}
\rho = \frac{V(\phi)}{\sqrt{1 - \dot{\phi}^2}},
\end{equation*}
\begin{equation*}
p = - V(\phi) \sqrt{1 - \dot{\phi} ^2}.
\end{equation*}

Substituting $\rho$ and $p$ in Einstein's field equations, one can write the field equations for tachyon, 
\begin{equation*}
H^2 = \frac{8 \pi G V(\phi)}{3 \sqrt{1 - \dot{\phi} ^2}},
\end{equation*}

\begin{equation*}
\frac{\ddot{a}}{a} = \frac{8 \pi G V(\phi)}{3 \sqrt{1 - \dot{\phi} ^2}} (1 - \frac{3}{2} \dot{\phi}^2). 
\end{equation*}

The wave equation takes the form 
\begin{equation*}
\frac{\ddot{\phi}}{1- \dot{\phi}^2} + 3 H \dot{\phi} + \frac{1}{V} \frac{dV}{d\phi} =0.
\end{equation*}

The condition for accelerated expansion of the universe is $\dot{\phi}^2 < \frac{2}{3}$. The equation of sate, $w_{\phi} = \frac{\rho}{p} = \dot{\phi}^2 -1$. From the field equations, $0< \dot{\phi}^2 < 1$ , so $w_{\phi}$ varies between -1 and 0. From continuity equation the scalar field density $\rho \propto a^{-m}$ , with $0<m<3$.

By considering slow roll approximation, like quintessence scalar field models, $\epsilon = - \frac{\dot{H}}{H} << 1$, one can write the potential and the field in terms of $\dot{H}$ and $H$ as

\begin{equation*}
V = \frac{3 H^2}{8 \pi G} (1 +  \frac{2 \dot{H}}{3 H^2}),
\end{equation*}

\begin{equation*}
\phi = \int dt (- \frac{2 \dot{H}}{3 H^2})^{\frac{1}{2}}.
\end{equation*}

For example, if we consider power law expansion of the universe, i,e. $a \propto t^p$, potential has the form 
$V(\phi) = \frac{2 p}{4 \pi G} (1 - \frac{2}{3 p})^{\frac{1}{2}} \phi^{-2}$ and $\phi = \sqrt{\frac{2}{3p}} t$. There are many work on tachyonic dark energy models, see ref \cite{c} for more comprehensive study.

\subsubsection{IV. Phantom field}
Phantom field was first introduced by Hoyle \cite{c34, c35} in his steady state theory of the universe. Later it was used by Caldwell \cite{c35} as a dark energy candidate to drive the accelerated expansion of the universe. Phantom field has negative kinetic energy, so it roll up the potential. Motivation of phantom field comes from s-brane construction and the action is given as
\begin{equation}
S = \int d^4 x ~ L(\phi, X) ,
\end{equation}

$L(\phi, X) = -X - V(\phi)$, as it is mentioned above, the K.E of phantom field is negative. This is the difference of the phantom field models with the quintessence models of dark energy. The energy density and pressure of the phantom field are $\rho = - \frac{\dot{\phi}^2}{2} +{V(\phi)}$ and $p = - \frac{\dot{\phi}^2}{2} - {V(\phi)}$ respectively. The EOS parameter $w_{\phi} = \frac{\dot{\phi}^2 + 2 V(\phi)}{\dot{\phi}^2 - 2 V(\phi)}$. If $2 V(\phi) >> \dot{\phi}^2 $, $w_{\phi} < -1$. Phantom field models of dark energy predict different type of ultimate fate of the universe. The phantom field can roll up the potential due to it's negative K.E. But if the potential do not has any maxima then the universe will expand to infinity very rapidly in finite time, so each and every thing will rip apart and this scenario is known as Big Rip. With the potential having minima this kind of situation can be avoided. For example, $V(\phi) = V_0 [\cosh (\frac{\alpha \phi}{m_{pl}})]^{-1} $, where $\alpha$ is a constant has been suggested to avoid Big Rip. The field reaches maxima of the potential at $\phi = 0$ and after a damped oscillation it rests there as it rests, $\dot{\phi} = 0$ and $w_{\phi} = -1$.

 \subsubsection{C. Chaplygin Gas}
Kamenshchik, Moschella and Pasquir \cite{c37} in 2001 introduced Chaplygin gas in cosmology as a candidate for dark energy . Originally it was introduced by Chaplygin in 1904 in aerodynamics. The simplest Chaplygin gas has equation of state $p = \frac{A}{\rho}$, where $A$ is a positive constant. One of the initial attempts towards using such a gas as a dark energy was by Bento, Bertolami and Sen \cite{c50}. For generalized Chaplygin gas the equation of state is written as $ p = - \frac{A}{\rho^{\alpha}}$, $0< \alpha \leq 1$. Using the continuity equation and the generalized equation of state for Chaplygin gas the expression of density is $\rho= [A + \frac{B}{a^{3 (1 + \alpha)}}]^{\frac{1}{(1+\alpha)}}$. Here B is integration constant. For $\alpha =1$, the Chaplygin gas has very interesting asymptotic behaviour.  At the beginning when $a$ was small $\rho \sim \frac{\sqrt{B}}{a^3}$, $a << (\frac{B}{A})^{\frac{1}{6}}$, Chaplygin gas behaves as a pressure less dust. When $a >> (\frac{B}{A})^{\frac{1}{6}}$, $\rho \sim -p \sim \sqrt{A}$, so it has constant negative pressure and it resembles the form of the cosmological constant. Interestingly one can 
see unification of dark energy and dark matter in Chaplygin gas models. But Chaplygin gas models are throttled by the CMB anisotropy data \cite{c38, c39} . For generalized Chaplygin gas this problem can be solved but for a narrow parameter domain, $0  \leq \alpha \leq 0.2$ \cite{c38}. Further details of generalized Chaplygin gas models can be found in ref \cite{c} and the references therein. 

\section{Dynamical systems approach} \label{dy}
In this section we will briefly discuss the basics of dynamical systems analysis \cite{c40, c41, r13, r19}, focusing on relevant parts which we used in our works. 

Differential equations are the relation between functions and their derivatives. It was first discovered by Newton  in the middle of seventeenth century. Newton applied it in his theory of gravitation and found out solution for two body system i,e. movement of sun and earth. The three body problem i,e. movement of sun, earth and moon was a long standing issue and appeared to be impossible to solve. The break through came in late nineteenth century, when Poincaré found out a geometrical approach to qualitatively study a system rather that studying it quantitatively. This was the birth of a new subject called dynamical systems analysis.

A differential equation is said to be an ordinary differential equation (ODE) if there is only one independent variable. We will discuss ODE only, because in our work we shall encounter ODE.

Let us consider an ODE of the form

\begin{equation*}
\dot{x} = f(x),
\end{equation*}

where $\dot{x} \equiv \frac{dx}{dt}, x = (x_1,x_2......x_n) \in \mathbb{R}^n, f: \mathbb{R}^n \longrightarrow \mathbb{R}^n$. When a system of differential equations does not explicitly depends on time, the system is called an autonomous system. We will along with focus our discussion on autonomous systems. A non autonomous system can also be treated as an autonomous system by considering time(t) as a new variable i,e. $x_{n+1} = t$ and $\dot{x}_{n+1} = 1$. This will increase the dimension of the system by one. Though in most of the cases $f(x)$ is non-linear but we will start our discussion on linear systems as it will help us to understand non-linear systems. A linear differential equation is written as 

\begin{equation*}
\dot{x} = f(x) = Ax,
\end{equation*} 

where A is an $n \times n$ matrix. With the initial condition $x(0) = x_0$, the solution of this linear ODE can be written as 
\begin{equation*}
x(t) = e^{At} x_0,
\end{equation*}

$ e^{At}$ is an $n \times n$ matrix and can be expressed in terms of its Taylor series. There are different methods to find the exponential of a matrix depending on it's eigenvalues. For more details we refer to \cite{c40}.

In an $n$ dimensional space i,e. $x_1,x_2,....x_n$, the solutions are the curves in this space. This space is called phase space and the curves are called phase trajectories. To find qualitative behaviour of a dynamical system it is very important to find the fixed points of the system. Fixed points are the points where the solutions are stationary. Mathematically, fixed points or equilibrium points are simultaneous solutions of the equation $f(x) = 0$. Depending on the stability, the fixed points are classified as stable, unstable and saddle. Stability of a fixed point can be determine by perturbing the system from the fixed point. If the system comes back to the fixed point then the fixed point is a stable fixed point and if the system never comes back, the fixed point is an unstable fixed point. But if these two behaviour depends on the direction of perturbation then the fixed point is a saddle one.

 If all the eigenvalues of the matrix $A$ at a fixed point are negative then the fixed point is an attractor or a stable fixed point. If all the eigenvalues are positive then then the fixed point is a repeller or an unstable one. If there is mixture of both positive and negative eigenvalues, the fixed point is essentially a saddle. For a 2D system it is easy to visualize the phase portraits, so we will discuss 2D system in more details. For a 2D system A is a $2 \times 2$ matrix and if $\lambda_1, \lambda_2$ are two eigenvalues of A, the system can be classified in to the following  classes,
 
 Case 1: If $\lambda_1 = \lambda_2 < 0$, the fixed point is an attracting focus. 
 
 Case 2: If $\lambda_1 < \lambda_2 < 0$, the fixed point is an attracting node.
 
 Case 3: If $\lambda_1 < \lambda_2 = 0$, attracting line.
 
 Case 4: If $\lambda_1 < 0 < \lambda_2$; then the fixed point is a saddle one.
 
 Case 5: If the eigenvalues are imaginary of the form $\lambda = a \pm i b$, then depending on $a$ the phase portrait near the fixed point is either spirally inward or outward. For $a<0$, spirally inward and for $a>0$ spirally outward. When $a=0$, the fixed point is a center.

The fixed points can be classified in two other classes, hyperbolic and non-hyperbolic fixed points. Hyperbolic fixed points are those which has $Re(\lambda) \neq 0, i = 1,2...n$. Otherwise the fixed points are non hyperbolic. 
 
  Now we will discuss phase space behaviour of non-linear systems. Let us write a non-linear system of differential equations as  
 \begin{equation} \label{1}
  \dot{x} = f(x),
\end{equation}

   where $f: E \longrightarrow \mathbb{R}^n$ and $E$ is an open subset of $\mathbb{R}$. For a non-linear system the differential equations can not be written in a matrix form like the linear system. But near a hyperbolic fixed point, a non-linear system can be linearized as 
  \begin{equation} \label{2}
   \dot{x} = A x.
  \end{equation}
 Here, $A=Df(x^*)$ is the Jacobian matrix of the system. 
 
 Let us consider $x^*$ be a fixed point and $\zeta(t)$ be the perturbation from the fixed point i,e. $\zeta(t) = x(t) - x^*$. We can find the time derivative of the perturbation from the time derivative of $\zeta$.
\begin{equation*}
\dot{\zeta} = \frac{d}{dt} (x(t) - x^*) = \dot{x} = f(x) = f(x^* + \zeta).
\end{equation*}
We can do Taylor's expansion of the term 

$f(x^* + \zeta) = f(x^*) + \zeta Df(x^*) + ..$

As $\zeta$ is very small so quadratic and higher order terms are neglected.

$Df(x) = (\frac{\partial f_i}{\partial x_j})$ , $i,j = 1,...n    .$

So $\dot{\zeta} = \zeta Df(x^*)$, as $f(x^*) = 0$ from the definition of the fixed point. This is a linear differential equation and it is called the linearisation  of the system near the fixed point. Stability of a fixed point can be found from the eigenvalues of the Jacobian matrix $A = Df(x) = (\frac{\partial f_i}{\partial x_j})$.

Before beginning our discussion, we will discuss some of the basic theorems and methods of non-linear dynamics \cite{c40}.
 
\subsection{Existence of Uniqueness Theorem}
 Let E be an open subset of $\mathbb{R}^n$ containing $x_0$ and assume that $ f \in C^1 (E)$. Then there exists an $a> 0$ such that the initial value problem
 
$\dot{x} = f(x)$ with $x(0) = x_0$, 

has a unique solution $x(t)$ on the interval $[-a,a]$.

\subsection{The Stable Manifold Theorem}
Let E be an open subset of $\mathbb{R}^n$ containing the origin, let $f \in  C^1(E)$, and let $\phi_t$ be the flow of the nonlinear system (\ref{1}). Suppose that $f(0) = 0$ and that $Df(0)$ has k eigenvalues with negative real part and $n-k$ eigenvalues with positive real part. Then there exists a k-dimensional differentiable manifold S tangent to the stable subspace $E^s$ of the linear system (\ref{2}) at 0 such that for all $t \geq 0$ , $\phi_t(S) \subset S$
and for all $x_0 \in S $.

$$\lim_{t \to \infty} \phi_t (x_0) = 0;$$

and there exists an $n-k$ dimensional differentiable manifold U tangent to
the unstable subspace $E^U$ of (\ref{2}) at 0 such that for all $t \geq 0, \phi_t (U) \subset U$ and for all $x_0 \subset U$

 $$\lim_{t \to \infty} \phi_t (x_0) = 0.$$

\subsection{The Center Manifold Theorem}
Let $f \in C^r (E)$ where E is an
open subset of $\mathbb{R}^n$ containing the origin and $r \geq 1$. Suppose that $f(x_0) = 0$ and that $Df(0)$ has k eigenvalues with negative real part, j eigenvalues with positive real part, and $m= n-k-j$ eigenvalues with zero real part. Then there exists an m-dimensional center manifold $W^c(0)$ of class $C^r$ tangent to the center subspace $E^c$ of the linear system (\ref{2}) at 0, there exists a k-dimensional stable manifold $W^s (0)$ of class $C^r$ tangent to the stable subspace $E^s$ of the linear system (\ref{2}) at 0 and there exists a j-dimensional unstable manifold $W^u(0)$ of class $C^r$ tangent to the unstable subspace $E^u$ of the linear system (\ref{2}) at 0; furthermore, $W^c(0)$, $W^s (0)$ and $W^u (0)$ are invariant under the flow $\phi_t$ of the non-linear system (\ref{1}).

\subsection{The Hartman-Grobman Theorem}
Let E be an open subset of $\mathbb{R}^n$ containing the origin, let $f \in C^1 (E)$, and let $\phi_t$ be the flow of the nonlinear system (\ref{1}). Suppose that $f(x_0) = 0$ and that the matrix $A = Df(0)$ has no eigenvalue with zero real part. Then there exists a homeomorphism H of an open set U containing the origin onto an open set V containing the origin such that for each $X_0 \in U$, there is an open interval $I_0 \subset R$ containing zero such that for all $x_0 \in U$ and $t \in I_0$

$H \circ \phi_t(x_0) = e^{At} H(x_0)$

i.e., H maps trajectories of (\ref{1}) near the origin onto trajectories of (\ref{2}) near the origin and preserves the parametrization by time.

\subsection{Liapunov function}

Finding out the stability of a non-hyperbolic fixed point is typically more difficult than a hyperbolic one. Liapunov discovered a function for deciding the stability of a non-hyperbolic fixed point. This function is known as Liapunov function and widely used to find the stability of non-hyperbolic fixed points.

Let E be an open subset of $\mathbb{R}^n$ containing $x_0$. Suppose that $f \in C^1 (E)$ and that $f(x_0) = 0$. Now, if there exists a real valued function $V \in C^1 (E)$, called the Liapunov function, satisfying $V(x_0) = 0$ and $V(x) > 0$ if $x \neq x_0$. Then (a) if $\dot{V}(x) \leq 0$ for all $x \in E$, $x_0$ is stable; (b) if $\dot{V}(x) < 0$ for all $x \in E \sim \lbrace{x_0}\rbrace$, $x_0$ is asymptotically stable; (c) if $\dot{V}(x) > 0$ for all $x \in E \sim \lbrace{x_0}\rbrace$, $x_0$ is unstable.

\par
~~~

According to Hartman Grobman theorem we can not use linear stability analysis to find the stability of a non-hyperbolic fixed point. There is a special kind of non-hyperbolic fixed point called normally hyperbolic fixed point \cite{r13} whose stability can be easily found out from its eigenvalues. If each point of a non-isolated fixed point has at least one zero eigenvalue then the set of non-isolated fixed point is called normally hyperbolic fixed point. The stability of a normally hyperbolic fixed point can be easily determine from the sign of the remaining eigenvalues of the fixed point. If remaining eigenvalues are negative, then the fixed point is an attractor otherwise a repeller. Stability of a fixed point can also be found out numerically. One can perturb the system from the fixed point to study the behaviour of the system, if it comes back to the fixed point then the fixed point is stable otherwise it is unstable.

\section{Application of dynamical systems analysis to Cosmology}
The system of equations of commonly used cosmological models are non-linear differential equations. It is not always possible to find exact solutions of a non-linear system. But the dynamical systems approach to study non-linear systems can help us to know the qualitative behaviour of the system. Usually normalized dimensionless new variables are introduced with a dimensionless time variable to write the system as an autonomous system. These variables are directly related to physically observable quantities and they are well behaved. By finding the fixed points of the system and their stability  one can qualitatively study the beginning and the possible ultimate fate of the universe. As any heteroclinic solution starts from an unstable fixed point and ends at a stable fixed point, so unstable fixed points have the possibility of being the beginning of the universe and stable fixed points would be the ultimate fate of the universe.

This type of analysis is not new in both general relativity and cosmology. Almost all important models of general relativity and cosmology has been analysed in the light of dynamical systems analysis. The incomplete list includes: modified gravity, scalar tensor theory, Bianchi type models and non-minimally coupled scalar field models. 

In modified gravity models Ricci scalar $R$ in Einstein-Hillbert's action is replaced by $f(R)$ i,e. by analytical function of $R$. Modified gravity models quite successfully explain accelerated expansion of the universe. Detailed dynamical systems study of generalized gravity in a homogeneous and isotropic de Sitter space has been done by Faraoni \cite{c42}. By considering equivalent scalar field description of $f(R)$ gravity, dynamical system analysis has been done by Guo and Frolov \cite{c43}. In a very recent work, Shabani and Farhoudi \cite{c44} investigated late time attractor solution for $F(R,T)$ cosmological models considering minimally, non-minimally and pure non-minimally Lagrangian.  

Dynamical systems analysis also has been used in scalar-tensor theory of gravity. Qualitative analysis of scalar-tensor theory with exponential potential shows the existence of initial and final inflationary behaviour and also suggested current universe as an attractor in the phase space \cite{c45}. S J kolitch and D M Eardley \cite{c46} analyse false vacuum as a special case in FRW back ground and showed existence of bifurcation in the system. In a general scalar tensor theory of gravity in FRW back ground only one fixed point is compatible with the solar system PPN constrains \cite{c47}.

%For a detail analysis of Bianchi type models we refer to books by AA Coley \cite{coley2003dynamical} and by Wainwright \cite{wainwright2005dynamical}.

\subsection{Dynamical System analysis of quintessence scalar field models}

Let us consider quintessence scalar field models minimally coupled to matter Lagrangian \cite{c}. Lagrangian density of the scalar field is given by 

$ L = \frac{1}{2} \epsilon \dot{\phi}^2 + V(\phi),$

where $V(\phi)$  is the scalar field potential and $\epsilon$ has been introduced to differentiate between ordinary scalar field models and phantom scalar field models. If $\epsilon = +1$, the scalar field is ordinary scalar field and if $\epsilon = -1$, the scalar field is phantom scalar field. The field equations are

\begin{equation} \label{ps1}
H^2 = \frac{k^2}{3} [\frac{1}{2} \epsilon \dot{\phi}^2 + V(\phi) + \rho_m]
\end{equation}

\begin{equation}
\dot{H} = - \frac{k^2}{3} [\epsilon \dot{\phi}^2 + (1 + w_m) \rho_m]
\end{equation}

and the wave equation; 

\begin{equation}
\epsilon \ddot{\phi} + 3 H \dot{\phi} + \frac{dV}{d\phi} = 0.
\end{equation}

To write the system as a set of autonomous system of equations, we introduce new dimensionless variables \cite{r13}, 

$x = \frac{k \dot{\phi}}{\sqrt{6} H}, y = \frac{k \sqrt{V}}{\sqrt{3} H}, \lambda = - \frac{V,_{\phi}}{k V}$ and $\Gamma = \frac{V V,_{\phi \phi}}{V^2,_{\phi}}$. Here $V_  {\phi} = \frac{dV}{d \phi} $ and the system reduces to the following autonomous system, 

\begin{equation*}
x^{\prime} = -3x + \frac{\sqrt{6}}{2} \epsilon \lambda y^2 + \frac{3}{2} x [(1 - w_m) \epsilon x^2 + (1 + w_m) (1-y^2)],
\end{equation*}

\begin{equation*}
y^{\prime} = - \frac{\sqrt{6}}{2} \lambda x y + \frac{3}{2} y [(1-w_m) \epsilon x^2 + (1+w_m) (1-y^2)] ,
\end{equation*}

\begin{equation*}
\lambda^{\prime} = - \sqrt{6} \lambda^2 (\Gamma -1 ) x,
\end{equation*}

where `prime' is the differentiation w.r.t $N = \ln a$. One can use (\ref{ps1}) to write the constrain equation as

\begin{equation*}
\epsilon x^2 + y^2 + \frac{k^2 \rho_m}{3 H^2} = 1.
\end{equation*} 

The equation of state $w_{\phi}$ and density parameter $\Omega_{\phi}$ are expressed in terms of new variable as

$w_{\phi} = \frac{p_{\phi}}{\rho_{\phi}} = \frac{\epsilon x^2 -y^2}{\epsilon x^2 + y^2},$

$\Omega_{\phi} = \frac{k^2 \rho_m}{3H^2} = \epsilon x^2 + y^2 $.

Total effective equation of state parameter;

\begin{equation*}
w_{eff} = \frac{p_{\phi} + p_m}{\rho_{\phi} + \rho_m} = w_m + (1-w_m) \epsilon x^2 - (1+w_m) y^2.
\end{equation*}

Accelerated expansion happens when $w_{eff} < -  \frac{1}{3}$. Depending on the value of $\Gamma$, the system can be classified in to two classes \cite{c}. When $\Gamma =1 $, the potential is exponential and when $\Gamma \neq 1$ the potential is  a non exponential function of $\phi$. 

First we will discuss normal scalar field models i,e. $\epsilon = +1$. 

\subsubsection{Exponential potential ($V(\phi) = V_0$)}

Effectively the system reduces to a 2D system. Fixed point of the system with their qualitative behaviour is given in the following table \cite{c48, r12},

%\squeezetable
\begin{tabular}{|c|c|c|c|c|c|c|}
\hline 
Name & $x$ & $y$ & Existence & Stability & $\Omega_{\phi}$ & $\gamma_{\phi}$ \\ 
\hline 
(a) & $0$ & $0$ & All $\lambda$ and $\gamma$ & Saddle($ 0 < \gamma < 2 $ )& $0$ & $--$ \\ 
\hline 
$(b_1)$ & 1 & 0 & All $\lambda$ and $\gamma$ & \makecell{Unstable Node($\lambda < \sqrt{6}$), \\ Saddle ($\lambda> \sqrt{6}$)} & 1 & 2 \\ 
\hline 
($b_2$) & -1 & 0 &  All $\lambda$ and $\gamma$ & \makecell{Unstable Node($\lambda > -\sqrt{6}$),\\ Saddle ($\lambda <  -\sqrt{6}$)} & 1 & 2 \\ 
\hline 
(c) & $\frac{\lambda}{\sqrt{6}}$ & $[1- \frac{\lambda^2}{\sqrt{6}}]^{\frac{1}{2}}$ & $\lambda^2 < 6$ & \makecell{Stable Node ($\lambda^2 < 3 \gamma$) \\ and \\ Saddle point ($3 \gamma < \lambda^2 < 6$)} & 1 & $\frac{\lambda^2}{3}$ \\ 
\hline 
(d) & $(\frac{3}{2})^{\frac{1}{2}} \frac{\gamma}{\lambda}$ & $[3(2-\gamma) \frac{\gamma}{2 \lambda^2}]^{\frac{1}{2}}$ & $\lambda^2 > 6$ & \makecell{Stable node $(3 \gamma < \lambda^2 < \frac{24 \gamma^2}{(9\gamma -2)})$ \\ and \\ stable spiral $(\lambda^2 >\frac{24 \gamma^2}{(9\gamma -2)})$} & $\frac{3 \gamma}{\lambda^2}$ & $\gamma$ \\ 
\hline 
\end{tabular} 

~~~~~~~~~~~

Eigenvalues of the fixed points are :

\textbf{Point (a):} 

$\mu_{1} = - \frac{3}{2} (2-\gamma), \mu_{2} = \frac{3}{2} \gamma$

\textbf{Point ($b_1$):}

$\mu_1 = 3- \frac{\sqrt{6}}{2} \lambda , \mu_2 = 3(2-\gamma)$

\textbf{Point ($b_2$):}

$\mu_1 = 3+ \frac{\sqrt{6}}{2} , \mu_2 = 3(2 - \gamma) $

\textbf{Point (c):}

$\mu_1 = \frac{1}{2} (\lambda^2 - 6), \mu_{2} = \lambda^2 - 3 \gamma$

\textbf{Point (d):}

$\mu_{1,2} = -\frac{3(2-\gamma)}{4} [1 \pm \sqrt{1-\frac{8 \gamma (\lambda^2 - 3\gamma)}{\lambda^2(2-\gamma)}}]$

As both the eigenvalue have opposite sign, fixed point (a) is a saddle point in a fluid dominated region $0 < \gamma < 2$. Depending on the value of $\lambda$ fixed point $b_1$ and $b_2$ are either unstable node or saddle point. For $\lambda^2 < 3 \gamma$, the fixed point (c) is stable node and for $ 3 \gamma < \lambda^2 < 6$ , it is saddle. One can find existence of scaling solution at the fixed point (d). The energy density of the scalar field is proportional to the barotropic fluid density. But this case is not physical as it violates the condition $\Omega_{\phi} \leq 1$. The fixed point is stable spiral in the region $\lambda^2 > 24 \gamma^2 /(9\gamma -2))$. Though the scaling solution of the fixed point (d) is not physically acceptable, but it can give us intermediate state in which the energy density of the scalar field decreases proportionally to background fluid. For more discussion on this we refer to \cite{c48}.

\subsubsection{Non exponential potential:}

When $\Gamma >1$, $\lambda$ decreases to zero. Therefore the slope of the potential becomes more flat. Potential $V(\phi)$ dominates over the kinetic term giving rise to late time accelerated expansion of the universe. This condition is the tracking condition in which scalar field density track the back ground fluid density.

When $\Gamma < 1$, the slope of the potential increases to infinity and we do not have any late time acceleration. 

In order to get complete dynamical behaviour of the system, we need to evolve the whole system simultaneously. Fixed point of constant $\lambda$ can be thought as instantaneous critical point \cite{r12}. More details can be found in ref \cite{c}.

\subsection{Phantom field} 

The phantom field (\cite{c} and references therein) has negative kinetic energy i,e. $\epsilon = -1$ and the Lagrangian of the phantom field looks like;

$L = - \frac{1}{2} \dot{\phi}^2 + V(\phi)$. 

Let us first consider exponential potential. The fixed points are given in the following table

~~~~~~~

%\squeezetable
\begin{tabular}{|c|c|c|c|c|c|c|}
\hline 
Name & $x$ & $y$ & Existence & Stability & $\Omega_{\phi}$ & $\gamma_{\phi}$ \\ 
\hline 
(a) & $0$ & $0$ & No for $0 \leq \Omega \leq 1 $ & Saddle point & $0$ & $-$ \\ 
\hline  
(b) & $-\frac{\lambda}{\sqrt{6}}$ & $[1+ \frac{\lambda^2}{\sqrt{6}}]^{\frac{1}{2}}$ & All values & Stable Node & $1$ & $-\frac{\lambda^2}{3}$ \\ 
\hline 
(c) & $(\frac{3}{2})^{\frac{1}{2}} \frac{\gamma}{\lambda}$ & $[-3(2-\gamma) \frac{\gamma}{2 \lambda^2}]^{\frac{1}{2}}$ & $\gamma < 0$ & Stable point for $\lambda^2 > - 3 \gamma$ & $\frac{- 3 \gamma}{\lambda^2}$ & $\gamma$ \\ 
\hline 
\end{tabular}

~~~~~~~~~~~~~~~~~~~~~~~~~`

One can note the disappearance of two fixed points $(x,y) = (\pm 1, 0)$, which exist in the quintessence models. The eigenvalues of the fixed point are,\\

\textbf{Point (a):} \\

$\mu_{1} = - \frac{3}{2} (2-\gamma), \mu_{2} = \frac{3}{2} \gamma$, corresponds to a saddle point.\\

\textbf{Point (b):}

$\mu_1 = - (\lambda^2 + 6)/2, \mu_2 = - \lambda^2 - 3 \gamma$. The fixed point is scalar field dominated and the equation of state parameter of phantom scalar field is given by $w_{\phi} = -1 - \frac{\lambda^2}{3}$, which is always less than $-1$. For $\gamma > 0$, the fixed point is stable. The fixed point (c) exist only when $\gamma<0$ and also it has the scaling behaviour. The eigenvalues of the matrix are

$\mu_{1,2} = - \frac{3 (2 - \gamma)}{4} [1 \pm \sqrt{1- \frac{8 \gamma (\lambda^2 + 3\gamma)}{\lambda^2(2 - \gamma)}}]$.

So the fixed point is saddle in nature for $\lambda^2 > -3 \gamma$.

For a dynamical $\lambda$ i,e. non exponential potential, the fixed points of constant $\lambda$ can be considered as instantaneous fixed point. In case of the potential of the form $V = V_0 [\cosh (\frac{\alpha \phi}{m_{pl}})]^{-1}$, where $\alpha$ is constant, $\lambda$ approaches to zero as the field settles on the top.

\section{Present work}

In this thesis we looked at stability of various dark energy models. The models of specific interest are quintessence scalar field models of dark energy, chameleon scalar field models and holographic dark energy models.

We have investigated stability of the quintessence scalar field solutions with and without a tracking condition.  Einstein's field equations for a spatially flat FRW metric are written as an autonomous system of equations.  In one part, standard dynamical systems analysis has been done to qualitatively understand the system near the fixed points with tracking. We also found that only one fixed point is of physical interest. Two specific potentials has been chosen as examples. These two potentials can drive the accelerated expansion of the universe if the system is in a neighborhood of the fixed point. For a region in parameter space this accelerated expansion is stable.

In the second part the tracking condition has been relaxed and the system has been allowed to evolve numerically for two type of specific potentials. Boundary values are chosen so that they are consistent with the present observations. In both cases, we found solutions starting from an unstable fixed point and evolving in to a stable fixed point. From this behaviour of the the solutions, we qualitatively studied the beginning, present accelerated expansion and ultimate fate of the universe. This discussion form the second chapter.

In the third chapter we provide a diagnostic of fixed points and their stability in  chameleon scalar field models for all combinations of the potential and the coupling of the scalar field with the fluid distribution. In chameleon mechanism chameleon field is coupled to matter Lagrangian in such a way that it's effective mass depends on the local matter density. We constructed an autonomous system of equations from Einstein's field equations for spatially flat FRW metric. Depending on the functional form of the potential and the coupling, we classified our system into four classes. Phase space analysis for each classes has been done. Nature
of the fixed points of the system and asymptotic behaviour of the system have also been investigated. 

In the fourth and the last chapter, the holographic dark energy models are analysed as an autonomous system. We have assumed the decay rate to be a function of $H$. The system is classified in to two different classes depending on the functional form of the decay rate.  The bifurcation in the system  has also been noted. When decay rate is a linear function of $H$, there is no transition from decelerated expansion phase to accelerated expansion phase corresponding to a stable solution. 
%*******************************************************************************
%****************************** Second Chapter *********************************
%*******************************************************************************

\chapter{Dynamical Systems Analysis of Quintessence Scalar Field Models }

\ifpdf
    \graphicspath{{Chapter2/Figs/Raster/}{Chapter2/Figs/PDF/}{Chapter2/Figs/}}
\else
    \graphicspath{{Chapter2/Figs/Vector/}{Chapter2/Figs/}}
\fi
\section{Introduction:}
The accelerated expansion of the universe is now a widely accepted reality. This strange behaviour has strong observational evidences \cite{r1}. It is also found that the acceleration is a recent phenomena, started well within the matter dominated regime\cite{r20}.  After a long stint of decelerated expansion, the universe has entered in to this accelerated expansion phase \cite{r2}. But the driver of this acceleration is neither detected observationally nor has any single firmly accepted theoretical model. A re-entry of the  cosmological constant $\Lambda$ does very well in explaining this recent acceleration, but $\Lambda$ suffers from serious discrepancy  between theoretical prediction and observational bound \cite{r3}. Amongst a host of alternatives, a quintessence field\cite{r4} is the most talked about option. A scalar field, minimally coupled to gravity, endowed with a potential, can supply the required negative pressure which drives the accelerated expansion. This type of scalar field is called quintessence scalar field. However neither an observational evidence nor any theoretically supported prediction strongly points towards a particular  form of quintessence potential, and this actually explains the existence of so many potentials in theoretical models. 

 Along with the problem of finding a suitable candidate of dark energy there are two more problems, namely, cosmic coincidence problem and fine tuning problem. The cosmic coincidence problem arises from the question ``why at the present epoch this dark energy bears a constant ratio, of order unity, to the dark matter ". If the cosmological constant is the solution to the dark energy problem, it should be extremely fine tuned to the density of matter or radiation at the very early stage\cite{r5}.  

 %Though the absence of a sound theoretical basis in favour of any particular quintessence potential has no serious remedy but in \cite{r}, it has been shown that two scalar fields with an arbitrary potential can drive the accelerated expansion of the universe.

Tracking quintessence scalar field, which `tracks' the matter density is one of the way to address the second problem. In these models the dark energy evolves at almost the same rate but below the level of the dark matter but slowly catches up so as to eventually lead the scenario only at a later stage \cite{r5, r6}. 

 The tracking behaviour of the quintessence scalar field has naturally attracted lot of attentions. Johri \cite{r7} looked for quintessence potentials which are trackers. A tracker solution which acts as a quintessence has been discussed by Urena-Lopez etal \cite{r8}. Reconstructing a quintessence potential, Sahlen, Liddle and Parkinson checked its tracking viability \cite{r9}. Dodelson, Kaplinghat and Stewart \cite{r10} showed that the coincidence problem can be resolved considering an oscillating potential as the quintessence potential. Using the observational data Wang, Chen and Chen \cite{r11} found that tracker potentials of the form $exp(\frac{M_p}{\phi})$ or [$exp(\frac{\gamma M_p}{\phi}) - 1 $] are clearly unsuitable. Amongst others, a use of two scalar fields can also address this problem, a wide range of initial conditions can lead to a late time acceleration \cite{r}. 

 Dynamical systems analysis of various cosmological models are not new. A volume edited by Ellis and Wainwright and work of Coley are two very comprehensive  reviews \cite{r13}. Ng, Nunes and Rosati \cite{r12} discussed attractor solutions in scalar field cosmology and some of their applications in an inflationary scenario and also in the quintessence scenario. A dynamical systems analysis for an FRW universe with a number of non minimally coupled scalar fields has been discussed by Lara and Castagnino \cite{r14}. Gunzig et al \cite{r15} presented dynamical systems analysis of a spatially flat FRW universe with a scalar field in the context of inflation at an early epoch. In the frame work of dynamical systems Carot and Collinge also discussed inflationary scalar field cosmologies \cite{r16}. This method has been employed in  phantom cosmology by Urena-Lopez \cite{r17}. Recently Fang et at \cite{r31} gave a dynamical systems study of Phantom, Tachyonic and K-essence fields including a discussion about  non-canonical quintessence field. This method also found application in axionic quintessence by Kumar, Panda and Sen\cite{r22} and thawing dark energy by Sen, Sen and Sami\cite{r23}. Brans-Dicke theory and also its equivalence with  a minimally coupled scalar field have also been discussed\cite{r18} in the framework of dynamical systems.
 
\section{The dynamical system}

For a spatially flat Robertson Walker Universe, the metric is given by
\begin{equation} \label{eq.metric}
ds^2 = dt^2 - a^2(t)(dr^2 + r^2 d\Omega ^2).
\end{equation}

If the universe is filled with a barotropic perfect fluid with an equation of state $P_B = (\gamma -1) \rho_B$ and a scalar field distribution, Einstein field equations are written as
\begin{equation} \label{H11}
H^2= \frac{8 \pi G}{3}(\rho_{B} + \frac{1}{2} \dot{\phi}^2 + V(\phi)),
\end{equation}
and
\begin{equation} \label{eq.dH}
\dot{H}=-\frac{8 \pi G}{2}(\gamma \rho_{B} + \dot{\phi}^2).
\end{equation}
Here $H = \frac{\dot{a}}{a}$ is the Hubble's parameter, $\rho_B$ and $P_B$ are energy density and pressure of the perfect fluid respectively, $\phi$ is the scalar field and $V(\phi)$ is the scalar potential. By varying the action of the quintessence scalar field with respect to $\phi$, one can get the wave equation 
\begin{equation} \label{eq.phi}
\ddot{\phi} + 3 H \dot{\phi} = - \frac{dV}{d\phi}.
\end{equation}

The conservation equation for the barotropic fluid is
\begin{equation} \label{eq.rho}
\dot{\rho}_{B}= -3 \gamma H \rho_{B}.
\end{equation}
Not all these equations are independent as any of the last two equations can be derived from remaining three. So out of these four equations, we chose (\ref{H11}), (\ref{eq.dH}) and (\ref{eq.phi}) as the system of equations to be solved. The energy density and pressure of the scalar field are respectively
\begin{equation} \label{eq.rhop}
\rho_{\phi}= \frac{1}{2} {\dot{\phi}}^2 + V(\phi),~~ p_{\phi}= \frac{1}{2} {\dot{\phi}}^2 - V(\phi).
\end{equation}
Energy density and pressure of the scalar field are connected, in analogy with a perfect fluid, by an equation of state $p_{\phi} = (\gamma_{\phi} - 1) \rho_{\phi}$. To write the system of equations as a system of autonomous equations, we introduce new dimensionless variables $x$ and $y$ as
\begin{equation} \label{eq.trans}
x^2=\frac{k^2 {\phi^{\prime}}^2}{6}, ~~ y^2 = \frac{k^2 V}{3 H^2} ,
\end{equation}
where a prime denote a differentiation with respect to $N=ln(\frac{a}{a_0})$ and $k^2=8 \pi G$. $a_0$ is the present value of the scale factor. Equations (\ref{H11}), (\ref{eq.dH}) and (\ref{eq.phi}) in terms of new variable can be written as a 3-dimensional autonomous system of equations,
\begin{equation} \label{eq.x}
 x^{\prime} = -3x + \lambda \sqrt{\frac{3}{2}} y^2 +\frac{3}{2} x [2 x^2 + \gamma (1-x^2-y^2)]~~~,
\end{equation}

\begin{equation} \label{eq.y}
y^{\prime} =- \lambda \sqrt{\frac{3}{2}} xy + \frac{3}{2} y [2 x^2 + \gamma (1-x^2-y^2)] ~~~,
\end{equation}
and 
\begin{equation} \label{eq.lambda}
\lambda^{\prime} = -\sqrt{6} \lambda^2 (\Gamma -1) x ~~~,
\end{equation} 
where $ \lambda = - \frac{1}{k V} \frac{dV}{d \phi}$ and $\Gamma =  V\frac{d^2 V}{d \phi^2} / (\frac{dV}{d \phi})^2$. $\Gamma$ is called the tracker parameter. For tracking, $\Gamma \simeq 1$, the scalar field energy density tracks the background fluid density. The equation of state parameter and density parameter of the scalar field are written in terms of new variables as 

\begin{equation} \label{eq.gamma}
\gamma_{\phi}=\frac{\rho_{\phi} + p _{\phi}}{\rho_{\phi}} = \frac{{\dot{\phi}}^2 }{\frac{{\dot{\phi}}^2}{2} + V} = \frac{2 x^2}{x^2 + y^2},
\end{equation}

\begin{equation} \label{eq.omega}   
\Omega_{\phi}=\frac{k^2 \rho_{\phi}}{3 H^2}= x^2 + y^2.
\end{equation}

For a spatially flat universe density parameter of barotropic fluid ($\Omega_B$) and density parameter of scalar field $(\Omega_{\phi})$ are restricted by the constraint equation $\Omega_B +\Omega_{\phi} =1 $. A near tracking situation i.e $\Gamma \simeq 1$ leads to $\lambda^{\prime} \simeq 0$ i.e $\lambda $ is nearly a constant. In the case our system effectively reduces to a 2-dimensional autonomous system, with equations (\ref{eq.x}) and (\ref{eq.y}). 
\section{Tracking Quintessence }
In this section we will discuss about the stability of the solutions for FRW cosmological models with a pressure-less background fluid and a quintessence scalar field, which tracks the background fluid. To find the qualitative behaviour of the system, the equations are written in the form of an autonomous system. The fixed points are found out. As we are interested in stable solutions, the fixed points which correspond to the physical requirements of the evolution of the universe are picked up.

\subsection{Systems of Equations in Polar Form }
Often a system of equations in the polar form has advantages of finding out the fixed point in an easier way. So another transformation of the variables is now effected with $x=r \cos\theta$, and $y= r \sin\theta$, where $0\leq r \leq \infty$ and $0 \leq \theta \leq 2\pi$. Here $r^2$ is actually scalar field energy density parameter and it is restricted in the region $0< r^2 <  1$. Equations ($\ref{eq.x}$) and ($\ref{eq.y}$) can now be written in terms of polar variables as 
\begin{equation}\label{eq.r}
r^{\prime} = (\frac{3 \gamma}{2} - 3 \cos^2 \theta) (1-r^2) r,
\end{equation}
\begin{equation}\label{eq.theta}
\theta^{\prime} = (3 \cos \theta - \sqrt{\frac{3}{2}} \lambda r) \sin \theta.
\end{equation}
To study the phase space behaviour of the system first we have to find fixed point of the system. The fixed points of the system are the simultaneous solutions of the equations $r^{\prime}=0$ and $\theta^{\prime} =0$. As we are interested in solution of the system near a fixed point so the extreme cases like $r^2=0$ (meaning $\Omega _{\phi} =0$) and $r^2=1$ (meaning $\Omega _{\phi}=1$) are excluded. The intention is also to have a blend of the dark energy and dark matter. As $r$ is radial coordinate and $0< r^2 < 1$ so fixed point $ (r, \theta)=(-\frac{\sqrt{3 \gamma}}{\lambda} , \cos^{-1}(-\sqrt{\frac{\gamma}{2}})) $ is excluded. Amongst all the fixed point, the only viable fixed point is $ (r, \theta)=(\frac{\sqrt{3 \gamma}}{\lambda} , \cos^{-1}(\sqrt{\frac{\gamma}{2}})) $. The old variables $x$ and $y$ at this fixed point are 
\begin{equation} \label{eq.fx}
x^2=\frac{3 \gamma^2}{2 \lambda^2},
\end{equation} 
 and
\begin{equation} \label{eq.fy}
y^2= \frac{3 \gamma^2}{2\lambda^2} (1-\frac{\gamma}{2}).
\end{equation}
 
\subsection{Stability of the solution in polar coordinates}
 As only one fixed point $ (r, \theta)= (\frac{\sqrt{3 \gamma}}{\lambda} , \cos^{-1}(\sqrt{\frac{\gamma}{2}})) $ is of physical interest, we want to analyze  the stability of this fixed point. In section (1.4) we have already discussed the stability analysis of a fixed point. 

Let us consider a 2D system of non-linear differential equations

    $~~~~~~~\dot{x}=f(x,y)$ 
    
and $~\dot{y}=g(x,y)$, 

where an overhead dot denotes differentiation with respect to some parameter ($N$ in the present case).

In the neighbourhood of a fixed point a non-linear system can be linearised. 
If $u$ and $v$ are the small disturbances from the fixed points  then the system can be written in the form 
\begin{equation*}
\begin{bmatrix}
\dot{u} \\
\dot{v}
\end{bmatrix} 
= A
\begin{bmatrix}
u \\ 
v
\end{bmatrix}
\end{equation*}

where
 \begin{equation*}
 A=
 \begin{bmatrix}
\dfrac{\partial f}{\partial x} & \dfrac{\partial f}{\partial y} \\
\dfrac{\partial g}{\partial x} & \dfrac{\partial g}{\partial y}
\end{bmatrix} ,
 \end{equation*}
 is the Jacobian matrix at the fixed point. For a 2D system the stability of a fixed point can be determined from the determinant ($ \bigtriangleup $) and trace ($ \tau $) of A at that fixed point. The eigenvalues are real and have opposite signs if $ \bigtriangleup < 0$, irrespective of $\tau$ and hence the fixed point is a saddle point. If $ \bigtriangleup > 0$ and $\tau < 0$ then both the eigenvalues have negative real part hence the fixed point is stable. When $ \bigtriangleup > 0$ and $\tau > 0$ then the fixed point is unstable. Nodes satisfy $\tau^2 - 4 \bigtriangleup > 0$ and spirals satisfy $\tau^2 - 4 \bigtriangleup < 0$ \cite{r19}.

\begin{figure}[H] 
\centering
\includegraphics[width=100mm]{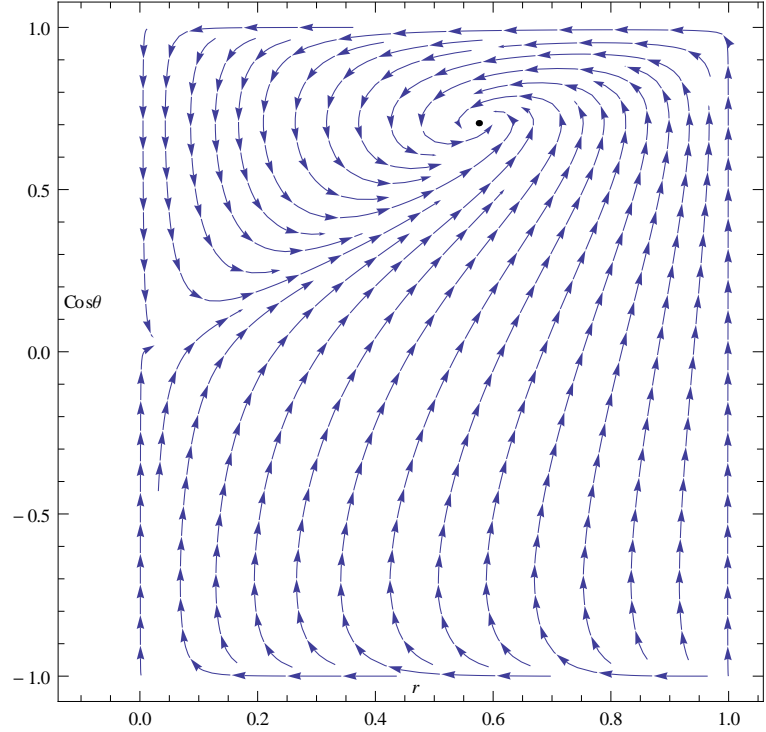}
\caption{Phase plot of the system. The point, P is the fixed point of our interest. The plot clearly shows that the fixed point is stable in nature.The plot is for $\gamma=1$ and $\lambda=3$.}
\label{fig: t1}
\end{figure}

For the fixed point $ (r, \theta)= (\frac{\sqrt{3 \gamma}}{\lambda}, \cos^{-1}(\sqrt{\frac{\gamma}{2}}))  $ the Jacobian matrix  
\begin{equation*}
A=
\begin{bmatrix}
0 & {3 \sqrt{6} \frac{\gamma}{\lambda} \sqrt{1-\frac{\gamma}{2}}(1-\frac{3 \gamma}{\lambda^2})}\\
-\frac{\sqrt{3}}{2} \lambda \sqrt{1-\frac{\gamma}{2}} & 3 (\gamma -1) - \frac{3 \gamma}{4}
\end{bmatrix} .
 \end{equation*}
The determinant and the trace of the matrix A are, $\bigtriangleup = \frac{9}{\sqrt{2}} \gamma (1- \gamma / 2) (1- \frac{3 \gamma}{\lambda^2})$ and $\tau = 3 (\frac{3 \gamma}{4} -1)$ respectively. For a matter dominated universe $\gamma = 1$ and $p=0$. It is known that the fixed point is stable only when $\bigtriangleup > 0$ and $\tau < 0 $. So $\lambda^2 > 3$ is the condition for which the fixed point is stable. To draw the phase portrait of the system, we have plotted  $r$ against $\cos \theta$ instead of  $\theta$ for $\lambda = 3$. From the phase plot(figure ~\ref{fig: t1}) it is clear that the fixed point is stable in nature. So any solution of the system around this fixed point will be a stable solution for a wide range of initial values. It is clear from the phase plot that $r=1$ and $r=0$ are two invariant sub-manifolds. Here we have considered a local subspace bounded by invariant sub-manifold $r=0$ and $r=1$. It also deserve mention that ($r=0, \cos\theta = 0$) is a saddle fixed point for our choice of parameter $\gamma=1, \lambda =3$. The value of $r$ has been restricted between 0 and 1 from the physical requirement that $\Omega_\phi = x^{2}+y^{2} = r^{2}$ and $0 < \Omega_\phi< 1 $.

 \subsection{Examples with specific potentials}  
To study the system near the fixed point with tracking we consider two specific types of potentials as examples. Among many possibilities \cite{r21}, we choose hyperbolic potential of the form $V=V_0 \cosh \alpha \phi,$ where $V_0$ and $\alpha$ are constants and exponential potential of the form $ V= e^{\alpha \phi} - \beta$, where $\alpha$, $\beta$ are positive constants and $\phi$ is the quintessence scalar field.
 
\textbf{(i) $V=V_0 \cosh \alpha \phi$}
 
The tracking parameter $\Gamma$ and the parameter  $\lambda$ in this case are given by

 $\Gamma =  V\dfrac{d^2 V}{d \phi^2} / (\dfrac{dV}{d \phi})^2 = \coth^2 \alpha \phi $

  and $ \lambda = -\frac{1}{k V} \dfrac{dV}{d \phi} = -\frac{\alpha}{k} \tanh \alpha \phi $ 
  
respectively. If the quintessence field density tracks the background matter density then $\Gamma \approx 1 $. A near tracking scenario can be achieved by assuming $\Gamma =1 + \delta $, where $\delta$ is very small. This would imply $\coth^2 \alpha \phi = 1 + \delta$, i.e, scalar field $\alpha \phi$ has a very high value. As $\delta$ is very small we have neglected the higher powers of $\delta$. The parameter $\lambda$ will have a near constant value as for $\delta\longrightarrow 0$, $\tanh \alpha \phi \longrightarrow (1+ \delta)^{-\frac{1}{2}} \approx 1-\frac{\delta}{2}$. So $\lambda$ can be written as $\lambda =- \frac{\alpha}{k} (1-\frac{\delta}{2})$.
 
At the fixed point $ (r, \theta)=(\frac{\sqrt{3 \gamma}}{\lambda} , \cos^{-1}(\sqrt{\frac{\gamma}{2}}))$ for the potential ($V=V_0 \cosh \alpha \phi$), using  equations (~\ref{eq.fx}) and (~\ref{eq.fy}), one has
 
\begin{equation} \label{xcosh}
 x^2=\frac{3 \gamma^2}{2 \lambda^2} = \frac{3 \gamma^2}{2} \frac{k^2}{\alpha^2} \coth^2 \alpha \phi = \frac{3 \gamma^2}{2} \frac{k^2}{\alpha^2} (1+ \delta) ,
\end{equation}
and also 
\begin{equation} \label{ycosh}
y^2 = \frac{3 \gamma}{\lambda^2} (1- \frac{\gamma}{2}).
 \end{equation} 
 
  From the definition of $x$ and $y$ (equation(~\ref{eq.trans})), one can write $\phi$ and H as 

 \begin{equation} \label{phicosh}
\phi=ln(B a^A),
 \end{equation}
 
 \begin{equation} \label{Hcosh}
 H^2 =\frac{V_0}{3} \frac{\alpha^2}{3 \gamma (1-\frac{\delta}{2}) (1+ \delta)} \frac{B^{\alpha} a^{\alpha A} +B^{-\alpha} a^{-\alpha A}}{2},
 \end{equation}
 
  where $A=3(1+ \delta /2) \gamma / \alpha $ and B is arbitrary integration constant.This last equation can be utilized to write the deceleration parameter
  
  \begin{equation*}
   q = - \frac{\ddot{a}/a}{\dot{a}^2 / a^2} = - \frac{\dot{H} + H^2}{H^2}
  \end{equation*}
  
   as
   
 \begin{equation} \label{qcosh}
 q =-1-\frac{\alpha A}{2}  \frac{B^{2\alpha } a^{2 \alpha A} -1}{B^{2\alpha } a^{2 \alpha A} +1}.
 \end{equation}
 
  \begin{figure}[htbp!]
  \begin{center}
 \includegraphics[scale=0.4]{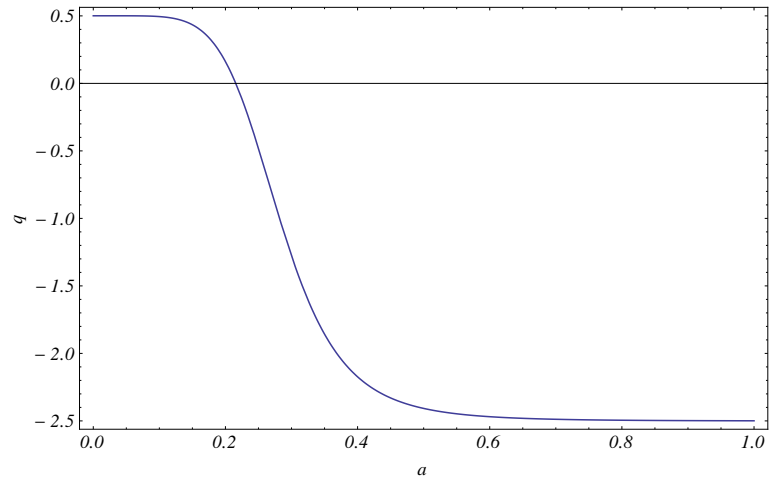} 
\caption{Plot of `q' against `a'. The plot is for the parametric value of $B^{\alpha} = 44.73$, $\delta=0.001$ and $\gamma = 1$.}
 \label{t2}
 \end{center}
 \end{figure}
 
Now using equation(~\ref{qcosh}) `q' can be plotted against `a' (in units of $a_0$, the present value of a).  As we  are looking for a tracking solution, $\Gamma \approx 1$, i.e. $\coth^2 \alpha \phi \approx 1$, so `$\phi$' is severely restricted.  With the help of the equation (\ref{phicosh}) and the tracking condition, one can estimate the constant $B^{\alpha}$ to be $B^\alpha = 0.022$ and $B^\alpha = 44.73$. For a matter dominated universe i,e. $\gamma = 1$, $B^\alpha = 44.73$ yields a `q' which at least qualitatively resembles the present acceleration which is an observational quantity (see figure\ref{t2} ). The problem is that the acceleration sets in quite early in the matter dominated era (near z=4). The other value, $B^\alpha = 0.022$, predicts an acceleration at a distant future and hence not discussed. 

\textbf{(ii) $V=e^{\alpha \phi} - \beta$ } ~~~~~

 \vspace{-0.65 cm}
 
In this case $\Gamma =  V\dfrac{d^2 V}{d \phi^2} / (\dfrac{dV}{d \phi})^2 = \frac{e^{\alpha \phi} - \beta}{e^{\alpha \phi}}$ and $\lambda = -\frac{1}{kV} \dfrac{dV}{d \phi} = -\frac{\alpha e^{\alpha \phi}}{e^{\alpha \phi}-\beta}$. To satisfy tracking condition, $\Gamma \approx 1$, $\beta \rightarrow 0$ is only one option. This condition also would imply $\lambda$ has a near constant value. Near the fixed point $x^2=\frac{3}{2} \frac{\gamma^2}{\lambda^2}$ and $y^2=\frac{3 \gamma^2 (1- \frac{\gamma}{2})}{2 \lambda^2}$, it can be shown that 
\begin{equation} \label{phiexp}
e^{\alpha \phi} - \beta= B a^A
\end{equation}
 and
\begin{equation} \label{Hexp}
H^2=\frac{\alpha^2}{A^2} \frac{\gamma}{1-\gamma/2} \frac{(B a^A + \beta)}{B a^A}.
\end{equation} 

Here B is an integration constant and $A=\frac{3 \gamma}{k}$. The deceleration parameter `q' can be written as 
 
 \begin{equation} \label{qexp}
 q=-1-A \frac{B a^A - \beta}{B a^A + \beta}
 \end{equation}
 
\begin{figure}[H]
\begin{center}
\includegraphics[scale=0.7]{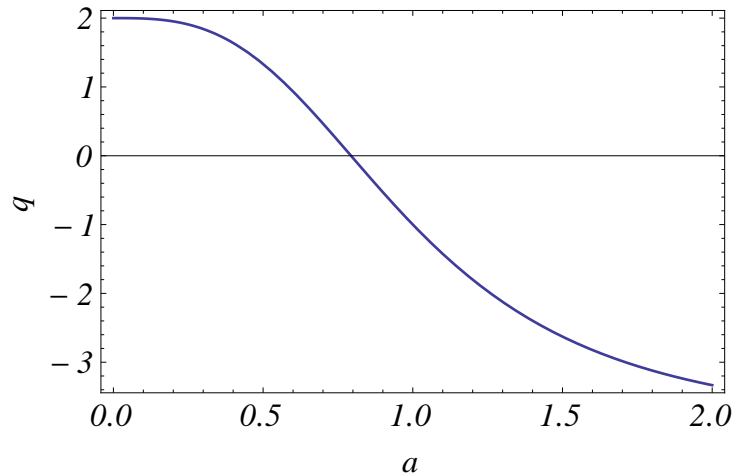} 
\caption{The plot of q vs $a$ for $A=3$, $\gamma=1$ and $\beta=0.00001$.}
\label{t3}
\end{center}
\end{figure}

By considering present value of deceleration parameter `q' and $a=1$, the integration constant B can be expressed in terms of A and $\beta$ as  $B=\frac{A - 0.3}{A + 0.3} \beta$. If we take k=1 and a matter dominated universe $\gamma = 1$ then $A= 3$. Plot of q vs $a$ for this potential agree well with the present accelerated expansion of the universe (see figure: \ref{t3}).

\subsection{Discussion}
The stability of tracking quintessence models for the universe has been investigated in the present section. With a general tracking condition, the fixed point solutions are not too many. In fact there is one physically relevant generic fixed point, with tracking conditions, leading to a stable solution. The condition for stability, in terms of the fractional rate of change of the quintessence potential, is found out to be $\lambda^2 > 3$ where $ \lambda = - \frac{1}{k V} \frac{dV}{d \phi}$.

Two specific potentials, giving rise to the present acceleration, are worked out as examples. It is found that for $V=V_0 \cosh \alpha \phi$, the scalar field is severely restricted to a zone such that $\coth^2 \alpha \phi$ is close to unity. For the other example, $V=e^{\alpha \phi} - \beta$, the conditions do not restrict the scalar field, but rather fine tunes the constants in the potential(eg. $\beta$ must have a very small value).
\par

\section{Quintessence Scalar Field: With Tracking Condition Relaxed}
This section deals with the study of the possible stable solutions for some quintessence models when the tracking condition is relaxed. There is a host of quintessence potentials with some merits for each of them. For a  comprehensive review, refer to( \cite{r21}). Two well known potentials will be used as examples in the present work. The autonomous system of equations (\ref{eq.x},\ref{eq.y} and \ref{eq.lambda}) is a three dimensional one for a general potential $V=V(\phi)$. For the exponential potential, the system effectively reduces to a two dimensional one as $\lambda$ is constant. Where as the system is actually three dimensional for a power-law potential. The numerical solutions are found out. Some of the fixed points are non-hyperbolic so the stability of these fixed points can not be analysed by using linear stability analysis. The stability of these non-hyperbolic fixed points is analysed with a different strategy, namely by a small perturbation around the fixed point. 

\subsection{Estimation of boundary values:} 
From equation (\ref{eq.dH}) the deceleration parameter can be express in terms of $x$ and $y$ as
\begin{equation} \label{eq.q}
q= \frac{3 \gamma}{2} (1- x^2 -y^2) + 3x^2 -1.
\end{equation}
For a spatially flat universe,
\begin{equation} \label{eq.omega}
\Omega_{\phi}=\frac{k^2 \rho_{\phi}}{3 H^2}= x^2 + y^2,
\end{equation}
which, in view of the constraint equation (\ref{H11}), is restricted as 
\begin{equation}
 \Omega_{B} + \Omega_{\phi}=1 , 
 \end{equation}
 where $\Omega_{B}$ is cold dark matter energy density parameter. This implies $0\leq \Omega_{\phi} \leq 1$. 
  
  As the dark matter is expected to be cold, the subsequent discussion assumes that the corresponding pressure $p_{B}$ to be zero which implies $\gamma =1$. The numerical solutions for the system can be found out by fixing the boundary values. The boundary values of $x, y$ etc has been estimated using the present ( i.e., at $N = ln a = 0$) values of  $\Omega_{B},~ q$ etc as suggested by observations. Present observational value of $\Omega_{B}= 0.27$, so
\begin{equation} \label{eq.boun}
 \Omega_{\phi} = x_{0}^2 + y_{0}^2=0.73 , 
 \end{equation}
  where $x_{0}$ and $y_{0}$ are the present values of $x$ and $y$ respectively. Using equation ({\ref{eq.boun}}) in equation ({\ref{eq.q}}), a lower bound is set for present value of deceleration parameter ($q_{0} > -0.595$) so as to make $x_{0}$ real. 
  
   To make it consistent with observations (see ref.\cite{r29} and references therein) and the lower bound of the deceleration parameter set for the present work, we set $q_{0}=-0.53$, then equations (\ref{eq.q}) and (\ref{eq.boun}) would yield $x_{0}=0.147$ and $y_{0}=0.842$. From the definition of $\Gamma$, it is easy to check that when $\Gamma = 1$ the potential is an exponential function of $\phi$ and for other values of $\Gamma$, the potential has the form of power law. The present value of $\lambda$ has been chosen to fit the observational results. Interestingly the numerical solutions evolve from a fixed point (asymptotic as $N \longrightarrow -\infty$) of the system and attracted towards  another fixed point (asymptotic as $N \longrightarrow \infty$) of the system. The solutions are heteroclinic orbits in phase space as they connect two different fixed points. Thus, the beginning and the ultimate fate of the universe can be qualitatively investigated by finding the behaviour of the fixed points. 

\subsection{Examples with specific potentials:}
\textbf{(i) Exponential Potential:}

 When $\Gamma =1$, the definition of $\Gamma$ indicates that the potential has the form 
\begin{equation}\label{exp.pot}
 V(\phi) = A e^{\alpha \phi},
\end{equation}
 where A and $\alpha$ are constants. From equation (\ref{eq.lambda}), $\lambda={\lambda}_{0}$ is a constant. So effectively the system becomes a 2-dimensional one.
 
  By numerically solving the equations (\ref{eq.x}) and (\ref{eq.y}), the plots for $x$ and $y$ are obtained. Plots for observationally relevant cosmological parameters have also been obtained. For these plots, we have taken $\lambda_{0} =1$. This also yields $\alpha=-k$. It has been checked that for some other values of $\lambda$ the analysis is possible, but there is hardly any qualitative difference, so we do not include them.\\
 
\begin{figure}[H]
\centering
\includegraphics[scale=0.6]{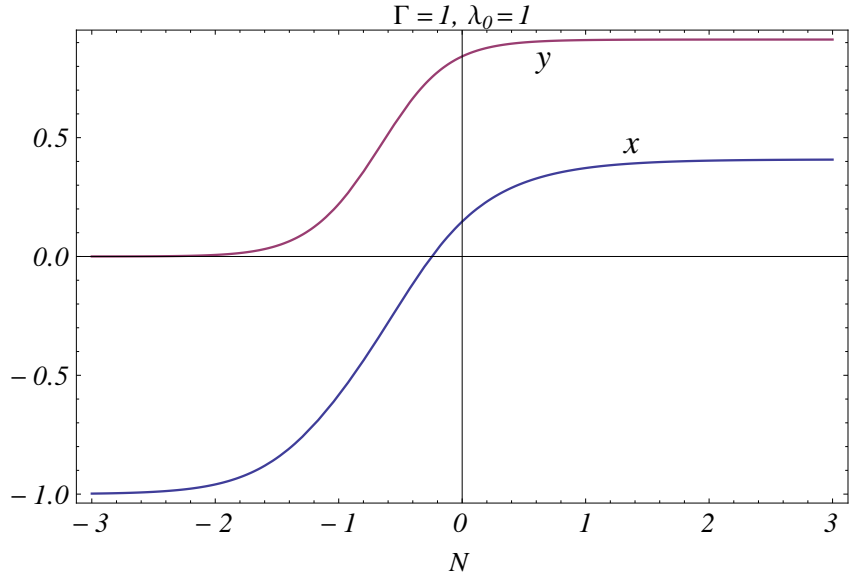}
\caption{The plot of $x$ and $y$ vs $N$ for exponential potential.}
\label{q1}
\end{figure}

\begin{figure}[H]
\centering
\includegraphics[scale=0.6]{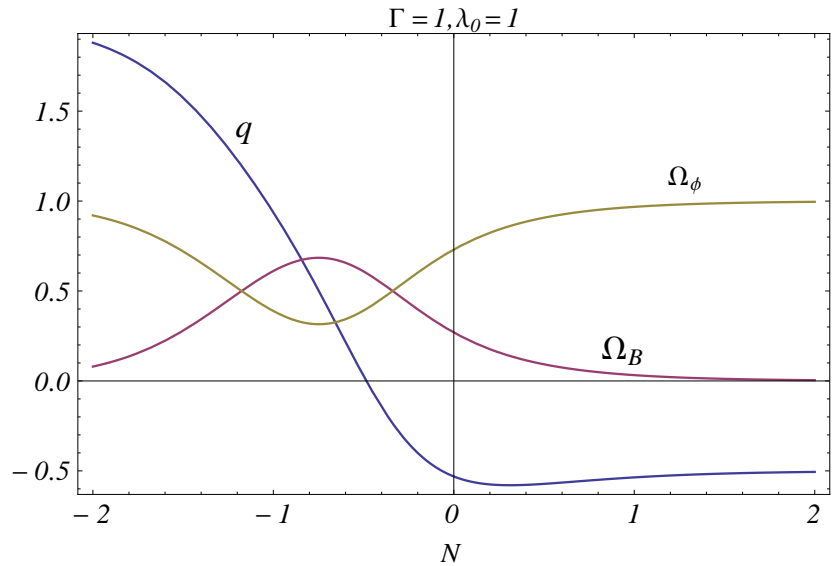}
\caption{The plot of $q, \Omega_{B}$ and $\Omega_{\phi}$ vs $N$ for exponential potential.}
\label{q2}
\end{figure}

Figure (\ref{q1}) shows the evolution of $x,y$ against $N$. The plots of cosmological parameters $q,~ \Omega_{\phi}$ and $\Omega_{B}$ against $N$ are shown in figure (\ref{q2}). The fixed points($x_{0},y_{0}$) of the system when $\lambda_{0} = 1$ and $\gamma=1$ are (1.224,-1.224), (0.408,-0.912), (1,0), (-1,0), (1.224,1.224), (0.408,0.912) and (0,0). From figure (\ref{q1}), it is easy to find out the asymptotic behaviour of the system. We see that ($x,y$) is asymptotic to (-1,0) as $N \longrightarrow -\infty$ in the past and to (0.408,0.912) as $N \longrightarrow \infty$ in the future. Determinant($\bigtriangleup$) and trace($\tau$) of the Jacobian matrix at (-1,0) are $\bigtriangleup = 3 (3 +\sqrt{\frac{3}{2}})$ and $\tau = 6 + \sqrt{\frac{3}{2}}$. As both these are positive, the point (-1,0) is an unstable fixed point. Determinant($\bigtriangleup$) and trace($\tau$) of the Jacobian matrix at (0.408,0.912) are $\bigtriangleup = 5.123$ and $\tau =-4.55$, so this point is a stable fixed point. The fixed point (-1,0), which is past time attractor of our solution has the qualitative features like $a \longrightarrow 0, \phi \longrightarrow \infty, V(\phi) \longrightarrow 0, H \longrightarrow \infty $ as $N \longrightarrow -\infty$. The qualitative features of the fixed point (0.408,0.912), which is a future time attractor, are $a \longrightarrow \infty, \phi \longrightarrow \infty, V(\phi) \longrightarrow 0$ and $q \longrightarrow -0.5$ as $N \longrightarrow \infty$.

\begin{figure}[H]
\centering
\includegraphics[scale=0.6]{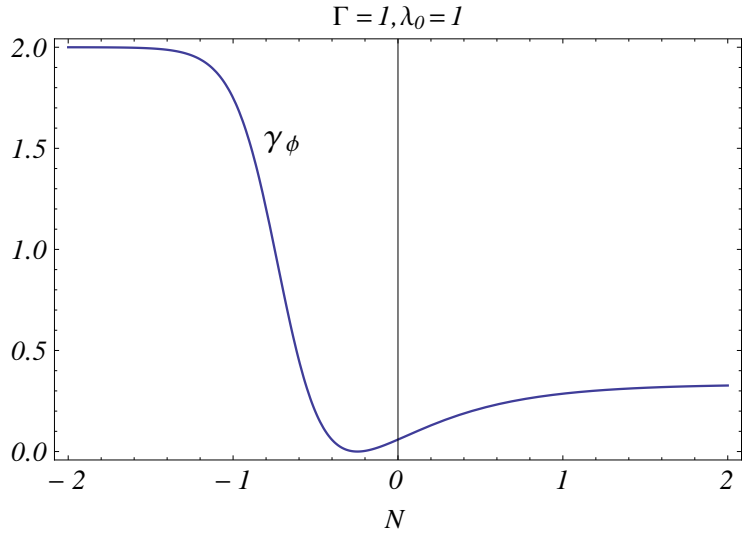}
\caption{The plot of $\gamma_{\phi}$ vs $N$ for exponential potential.}
\label{q3}
\end{figure}

This analysis, along with reasonable boundary values (i.e. those indicated by observations) allows us to qualitatively study the beginning and the possible ultimate fate of the universe. It is seen that the present accelerated expansion is indeed a stable solution for the universe with an exponential potential for the quintessence scalar field. As the fixed point $(x,y)=(-1,0)$ is an unstable fixed point thus, given a small perturbation from it, the universe evolves from the state when the linear size ($a$) was zero, the value of the scalar field infinity and the potential was zero. As the solutions connects two different fixed points thus, the solutions form heteroclinic orbits in phase space.At the beginning it expanded very rapidly ($H \longrightarrow \infty$) but at a decelerated rate. After  this phase of decelerated expansion the universe has entered in to an accelerated expansion phase which is shown in figure(\ref{q2}).  The accelerated expansion starts at about $z \simeq 0.49$,  which is quite consistent with the observations. From figure (\ref{q2}) , it is also seen that in recent past $\Omega_{\phi} < \Omega_{B}$, but in remote past, $\Omega_{\phi}$ dominates over $\Omega_{B}$ though at that time the universe was undergoing a decelerated expansion.This is indeed intriguing.  We see that in the remote past $\gamma_{\phi} > 1$ ( figure \ref{q3}) and the pressure and the density of the quintessence field are related as $p_{\phi} = (\gamma_{\phi}-1) \rho_{\phi}$. This indicates that the contribution to the pressure sector from the quintessence field had been positive and hence the quintessence field failed to drive an acceleration. But as the universe evolves, the value of $\gamma_{\phi}$ drops below unity and the effective pressure becomes negative so as to drive an accelerated expansion.

As the universe asymptotically approaches the fixed point (0.408,0.912), so the future behaviour of the universe in this model can be described by qualitatively analysing it. This fixed point is a future time attractor as discussed before. So as $N \longrightarrow \infty$, the universe settles to a state where ${\Omega}_{B} \longrightarrow {0}$ and ${\Omega}_{\phi}$ attains a constant value. The universe expands with a nearly constant acceleration.

\begin{figure}[H]
\centering
\includegraphics[scale=0.5]{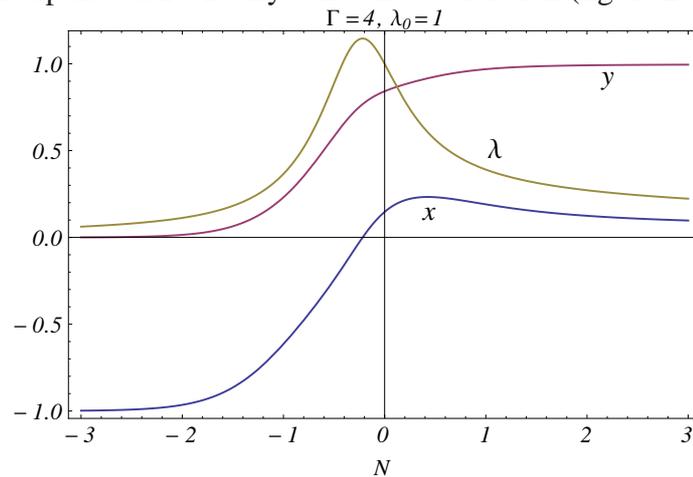}
\caption{The plot of $x,y, \lambda $ vs $N$ for power law type potentials.} 
\label{q4}
\end{figure}

\begin{figure}[H]
\centering
\includegraphics[scale=0.5]{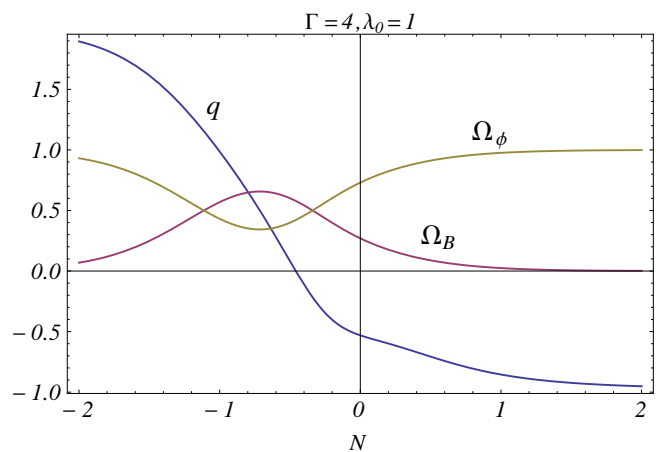}
\caption{The plot of $q,\Omega_{B},\Omega_{\phi}$ vs $N$ for power law type potentials.}
\label{q5}
\end{figure}

\begin{figure}[H]
\centering
\includegraphics[scale=0.5]{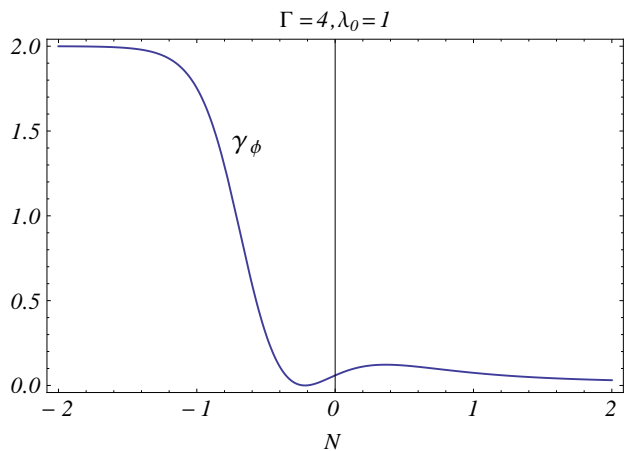}
\caption{The plot of $\gamma_{\phi}$ vs $N$ for power law type potentials.}
\label{q6}
\end{figure}

\textbf{(ii) Power law type potentials:}  
When $\Gamma =m$, a constant but $m \neq 1$, the form of the potential can be found out from definition of $\Gamma$ as
\begin{equation}\label{pwrlaw.pot}
 V(\phi) = (A \phi + B)^{\frac{1}{1-m}},
\end{equation}
  where A,B and m are constants. In this case $\lambda$ has a dynamics. It is not necessarily a constant, and we plot $x,y,\lambda$ (fig \ref{q4}) and also the cosmological parameters like $q, \Omega_{B}, \omega_{\phi}, \gamma_{\phi}$ (fig: \ref{q5} and \ref{q6} respectively), all against $N$. As an example, we take $\Gamma = 4$. The fixed points ($x_{0},y_{0},\lambda_{0}$) of the system for power law type potentials are (-1,0,0), (0,0,0), (1,0,0), (0,-1,0) and (0,1,0). These fixed points are independent of $\Gamma$. From figure \ref{q4} we see that ($x,y,\lambda$) curves are asymptotic to (-1,0,0) as $N \longrightarrow -\infty$ and asymptotic to (0,1,0) as $N \longrightarrow \infty$. The eigenvalues of the Jacobian of the system at (-1,0,0) is (3,3,0) and those at (0,1,0) is (-3,-3,0). As these fixed points are non-hyperbolic, we can not use linear stability analysis. The fixed point (-1,0,0) has two positive eigenvalues and one zero eigenvalue so it is an unstable fixed point. However, the fixed point (0,1,0) has two negative eigenvalues and one zero eigenvalue, so its being a stable fixed point cannot be ruled out before further investigation.
 
\begin{figure}[H]
\centering
\includegraphics[scale=0.4]{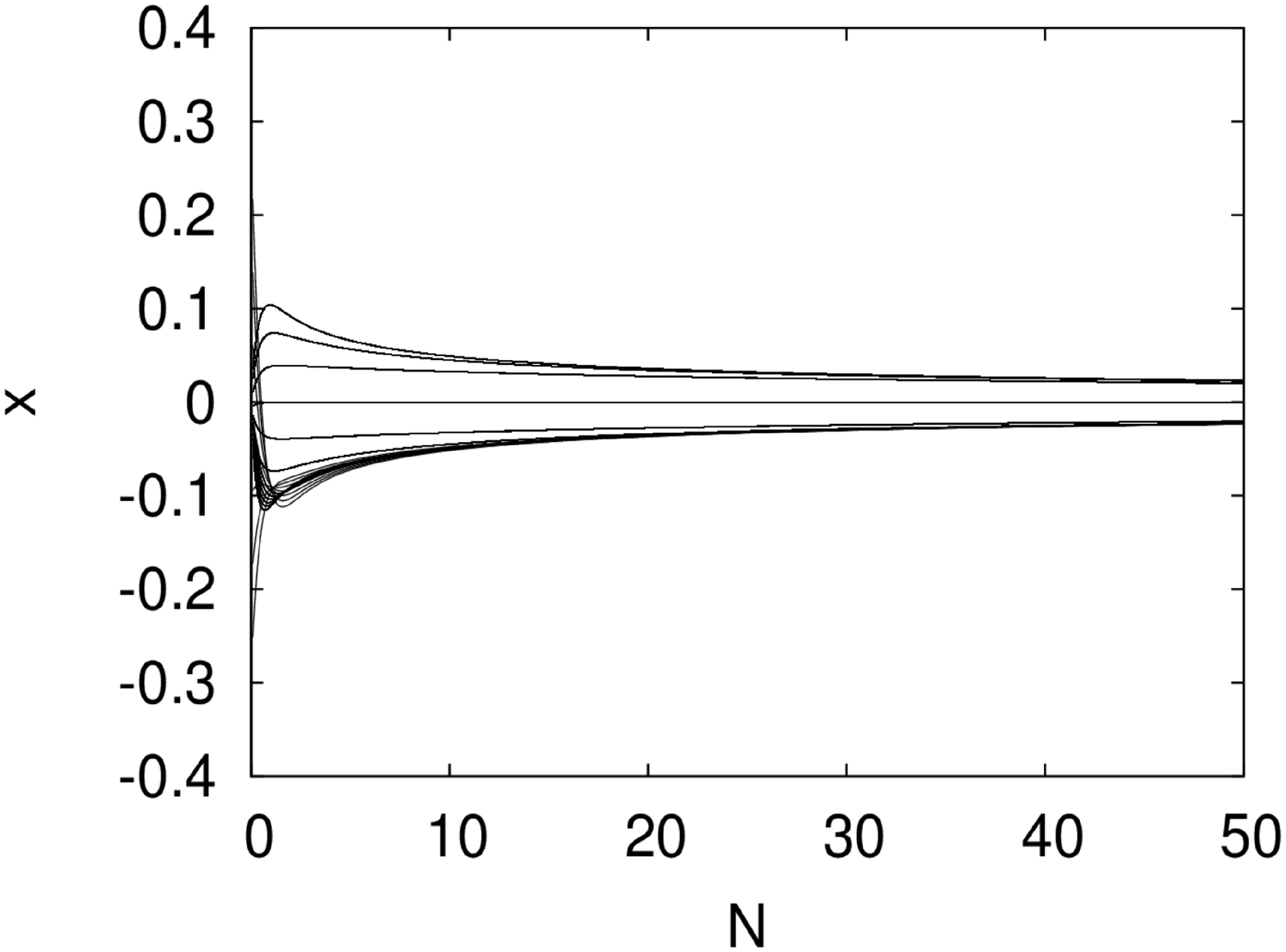}
\caption{The plot of $x$ vs $N$ for perturbation near $(0,1,0)$.} 
\label{q7}
\end{figure}

\begin{figure}[H]
\centering
\includegraphics[scale=0.8]{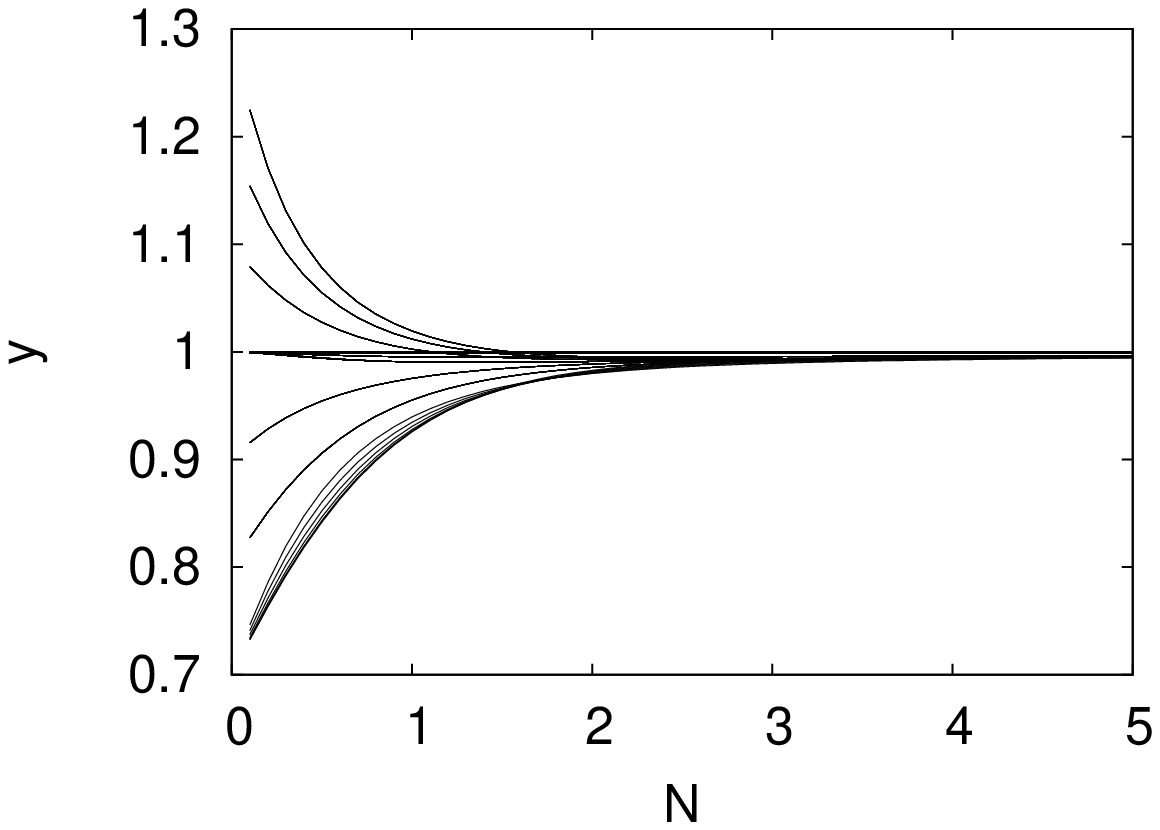}
\vspace{10 mm}
\caption{The plot of $y$ vs $N$ for perturbation near $(0,1,0)$.} 
\label{q8}
\end{figure}

\begin{figure}[H]
\centering
\includegraphics[scale=0.4]{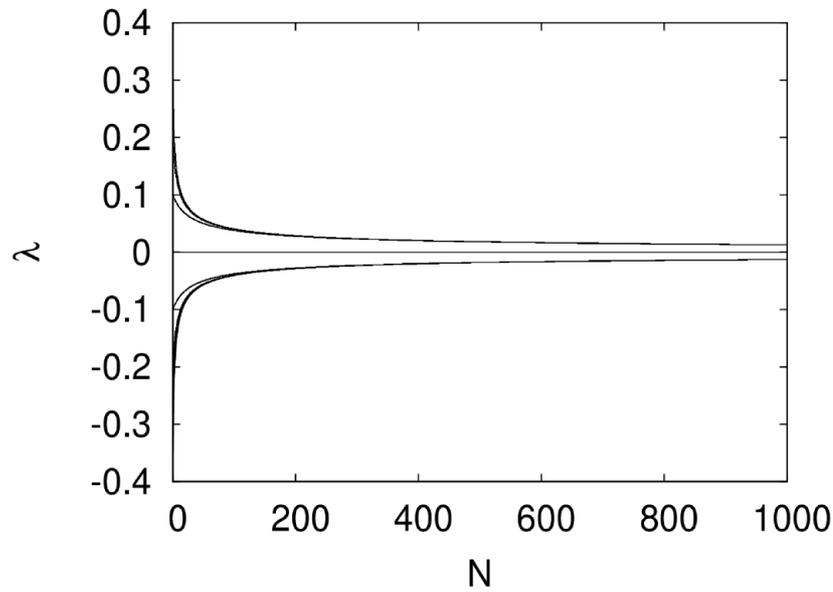}
\caption{The plot of $\lambda$ vs $N$ for perturbation near $(0,1,0)$.} 
\label{q9}
\end{figure}

In order to check the stability of the latter fixed point,  we perturb the system in every direction near the fixed point and numerically solve the system of equations to find its asymptotic behaviour as $N \longrightarrow \infty$. The 3D phase portrait looks obscure and it is difficult to figure out the behaviour. So we have plotted $x, y$ and $\lambda$ against $N$ in figure \ref{q7}, \ref{q8} and \ref{q9}. From figure \ref{q7}, we see that $x(N)$ approaches $x=0$ as $N \longrightarrow \infty$. Also from figure \ref{q8} and figure \ref{q9}, one concludes that y(N) approaches $1$ and $\lambda(N)$ approaches $0$ respectively as $ N \longrightarrow \infty$. So the system approaches the fixed point (0,1,0) as $N \longrightarrow \infty$ against perturbation near (0,1,0). The fixed point is thus  a stable one. Thus, for a power law type potential also, a qualitative analysis can be looked at. 
 
 \vspace{-0.2 cm }
 The universe evolves from the unstable fixed point (-1,0,0) representing the physical state ($a \longrightarrow 0, \phi \longrightarrow \infty, V(\phi) \longrightarrow$ constant and finite, $H \longrightarrow \infty $). As it is an unstable state, a small perturbation starts the evolution. After a long stint of decelerated expansion, the system enters into the phase of  accelerated expansion and is attracted towards the stable fixed point (0,1,0). This final stable state will correspond to the physical state ($a \longrightarrow \infty, \phi \longrightarrow$ constant, $ V(\phi) \longrightarrow$ constant) and it will expand with a nearly constant acceleration (figure \ref{q5}). 

\vspace{-0.4 cm }

\vspace{-0.55 cm }

\vspace{-20 mm}

\subsection{Discussion:}
This present section deals with a dynamical systems analysis of a quintessence scalar field that drives the recent accelerated expansion of the universe.  An exponential potential and a power law type potential are worked out as examples. The boundary values of $x$ and $y$ are estimated from the recent observations and the boundary value of $\lambda$ is chosen so that the result is consistent with the present observations. In both cases, it is possible to find a scenario where the universe in fact starts from an unstable fixed point (hence apt to move away from that)  and evolves to a future attractor through the present phase of accelerated expansion. Amongst the several possible fixed points, only those are chosen which are past and present asymptotes for the set of solutions.

The observational value of density parameters of quintessence scalar field and background fluid are  ${\Omega}_{\phi}=0.73$  and ${\Omega}_{B}=0.27$ respectively.
This, in fact, is put into the system as a boundary value. It is intriguing to note that evolution suggests that ${\Omega}_{\phi}$ actually dominates over its dark matter counterpart in the past as well except for a brief interval in the recent past. This feature is there for both the examples, i.e., an exponential or a power law potential. The reason that the universe decelerated in the past is the fact that the equation state parameter ${\gamma}_{\phi}$ had been more than unity so that the quintessence field could not generate the all important negative pressure.

Both the examples lead to a situation where the universe in future will be governed completely by the quintessence field (${\Omega}_{\phi}=1$ and ${\Omega}_{B}=0$) and will be steadily accelerating. Although we have worked out for two examples, it appears from equation(\ref{eq.lambda}), the method can, in principle extended for any potential for which $\Gamma$ can be expressed as a function of $\lambda$.

It is important to note that a small change in the initial conditions does not inflict any major or qualitative change in the system. For instance, the system does not show any chaotic behaviour. This has been carefully checked.

\chapter{Stability of chameleon scalar field models} 

% **************************** Define Graphics Path**************************
\ifpdf
    \graphicspath{{Chapter3/Figs/Raster/}{Chapter3/Figs/PDF/}{Chapter3/Figs/}}
\else
    \graphicspath{{Chapter3/Figs/Vector/}{Chapter3/Figs/}}
\fi

\section{Introduction:}
Theoretically mass-less scalar fields are abundant in string theory and supergravity theory. But these mass-less scalar fields predict very large violation of the Equivalence Principle(EP), which is physically unacceptable. Another interesting observational fact from the measurement of absorption line in quasar spectra is that the evolution of the fine structure constant $\alpha$ is of the order of one part of $10^5$ over the redshift interval $0.2 < z < 3.7$ \cite{webb}.  To model a time varying coupling constant $\alpha$ with a rolling scalar field the mass of the scalar field has to be of the order of present value of Hubble parameter ($H_0$).   

In \cite{justin1, justin2} , Khoury {\it et al} introduced the concept of chameleon scalar field . Unlike the common non-minimally coupled scalar field theories, a chameleon field is coupled to matter sector rather than the geometry. As a consequence of this the scalar field acquire a mass whose magnitude depends on the ambient matter distribution. In a region of high density, such on Earth, the mass of the scalar field is very large and the EP violations are suppressed. But in a region of very low density i.e. on cosmological scale the mass of the scalar field can be of the order of $H_0$. So in cosmological scale the field can satisfy the bound on the violations of EP  \cite{khoury}. This change of properties of the field acts as a buffer in connection with the observational bounds set on the mass of the scalar fields coupled to the matter sector  \cite{mota1}.

A strongly coupled chameleon scalar field has the possibility of being detected by carefully designed experiments \cite{mota2, shaw1}. Possible effect of the chameleon field on the cosmic microwave background with possible observational imprints has been estimated by Davis, Schelpe and Shaw \cite{shaw2} and that on the rotation curve of galaxies by Burikham and Panpanich \cite{buri}. One very attractive feature of a chameleon field is that if it is coupled to an electromagnetic field \cite{brax1} in addition to the fluid, the fine tuning of the initial conditions on the chameleon may be resolved to a large extent \cite{mota3}. The remarkable features of the chameleon field theories are comprehensively summarized by Khoury \cite{khoury}. 

Brax {\it et al} used this chameleon field as a dark energy\cite{brax}. This kind of interaction between the dark matter and the dark energy was investigated in detail by Das, Coarsaniti and Khoury\cite{das} in the context of the present acceleration of the universe. It was also shown that with a chameleon field of this sort, it is quite possible to obtain a smooth transition from a decelerated to an accelerated expansion for the universe\cite{nbsdkg}.

The success of the chameleon field in explaining the current accelerated expansion and its lucrative properties which open up the possibility to evade a fine tuning of initial conditions, and its possible observational imprints inevitably attracted a lot of attention. The possibility of a scalar field nonminimally coupled to gravity, such as the Brans-Dicke scalar field, acting as a chameleon was discussed by Das and Banerjee \cite{sdnb}. Brans-Dicke scalar field acting as a chameleon with an infrared cut-off as that in the holographic models was discussed by Setare and Jamil \cite{jamil}. The field profile of a chameleon was discussed by Tsujikawa, Tamaki and Tavakol \cite{tavakol}. 

The aim of the present chapter is to thoroughly investigate the stability criteria of the chameleon models in a spatially homogeneous and isotropic cosmology. There are two arbitrary functions of the chameleon field $\phi$ to start with, namely $V=V(\phi)$ and $f=f(\phi)$. Here $V$ is the dark energy potential and $f$ determines the coupling of the chameleon field with the matter sector. We broadly classify the functions into two categories, exponential and non-exponential. So there are four combinations in all. We investigate the conditions for having a stable solution for the evolution for each of these categories. We find that there are possibilities of finding a stable evolution scenario where the universe may settle into a phase of accelerated expansion. However, if both $V$ and $f$ are exponential functions of $\phi$, the stability is very strongly dictated by the model parameters. In fact it is noted that in this latter case there is a possibility of a transient acceleration at the present epoch but the final stable configuration of the universe is that of a decelerated expansion.

The method taken up is the dynamical systems study. The field equations are written as an autonomous system and the fixed points are found out. A stable fixed point indicates a sink and thus marks the possible stable final configuration of the universe whereas an unstable fixed point, indicating a source, may describe the possible beginning of the evolution. Application of dynamical systems in cosmological problems, mostly for scalar field distributions, is already there in the literature \cite{gunzig}. For detailed discussions on some early work on such investigations, we refer to the monograph by Coley \cite{r13}. 

%The chapter is organized as follows. Section \ref{s2} deals with a chameleon scalar field model in a spatially flat homogeneous and isotropic universe. In section \ref{s3} the system of equations given in section \ref{s2} are written as an autonomous system. This section also includes a brief discussion of the method of the stability analysis that is used in the present work. The actual stability analysis in the four categories as mentioned is given in section \ref{s4}. The section \ref{s5} presents an example of a chameleon field where the chameleon mechanism together with the constraints imposed by laboratory experiments does not have any contradiction with the cosmological requirement of the decelerated expansion entering into an accelerated phase at a redshift of 0.74 \cite{farooq}. The final section \ref{s6} summarizes and discusses the results.

\section{A chameleon scalar field:} 
\label{s2}
The relevant action in gravity along with a chameleon field $\phi$ is given by
\begin{equation} \label{action}
A = \int\left[\frac{R}{16 \pi G} + \frac{1}{2}{\phi}_{,\mu}{\phi}^{,\mu} 
                 -V(\phi) + f(\phi) L_{m}\right]{\sqrt{-g}}~d^4x,
\end{equation}
where R is Ricci scalar, G is the Newtonian constant of gravity, $V(\phi)$ is the potential. Here $f(\phi)$ is a function of the chameleon field and determines the non-minimal coupling of the chameleon scalar field with the matter sector which is given by $L_{m}$. In what follows, $L_{m}$ is given by $-\rho_m$ alone. This indeed is not a standard description. However, for a collection of collision-less particles with non-relativistic speed, the rest mass of the particles completely dominates over the kinetic energy, the contribution from the pressure is negligible compared to the rest energy. The density is defined to be the contribution to that by these collision-less particles. This is thus quite a legitimate choice\cite{harko1, harko2} for a pressure-less fluid, which would be relevant for the dark matter distribution.

By varying the action with respect to the metric tensor components, one can find the field equations. For a spatially flat FRW spacetime given by the line element

\begin{equation} \label{metric}
ds^2 = dt^2 - a^2(t) \left( dr^2 + r^2d{\theta}^2 + r^2\sin^2{\theta} 
                      d{\phi}^2 \right),
\end{equation}

the field equations are written as
\begin{equation} \label{cons}
3\frac{{\dot{a}}^2}{a^2} = \rho_{m}f + \frac{1}{2}{\dot{\phi}}^2 + V(\phi),
\end{equation}
\begin{equation} \label{H}
2\frac{\ddot{a}}{a} + \frac{{\dot{a}}^2}{a^2} = -\frac{1}{2}{\dot{\phi}}^2 
                                                 + V(\phi),
\end{equation}
where the units are so chosen that $8 \pi G=1$. The fluid is taken in the form of pressureless dust $(p_m=0)$ consistent with a matter dominated universe and $ \rho_{m} $ is the matter density. Overhead dots denote differentiation with respect to the cosmic time $t$.

By varying the action with respect to the chameleon field $\phi$,  one can also find the wave equation as,
\begin{equation} \label{wave}
\ddot{\phi} + 3 H \dot{\phi} = -\frac{dV}{d\phi} - \rho_{m} \frac{df}{d \phi}.
\end{equation}

Equations (\ref{cons}), (\ref{H}) and (\ref{wave}) can be used to yield the equation

\begin{equation} \label{rho}
(\rho_{m} f)\dot{~} + 3 H (\rho_{m} f) = \rho_{m} \dot{\phi} \frac{df}{d \phi},
\end{equation}
which is actually the matter conservation equation.
On integration, the equation (\ref{rho}) yields
\begin{equation} \label{rhom}
\rho_{m} = \frac{\rho_{0}}{a^3}.
\end{equation} 
All of these equations (\ref{cons}), (\ref{H}), (\ref{wave}) and (\ref{rho}) are not independent. Any one of the last two can be derived from the other three as a consequence of the Bianchi identities. We take  (\ref{cons}), (\ref{H}) and (\ref{wave}) to constitute the system of equations. There are, however, four unknowns, namely $a, \phi, V, f$. The other variable, ${\rho}_{m}$, is known in terms of $a$ via the equation (\ref{rhom}). It is intriguing to note that even with the nonminimal coupling with the scalar field, the matter energy density itself still redshifts as $\frac{1}{a^3}$ (see reference \cite{nbsdkg})
\section{The Autonomous system:}
\label{s3}
By the introduction of following dimensionless variables  

$x= \frac{\dot{\phi}}{\sqrt{6} H}$, ~$y = \frac{\sqrt{V}}{\sqrt{3} H}$, ~ $z=\frac{\sqrt{\rho_{m} f}}{\sqrt{3} H}$,~ $\lambda=-\frac{1}{V}\frac{dV}{d\phi}$,~ $\delta=-\frac{1}{f}\frac{df}{d\phi}$, ~
$\Gamma = {V \frac{d^2 V}{d \phi^2}}/{(\frac{dV}{d \phi})^2}$ and $\tau = {f \frac{d^2 f}{d \phi^2}}/{(\frac{df}{d \phi})^2}$,
\\

the system of equations reduces to the following set,
\begin{equation}
x^{\prime} = -3 x + \sqrt{\frac{3}{2}} \lambda y^2 + \sqrt{\frac{3}{2}} \delta z^2 + x (3 x^2 + \frac{3}{2} z^2),
\end{equation}
\begin{equation}
y^{\prime} = - \sqrt{\frac{3}{2}} \lambda x y + y (3 x^2 + \frac{3}{2} z^2),
\end{equation}
\begin{equation}
z^{\prime} = - \frac{3}{2} z - \sqrt{\frac{3}{2}} \delta x z + z (3 x^2 + \frac{3}{2} z^2),
\end{equation}
\begin{equation}
\lambda^{\prime} = - \sqrt{6} \lambda^2 (\Gamma -1) x,
\end{equation}
\begin{equation}
\delta^{\prime} = - \sqrt{6} \delta^2 (\tau -1) x.
\end{equation}

A `prime' indicates differentiation with respect to $N = \ln ({a}/{a_0})$ with $a_0$ chosen as unity. One can write equation (\ref{cons}) in terms of these new variables as 
\begin{equation}
x^2 + y^2 + z^2 =1.
\end{equation} 
We use this equation as a constraint equation and the system which now effectively reduces to, 
 \begin{equation} \label{cx}
x^{\prime} = -3 x + \sqrt{\frac{3}{2}} \lambda (1-x^2-z^2) + \sqrt{\frac{3}{2}} \delta z^2 + x (3 x^2 + \frac{3}{2} z^2),
\end{equation}

\begin{equation} \label{z}
z^{\prime} = - \frac{3}{2} z - \sqrt{\frac{3}{2}} \delta x z + z (3 x^2 + \frac{3}{2} z^2),
\end{equation}

\begin{equation} \label{lambda}
\lambda^{\prime} = - \sqrt{6} \lambda^2 (\Gamma -1) x,
\end{equation}

\begin{equation} \label{delta}
\delta^{\prime} = - \sqrt{6} \delta^2 (\tau -1) x.
\end{equation}

We now use standard linear stability analysis to qualitatively analyse the system. It can be seen that some of the fixed points are non-hyperbolic in nature. For non hyperbolic fixed points, one can not use linear stability analysis. In the absence of a proper analytical process, a different strategy in such cases may be adopted. The solutions are numerically perturbed around the fixed points to check the stability in non hyperbolic cases. If the perturbed solutions asymptotically approach the fixed points, the corresponding fixed points are considered stable. This approach is quite standard in nonlinear dynamics\cite{r19} and  has already been utilised quite recently in a cosmological scenario\cite{nrnb}.

\section{Stability analysis of the chameleon model:}
\label{s4}
From definition of $\Gamma$ and $\tau$, it can be easily shown that $\Gamma = 1$ corresponds to the exponential form of the potential and $\tau = 1$ corresponds to  exponential form of the coupling. For the sake of making the system of equations a bit more tractable, we classify our system into four classes as described below. 

Class I : When $\Gamma \neq 1$ and $\tau \neq 1$, i.e.,  both the potential $V(\phi)$ and the coupling $f(\phi)$ are non-exponential functions of the chameleon field.

Class II : When $\Gamma=1$ and $\tau \neq 1$, the potential $V$ is an exponential function of the chameleon field but the coupling $f$ is any function of the chameleon field excluding an exponential function.

Class III: When $\Gamma \neq 1$ and $\tau = 1$ , the potential $V$ is any function excluding an exponential function but the coupling is an exponential function of the chameleon field .

Class IV : When $\Gamma = 1$ and $\tau = 1$, both the potential $V$ and the coupling $f$  are exponential functions of the chameleon scalar field $\phi$.
\subsection{Class I :When both the potential and the coupling with matter are non exponential functions of the chameleon field.}
In this class, the system formed by the equations (\ref{cx}), (\ref{z}), (\ref{lambda}) and (\ref{delta}), has the fixed points as listed in Table \ref{ct1}.\\

\begin{table}[H] 
\caption{Fixed points of Class I type models}
\begin{center}
%\begin{ruledtabular}
\begin{tabular}{|c|c|c|c|c|c|}
\hline 
 Points & $p_1$ & $p_2$ & $p_3$ & $p_4$ & $p_5$ \\ 
\hline 
$x$ & 0 & 0 & 0 & -1 & +1 \\ 
\hline 
$z$ & 0 & +1 & -1 & 0 & 0 \\ 
\hline 
$\lambda$ & 0 & $\lambda$ & $\lambda$ & 0 & 0 \\ 
\hline 
$\delta$ & $\delta$ & 0 & 0 & 0 & 0 \\ 
\hline 
\end{tabular}
%\end{ruledtabular}
\end{center}
\label{ct1}
\end{table} 

\begin{figure}[H]
\centering
\includegraphics[scale=0.2]{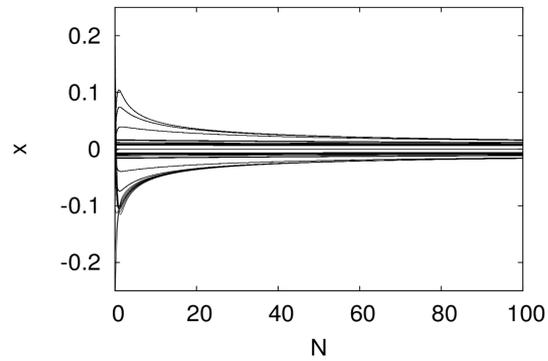} 
\caption{Plot of projections of perturbations on $x$ against N for Class I type models}
\label{c1}
\end{figure}

\begin{figure}[H]
\centering
\includegraphics[scale=0.3]{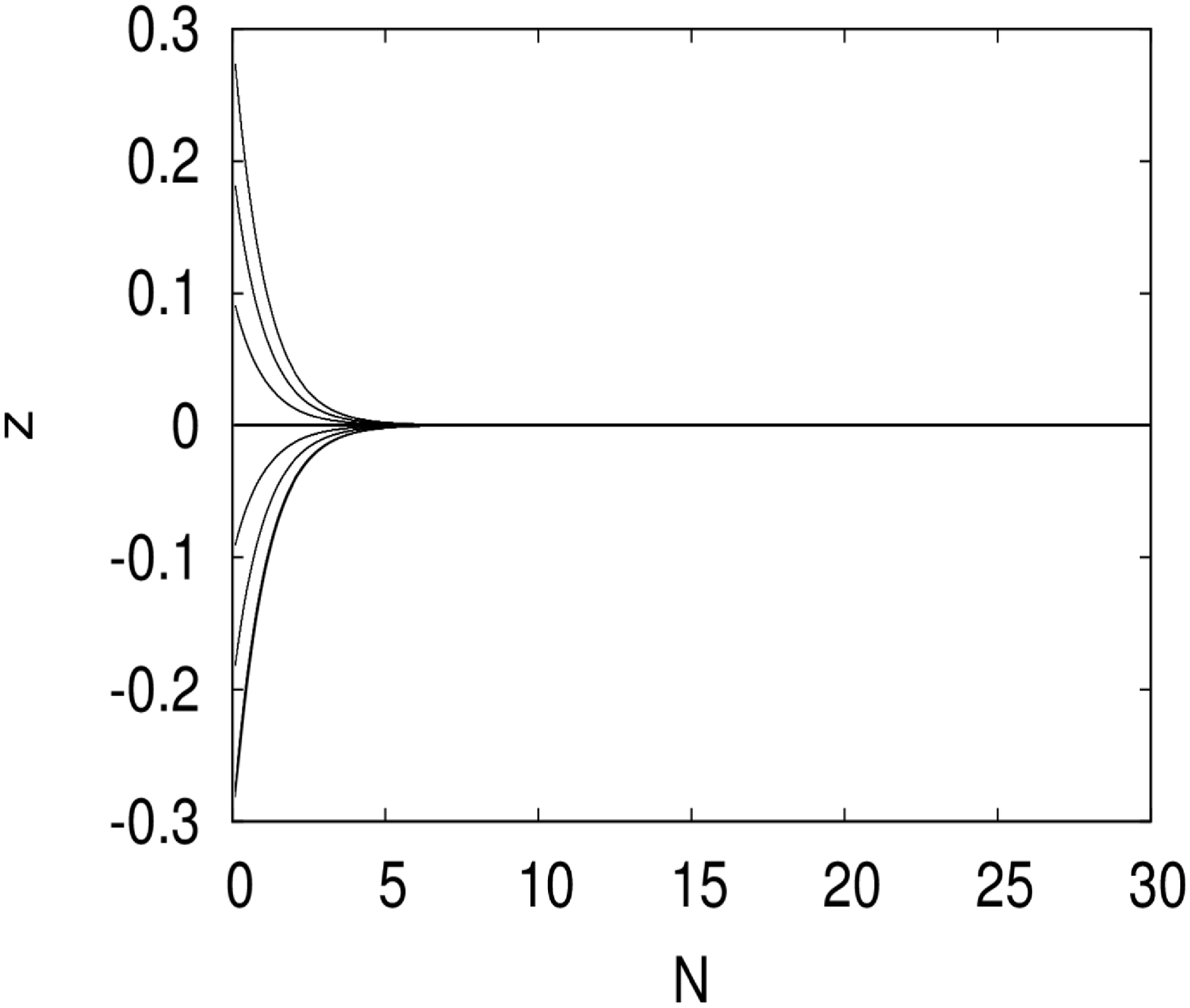}
\caption{Plot of projections of perturbations on $z$ against N for Class I type models}
\label{c2}
\end{figure}

\begin{figure}[H]
\centering
\includegraphics[scale=0.35]{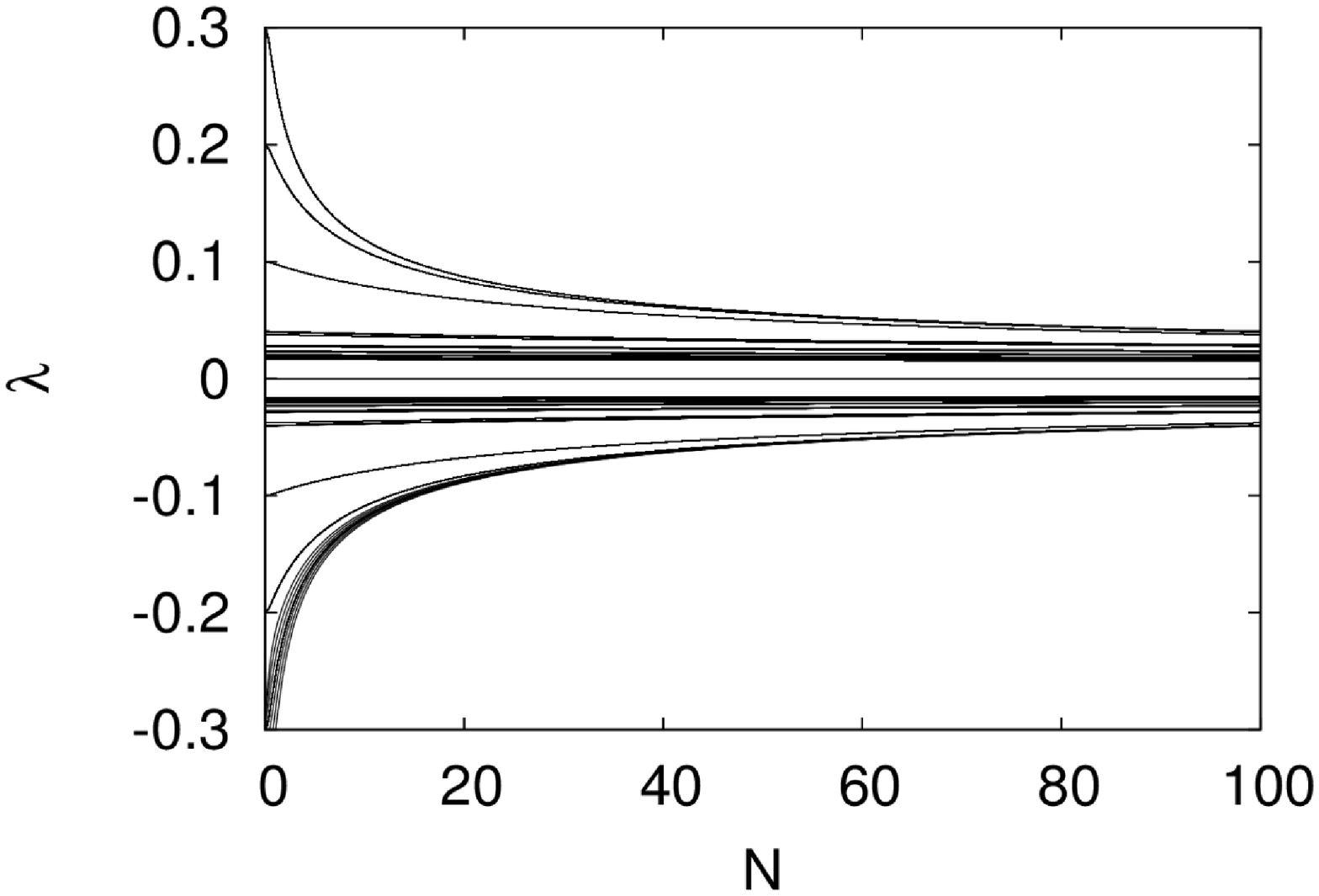}
\caption{Plot of projections of perturbations on $\lambda$ against N for Class I type models}
\label{c3}
\end{figure}

\begin{figure}[H]
\centering
\includegraphics[scale=0.4]{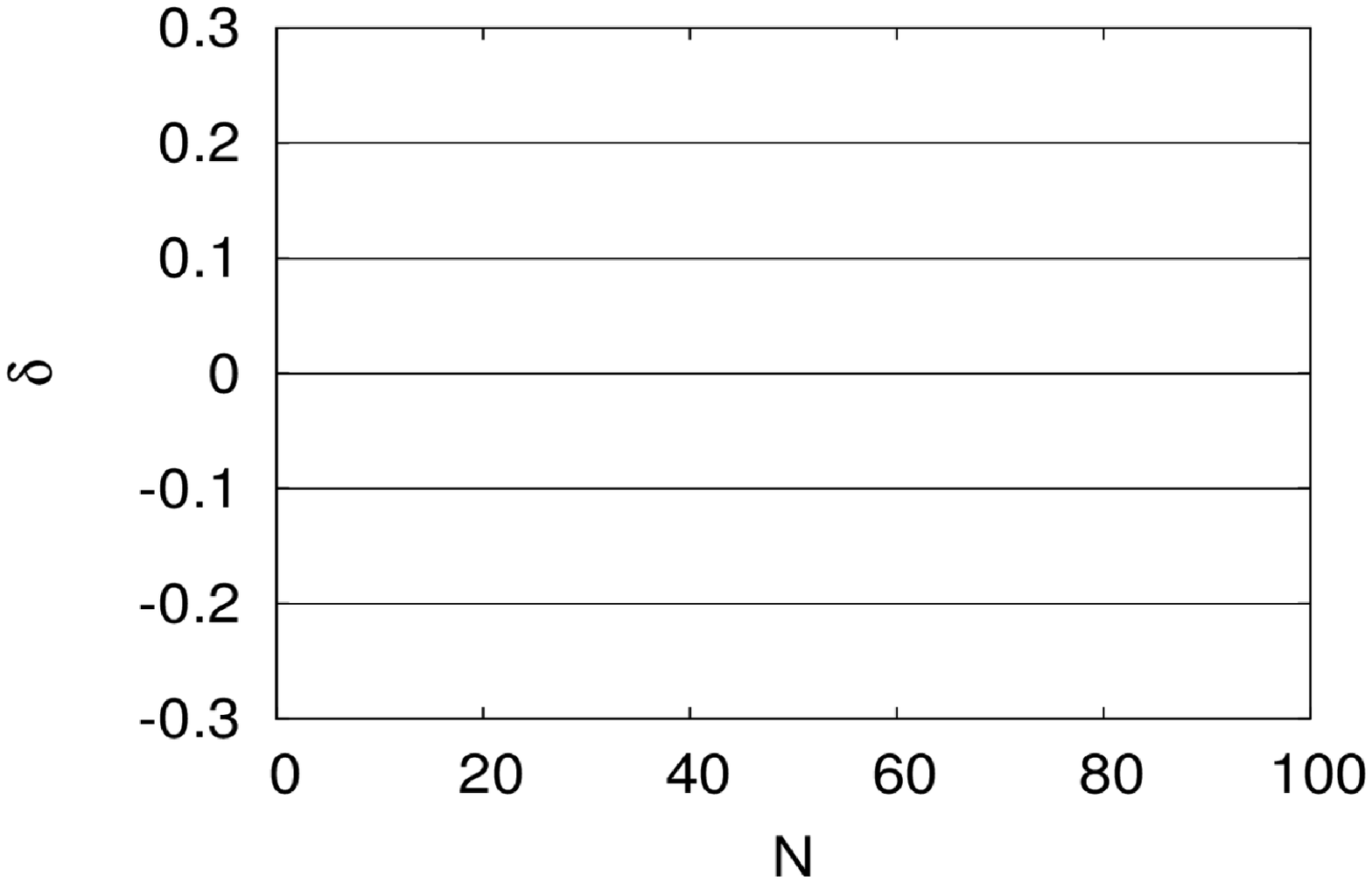}
\caption{Plot of projections of perturbations on $\delta$ against N for Class I type models}
\label{c4}
\end{figure}

In order to study the stability of the fixed points, the eigenvalues of the Jacobian matrix have been found out for all the fixed points which are shown in Table \ref{ct2}.
\begin{table}[H] 
\centering
\caption{Eigen values of the fixed points of Class I type models}
%\begin{ruledtabular}
\begin{tabular}{|c|c|c|c|c|}
\hline 
 Points & $\mu_1$ & $\mu_2$ & $\mu_3$ & $\mu_4$ \\ 
\hline 
$p_1$ & $-3$ & $-\frac{3}{2}$ & 0 & 0 \\ 
\hline 
$p_2$ & 3 & $-\frac{3}{2}$ & 0 & 0 \\ 
\hline 
$p_3$ & 3 & $-\frac{3}{2}$ & 0 & 0 \\ 
\hline 
$p_4$ & 6 & $\frac{3}{2}$ & 0 & 0 \\ 
\hline 
$p_5$ & 6 & $\frac{3}{2}$ & 0 & 0 \\ 
\hline 
\end{tabular} 
\label{ct2}
%\end{ruledtabular}
\end{table}
As the Jacobian matrix of the fixed points all have at least one zero eigen values, these are all non hyperbolic fixed points. One can not use the  linear stability analysis in this situation. Fixed points $p_2,p_3,p_4,$ and $p_5$ has at least one positive eigenvalue, so these fixed points are in fact unstable but the one at $p_1$ has two negative eigenvalues and two zero eigenvalues and thus requires further investigation. It must be made clear that $p_1$ is actually an infinite set of fixed points as the value of $\delta$ is arbitrary in this case, as indicated in Table \ref{ct1}. We perturb the system around the fixed points $p_1$. It is not possible to draw a 4D phase plot and the 3D phase plot looks too obscure to draw conclusions from. We adopt the following strategy. We plot the projection of perturbations on $x,z,\lambda$ and $\delta$ axes separately. In figure \ref{c1}, the whole $x=0$ line corresponds to the fixed point. For a small perturbation of the solution around the fixed point $p_1$, the evolution of the solution with $N$ is studied numerically. It is evident from figure \ref{c1} that the perturbed solutions asymptotically approach $x=0$ as $N \longrightarrow \infty$. Figure \ref{c2} and \ref{c3} show that the projection of perturbations around  $z=0$ and $\lambda=0$ approaches  $z=0$ and $\lambda=0$ respectively as $N \longrightarrow \infty$.  It is interesting to note that the results given by this perturbation technique is completely consistent with the finding that $\delta$ is arbitrary. Any perturbation of the system around any value of $\delta$ renders it remaining constant at the perturbed value without any further evolution (figure \ref{c4}). From this behaviour of the system near $p_1$, we can conclude that $\delta$ axis is an attractor line\cite{r19}. 

Any heteroclinic orbit in the phase space of the system starts from an unstable fixed point and ends at a stable fixed point. So the universe is apt to start evolving from one of these unstable fixed points $p_2$ to $p_5$ as a result of any small perturbation and approach towards the $\delta$ axis which is an attractor line as $N \longrightarrow \infty$. 
\vspace{-2em}
\vspace{-2em}
\vspace{-2em}
\vspace{-2em}
The set of fixed points $p_1$, given by $x=z=\lambda=0$ with an arbitrary $\delta$, is thus stable (an attractor) and gives the final state of the universe. The physical state of the universe in this case, as $N \longrightarrow \infty$ (i.e., $a \longrightarrow \infty$), is consistent with $\Omega_{\phi} \longrightarrow 1, ~  \Omega_{m}f \longrightarrow 0$. Here we define $\Omega_{\phi} = \frac{\frac{1}{2} \dot{\phi}^2 + V }{3 H^2}$ and $\Omega_m = \frac{\rho_m}{3 H^2}$.

It deserves mention that as $x=0$, the contribution from the kinetic part of the chameleon goes to zero and this complete domination of the chameleon field is entirely given by the potential part. From equation (\ref{H}), one can see that the universe settles to an accelerated phase with $q(- \frac{a \ddot{a}}{\dot{a}^2}) \longrightarrow -1$.

The unstable fixed points $p_{2} $ and $p_3 $ indicate states with $\rho_{m}\longrightarrow \infty$ and $\Omega_{\phi}\longrightarrow 0$. This is consistent with the physical requirement of the early phase of the evolution. However the fixed point $p_3$ will henceforth be disregarded as unphysical, as for a positive $H$, negative values of $z$ do not represent any physical state.
  
Like $p_2$, the other two fixed points $p_4$ and $p_5$ are also unstable fixed points and describe physical states with the relative magnitude of $\Omega_{m}f$ and $\Omega_{\phi}$ being reversed. However, it also deserves mention that  for both $p_4$ and $p_5$, $p_{\phi}$ remains positive (equal to $\rho_{\phi}$) and thus the scalar field does not really play the role of a dark energy. It deserves mention that $x$ and $\lambda$ as given by figures \ref{c1} and \ref{c3} might appear to be leading to a non-zero constant value. But the ordinates are not really parallel to the horizontal axis, they do vary, but only too slowly. It is checked that they indeed approach zero for a very large value of $N$.
\subsection{Class II: When the potential is an exponential function of the chameleon field but the coupling with matter is any function of the chameleon field except an exponential one:}
In this class $\lambda$ is a constant and equations (\ref{cx}), (\ref{z}) and (\ref{delta}) form the system of equations. The fixed points of the system are given in Table \ref{ct3} and the eigenvalues of the Jacobian matrix at the fixed points are given in Table \ref{ct4}.\\

\begin{table}[H] 
\centering
\caption{Fixed points of Class II type models}
%\begin{ruledtabular}
\begin{tabular}{|c|c|c|c|c|c|c|c|c|}
\hline 
Points & $q_1$ & $q_2$ & $q_3$ & $q_4$ & $q_5$ & $q_6$ & $q_7$ & $q_8$\\ 
\hline 
x & 0 & 0 & 0 & +1 & -1 & $\frac{\lambda}{\sqrt{6}}$ & $\sqrt{\frac{3}{2 \lambda^2}}$ & $\sqrt{\frac{3}{2 \lambda^2}}$\\ 
\hline 
z & 0 & +1 & -1 & 0 & 0 & 0 & $ \sqrt{1-\frac{3}{\lambda^2}} $ & $ -\sqrt{1-\frac{3}{\lambda^2}} $ \\ 
\hline 
$\delta$ & $\delta$ & 0 & 0 & 0 & 0 & 0 & 0 & 0 \\ 
\hline 
\end{tabular} 
%\end{ruledtabular}
\label{ct3}
\end{table}

\begin{table}[H] 
\centering
\caption{Eigen values of the fixed points of Class II type models}
%\begin{ruledtabular}
\begin{tabular}{|c|c|c|c|}
\hline 
{Points}  & $\mu_1$  &  $\mu_2$ & $\mu_3$ \\ 
\hline 
$q_1$ & $-3$ & $- \frac{3}{2}$ & 0 \\ 
\hline 
$q_2$ & 3 & $- \frac{3}{2}$ & 0 \\ 
\hline 
$q_3$ & 3 & $- \frac{3}{2}$ & 0 \\ 
\hline 
$q_4$ & 3 & $6 - \sqrt{6} \lambda$ &  0 \\ 
\hline 
$q_5$ & 3 & $6 + \sqrt{6} \lambda$ &  0 \\ 
\hline 
$q_6$ & $-3 + \frac{\lambda^2}{2}$ &  $- \frac{3}{2} + \frac{\lambda^2}{2}$ & 0 \\ 
\hline 
$q_7$ and $q_8$ & $\mu_{27}$ & $\mu_{37}$ & 0 \\ 
\hline 
\end{tabular} 
%\end{ruledtabular}
\label{ct4}
\end{table}

Here  the symbols, $\mu_{27}= \frac{3}{4} (-1-\frac{1}{\lambda} \sqrt{24- 7 \lambda^2})$ and $\mu_{37}= \frac{3}{4} (-1+\frac{1}{\lambda} \sqrt{24- 7 \lambda^2})$, are used for brevity.

 Existence of the fixed point at $q_1$ demands $\lambda =0$, indicating that $V$ is a constant in this case. This could have a bearing on the chameleon mechanism. This is a non isolated fixed point and has one zero eigenvalue at each point, defined by the value of $\delta$, of the equilibrium set. These type of fixed points are called normally hyperbolic fixed points (\cite{r13}). The stability of a normally hyperbolic fixed point depends on the signatures of the remaining eigenvalues. If sign of the remaining eigenvalues are negative then the fixed point is a stable fixed point otherwise the fixed point is an unstable one. For $q_1$, the remaining eigenvalues are $-3$ and $-\frac{3}{2}$, so $q_1$ is a stable fixed point.
 
Other fixed points are  non hyperbolic and not even normally hyperbolic. The fixed point $q_6$ can have one zero eigenvalue and two negative eigenvalues if $\lambda^2 < 3$ whereas $q_7$ can also have one zero eigenvalue with two negative eigenvalues if $\frac{24}{7} \leq \lambda^2 \leq 3$. So in the corresponding  parameter range $q_6$ and $q_7$ need further investigation as in case of Class I type of models. It is found that for a small perturbation about these fixed points ($q_6$, $q_7$ and $q_8$), the solutions do not approach towards the fixed points as $N \longrightarrow \infty$, so these fixed points are unstable. The fixed points $q_2, q_4$ and $q_5$ are also unstable fixed points as they have at least one positive eigenvalue. So $q_2$ to $q_8$ could be the beginning of the evolution and $q_1$  may be the possible ultimate fate of the universe. It also deserves mention that for $\lambda =0$, $q_6$ is in fact a subset of $q_1$. It is interesting to note that for $\lambda = 0$, the fixed point $q_7$ is realized only for $x \longrightarrow \infty$.  The fixed points $q_3$ and $q_8$ are not analysed on the plea that for a negative $z$, the state is not physical as it corresponds to a negative value of $H$.

Amongst the fixed points, potentially the beginning of the universe (i.e., unstable fixed points), $q_2$ and $q_3$ has the possibility of $\Omega_{m}f =1$ and $\Omega_{\phi} =0$ which is a desired situation. The only stable fixed point, $q_1$, yields the possibility of a final fate where $\Omega_{m} f =0$ and $\Omega_{\phi} =1$ with $q\longrightarrow -1$.

\subsection{Class III :When the coupling with matter is an exponential function of the chameleon field but the potential is any function of the same except an exponential function:}When the coupling of the chameleon with the matter field is an exponential function of $\phi$, one has $\delta =$ constant. The system of equations is formed by equations (\ref{cx}), (\ref{z}) and (\ref{lambda}). Fixed points of the system are given in Table \ref{ct5} and the eigenvalues of the Jacobian matrix at the fixed points are given in Table \ref{ct6}.

\begin{table}[H] 
\centering
\caption{Fixed points of Class III type models}
%\begin{ruledtabular}
\begin{tabular}{|c|c|c|c|c|c|c|c|}
\hline 
Points & $m_1$ & $m_2$ & $m_3$ & $m_4$ & $m_5$ & $m_6$ &$m_7$ \\ 
\hline 
x & 0 & 0 & 0 & 1 & -1 & $\sqrt{\frac{2}{3}} \delta$ & $\sqrt{\frac{2}{3}} \delta$ \\ 
\hline 
z & 0 & $+1$ & $-1$ & 0 & 0 & $\sqrt{1-\frac{2 \delta^2}{3}}$ & $-\sqrt{1-\frac{2 \delta^2}{3}}$ \\ 
\hline 
$\lambda$ & 0 & $\lambda$ & $\lambda$ & 0 & 0 & 0 & 0 \\ 
\hline 
\end{tabular} 
\label{ct5}
%\end{ruledtabular}
\end{table}

\begin{table}[H] 
\centering
\caption{Eigen values of the fixed points of Class III type models}
%\begin{ruledtabular}
\begin{tabular}{|c|c|c|c|}
\hline 
Points & $\mu_1$ & $\mu_2$ & $\mu_3$ \\ 
\hline 
$m_1$ & $-3$ & $-\frac{3}{2}$ & 0 \\ 
\hline 
$m_2$ & 3 & $-\frac{3}{2}$ & 0 \\ 
\hline 
$m_3$ & 3 & $-\frac{3}{2}$ & 0 \\ 
\hline 
$m_4$ & 6 & $\frac{3}{2} - \sqrt{\frac{3}{2}}\delta$ & 0 \\ 
\hline 
$m_5$ & 6 & $\frac{3}{2} + \sqrt{\frac{3}{2}}\delta$ & 0 \\ 
\hline 
$m_6$ & $\frac{1}{2} (-3+2 \delta^2)$ & $(3+2 \delta^2)$ & 0  \\ 
\hline 
$m_7$ & $\frac{1}{2} (-3+2 \delta^2)$ & $(3+2 \delta^2)$  & 0\\ 
\hline 
\end{tabular} 
%\end{ruledtabular}
\label{ct6}
\end{table}
%\end{center}
The existence of fixed point $m_2$ requires the condition that $\delta=0$ indicating that $f$ is a constant! This would require that the nonminimal coupling between the chameleon field and matter becomes trivial and certainly there is a breakdown of the chameleon mechanism. Unphysical states (negative values of $z$) given by the fixed points $m_3$ and $m_7$ are not discussed.

\begin{figure}[H]
\centering
\includegraphics[scale=0.6]{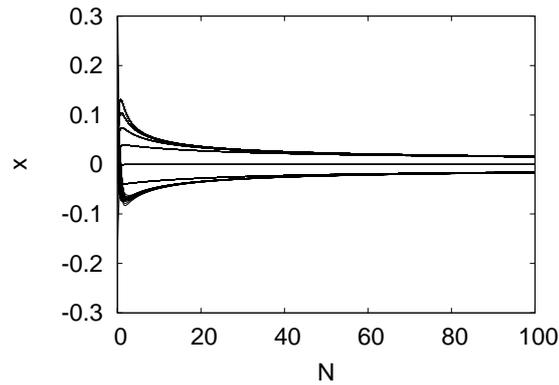}
\vspace{2em}
\caption{Plot of projections on $x$ vs $N$ for Class III type of models}
\label{c5}
\end{figure}

\begin{figure}[H]
\centering
\includegraphics[scale=0.6]{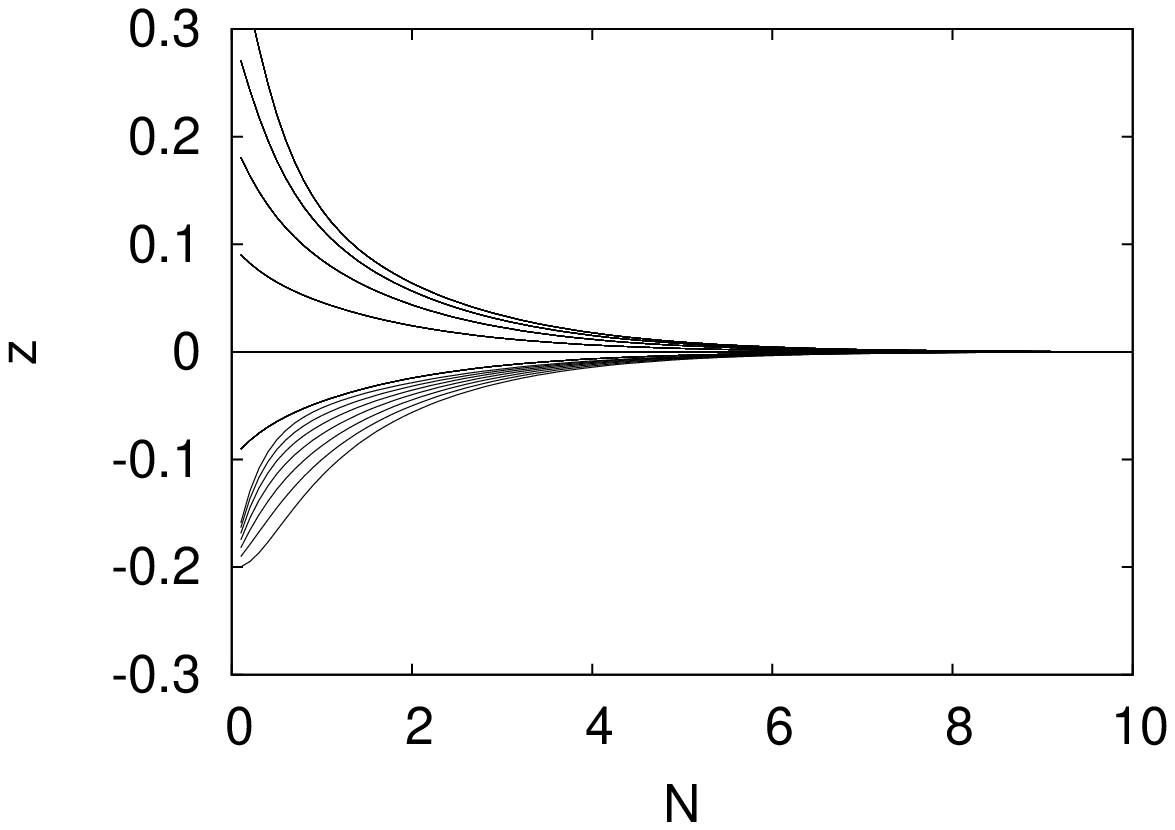}
\vspace{2em}
\caption{Plot of projections on $z$ vs $N$ for Class III type of models}
\label{c6}
\end{figure}

\begin{figure}[H]
\centering
\includegraphics[scale=0.6]{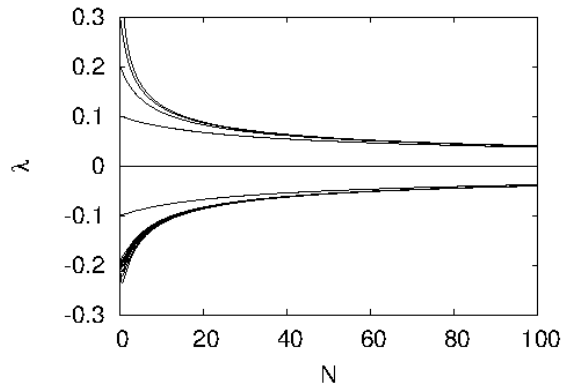}
\vspace{2em}
\caption{Plot of projections on $\lambda$ vs $N$ for Class III type of models}
\label{c7}
\end{figure}

 Fixed point $m_1$ has one zero eigenvalue and two negative eigenvalues. It is a non hyperbolic fixed point and thus one can not use the linear stability analysis in this case. So we perturbed the system from the fixed point and numerically solved the system of equations for  each perturbations. Plots of the projections of the perturbations on $x,z$ and $\lambda$ against $N$ in figures (\ref{c5}), (\ref{c6}) and (\ref{c7}) show quite convincingly that for large $N$, the solutions approach the fixed point which is thus an attractor. So $m_1$ is apt to represent the final state of the evolution.
 
  Other fixed points are unstable. In the beginning, the universe could be at any of these unstable fixed points and with a small perturbation from these unstable fixed points might start evolving towards the attractor $m_1$ which can be the possible final state of the universe for which $\Omega_{m}f \longrightarrow 0$ and $\Omega_{\phi} \longrightarrow 1$ are evident possibilities. The final stage is completely dominated by the potential $V$ which is a constant, effectively giving a cosmological constant, with $q\longrightarrow -1$.
  
   Amongst the unstable fixed point, $m_2$ is definitely favoured as they indicate $\Omega_{m}f \longrightarrow 1$ and $\Omega_{\phi}\longrightarrow 0$ for  the initial epoch. For the unstable fixed points $m_4$ and $m_5$, although $\Omega_{\phi}\longrightarrow 1$, the chameleon cannot act as a dark energy as $p_{\phi}$ is not negative. The fixed point  $m_6$ will indicate a matter dominated early epoch if $\delta \leq \frac{\sqrt{3}}{2}$ and could well be amongst the viable options. Like Case I, in this case also, $x$ and $\lambda$ actually approach zero which had been revealed on an  extension of the horizontal axis, roughly up-to 1000.
\vspace{-2em}
\vspace{-2em}
\vspace{-2em}
\vspace{-2em}
\subsection{Class IV :When the potential and the coupling with matter both are exponential functions of the chameleon field :} In this class of the model both $\lambda$ and $\delta$ are constants. Equations (\ref{cx}) and (\ref{z}) constitute the system of equations for the effectively two-dimensional problem. The fixed points of the system are given in Table \ref{ct7}.
%\squeezetable
\begin{table}[H] 
\centering
\caption{Fixed points of Class IV type models}
%\begin{ruledtabular}
\begin{tabular}{|c|c|c|c|c|c|c|c|}
\hline 
Points & $n_1$ & $n_2$ & $n_3$ & $n_4$ & $n_5$ & $n_6$ & $n_7$ \\ 
\hline 
x & -1 & 1 & $\frac{\lambda}{\sqrt{6}}$ & $\sqrt{\frac{2}{3}} \delta$ & $\sqrt{\frac{2}{3}} \delta$ & $-\sqrt{\frac{3}{2}} \frac{1}{\delta - \lambda}$ & $-\sqrt{\frac{3}{2}} \frac{1}{\delta - \lambda}$ \\ 
\hline 
z & 0 & 0 & 0 & $-\sqrt{1- \frac{2 \delta^2}{3}}$ & $\sqrt{1- \frac{2 \delta^2}{3}}$ & $- \frac{\sqrt{-3- \delta \lambda + \lambda^2}}{\delta - \lambda}$ & $ \frac{\sqrt{-3- \delta \lambda + \lambda^2}}{\delta - \lambda}$ \\ 
\hline 
\end{tabular} 
\label{ct7}
%\end{ruledtabular}
\end{table}

This is a 2D system, so the stability of the fixed points can be checked by both the methods, like from sign of eigen values of the Jacobian matrix or from trace and determinant of the Jacobian matrix at the fixed point. Depending on the degree of simplicity they offer, one of these two methods are used for different fixed points. There are seven fixed points, $n_1$ to $n_7$. We analyse the stability of the fixed points $n_1$ to $n_5$ by the signature of the  eigenvalues of corresponding Jacobian matrix (see Table \ref{ct8}). For $n_6$ and $n_7$, the trace and determinant of the Jacobian matrix are looked at. For the present case when both $V$ and $f$ are exponential functions of $\phi$, both of $\lambda$ and $\delta$ are constants and the fixed points are found in terms of these two constants of the theory. The stability criteria crucially depend on the values of $\lambda$ and $\delta$. Depending on the values of these two constants, there are special cases. One of them, namely $n_3$ with $\lambda =0$, which results in $x=0, z=0, \lambda =0$, has an interesting feature that it actually coincides with $p_1, q_1$ and $m_1$. This fixed point will be taken up in the last section.\\

%\squeezetable
\begin{table}[H] 
\centering
\caption{Eigen values of the fixed points of Class IV type models}
%\begin{ruledtabular}
\begin{tabular}{|c|c|c|c|}
\hline 
Points &  $\mu_1$ & $\mu_2$ & Conditions of stability \\ 
\hline 
$n_1$ & $\frac{1}{2} (3 + \sqrt{6} \delta)$ & $6+\sqrt{6} \lambda$ & $\lambda<-\sqrt{6}, \delta < - \sqrt{\frac{3}{2}}$ \\ 
\hline 
$n_2$ & $\frac{1}{2} (3 - \sqrt{6} \delta)$ & $6-\sqrt{6} \lambda$ & $\lambda>\sqrt{6}, \delta > \sqrt{\frac{3}{2}}$ \\ 
\hline 
$n_3$ & $\frac{1}{2}(-6+\lambda^2)$ & $\frac{1}{2}(-3-\lambda \delta+\lambda^2)$ & $\lambda^2 < 6, \delta> \frac{\lambda^2-3}{\lambda}$ \\ 
\hline 
$n_4$ & $-\frac{3}{2} + \delta^2$ & $3+ 2 \delta^2 - 2 \delta \lambda$ & $\delta^2 < \frac{3}{2}, \lambda > \frac{3+ 2 \delta^2}{2 \delta}$ \\ 
\hline 
$n_5$ & $-\frac{3}{2} + \delta^2$ & $3+ 2 \delta^2 - 2 \delta \lambda$ & $\delta^2 < \frac{3}{2}, \lambda > \frac{3+ 2 \delta^2}{2 \delta}$ \\ 
\hline 
\end{tabular} 
\label{ct8}
%\end{ruledtabular}
\end{table}

Trace and determinant of the Jacobian matrix at $n_6$ and $n_7$ are $TrA = \frac{-6 \delta + 3 \lambda}{2(\delta - \lambda)}$ and $DetA= \frac{-3(3+2\delta (\delta - \lambda))(3 + (\delta - \lambda)) \lambda}{2(\delta - \lambda)^2}$ respectively. Conditions for stability of $n_6$ and $n_7$ are,\\

1. $\lambda < -\sqrt{6} , \delta > \frac{\lambda}{2} + \frac{1}{2} \sqrt{\lambda^2 - 6}$.

2. $\lambda = - \sqrt{6}, \delta > - \sqrt{\frac{3}{2}}$.

3. $-\sqrt{6} < \lambda < 0, \delta > \frac{-3 + \lambda^2}{\lambda}$.

4. $0< \lambda < \sqrt{6}, \delta < \frac{-3 + \lambda^2}{\lambda}$.

5. $\lambda = \sqrt{6}, \delta < \sqrt{\frac{3}{2}}$.

6.$\lambda> \sqrt{6}, \delta < \frac{\lambda}{2} - \frac{1}{2} \sqrt{ 6- \lambda^2} $.\\

\begin{figure}[H]
\centering
\includegraphics[scale=0.5]{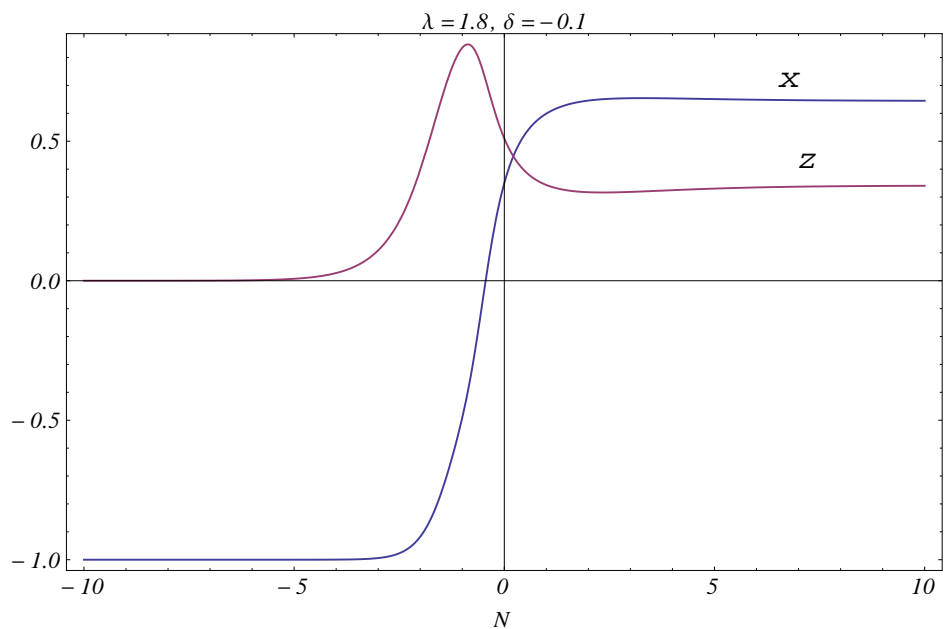}
\caption{Plot of x and z against N for Class IV models}
\label{c8}
\end{figure}

\begin{figure}[H]
\centering
\includegraphics[scale=0.4]{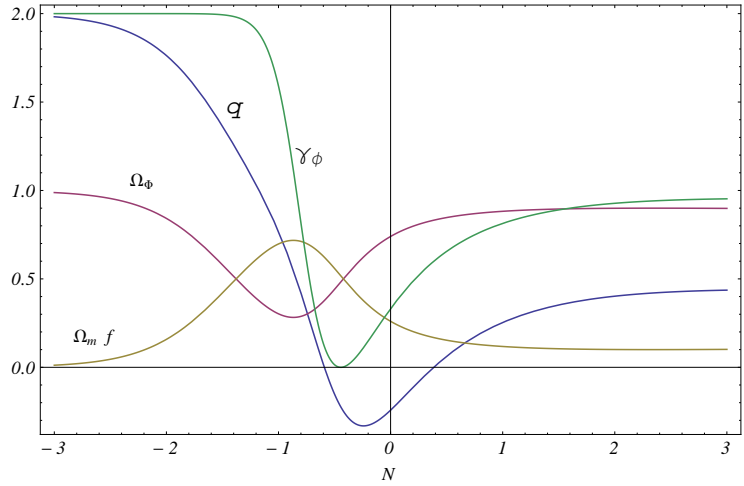}
\caption{Plot of physical parameters q, $\gamma_\phi, \Omega_{\phi} $ and $\Omega_{\phi} f$ against N for Class IV models for $\lambda = 1.8$ and $\delta = -0.1$}
\label{c9}
\end{figure}

  The fixed point $n_4$ is irrelevant as it requires a negative $z$ and thus a negative $H$ and $n_6$ is relevant only in cases where $\delta - \lambda$ is negative for the same reason. In what follows, an example of an unstable and that of a stable fixed point of this class are given. We choose $\Omega_{\phi}=0.73$ and the deceleration parameter $q=0.53$ \cite{r29} in order to pick up some relevant values of the model parameters $\lambda$ and $\delta$. With this value of $\Omega_{\phi}$, it is easy to see from the definitions of $x$ and $y$ that $x_0^{2} + y_0^{2} = 0.73$ and thus from the constraint $x^{2}+y^{2}+z^{2} = 1$, one has $ z_0^{2} = 0.27$, i.e., $z= 0.51$. From the field equations (\ref{cons}) and (\ref{H}), one can write 
\begin{equation}
 q=3x^{2}+\frac{3}{2}z^{2}-1,
\end{equation}
which yields $x=\pm 0.346$. Now from the Table \ref{ct7}, Table \ref{ct8} and list of the stability conditions of $n_6$ and $n_7$, one can check that for $\lambda = 1.8$ and $\delta = -0.1$, one has $n_1$ as an unstable fixed point and $n_6$ as a stable one. So the universe is apt to start its evolution from $n_1$ and settle into a final configuration at $n_6$. The behaviour of $x, z$ against $N$ and that of the cosmological parameters $q$, $\Omega_m$, $\Omega_{\phi}$ and $\gamma_{\phi}$ against $N$ are given in figures \ref{c8} and \ref{c9} respectively. It is easy to check that the universe starts with a deceleration ($q=2$) but with $\Omega_{\phi}=1$ and $\Omega_m =0$ and settles down to final phase of decelerated expansion with $q=0.42$ and again the same state of  $\Omega_{\phi}=1$ and $\Omega_m f =0$. There is an accelerated expansion in between, around the present stage of evolution (see figure \ref{c9}). So this is a nice example of a transient acceleration.

\begin{figure}[H]
\centering
\includegraphics[scale=0.8]{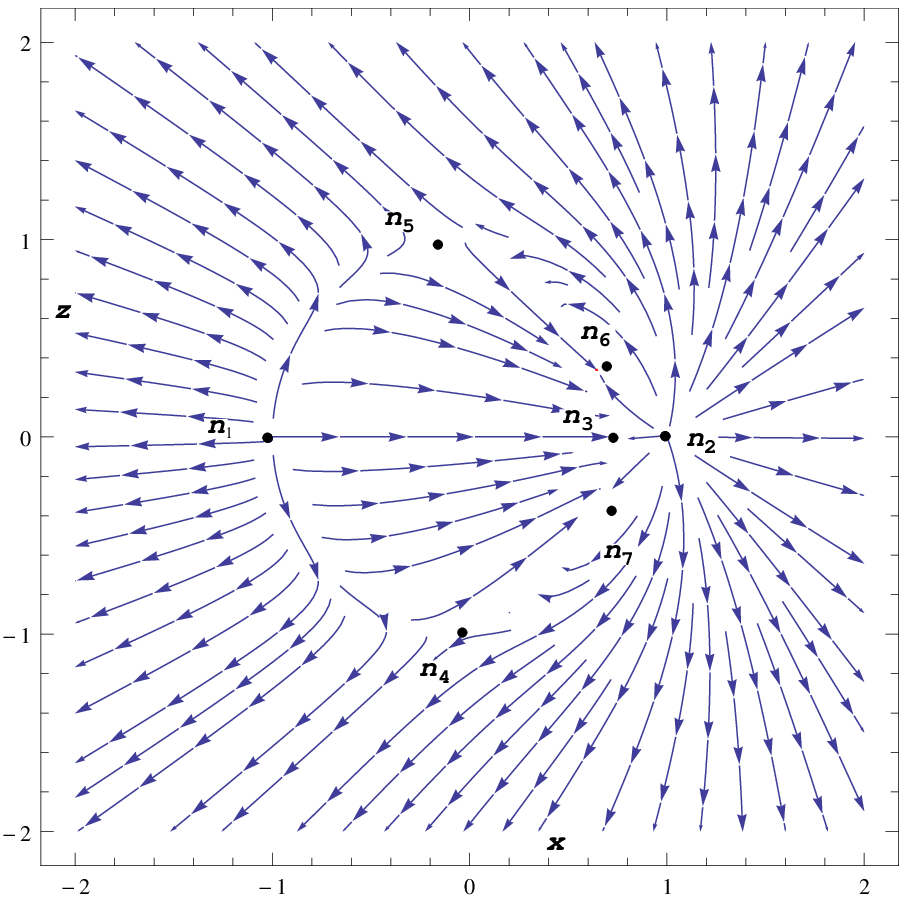}
\caption{Phase plot of the Class IV type models}
\label{c10}
\end{figure}

\begin{figure}[H]
\centering
\includegraphics[scale=0.8]{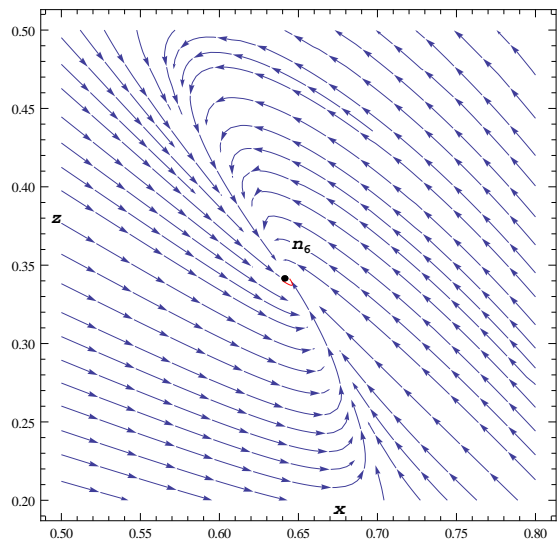}
\caption{Phase plot of the Class IV type models near $n_6$}
\label{c11}
\end{figure}

As this case effectively reduces to a 2-dimensional problem, one can draw the phase plot in $x$ and $z$, for the given values of $\lambda$ and $\delta$, 1.8 and -0.1 respectively. The plot is shown in figure \ref{c10}. In figure \ref{c11}, we zoom the plot around $q_6$ in order to understand the stability a bit more clearly.

\vspace{-1 cm}

\section{Chameleon mechanism and acceleration of the universe}
\label{s5}
Khoury discussed the restriction of the mass of the chameleon field vis-a-vis the coupling of the chameleon field with matter considering the compatibility of adaptation of the chameleon with the ambience and the requirement of the laboratory based experiments\cite{khoury}. The example taken up is the one for which $ V(\phi) = \frac{M^{4+n}}{\phi^n}$ and $f(\phi)= \xi \frac{\phi}{M_{pl}}$. Here $M$ and $n$ are constants and $\xi$ determines the strength of the coupling $f$. The allowed band in the $m_\phi$ vs $\xi$ plot was worked out. 

\begin{figure}[H]
  \centering
 \includegraphics[scale=0.6]{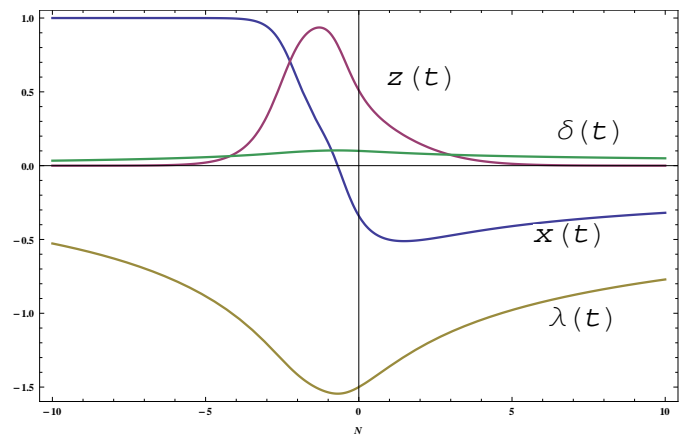}
 % sol1.eps: 0x0 pixel, 300dpi, 0.00x0.00 cm, bb=0 0 407 263
 \caption{Plot of $x,z, \lambda$ and $\delta$ against $N$ for $\lambda_0 = -1.5$ and $\delta_0 = 0.1$.}
 \label{c12}
 \end{figure}

\begin{figure}[H]
\centering
 \includegraphics[scale=0.6]{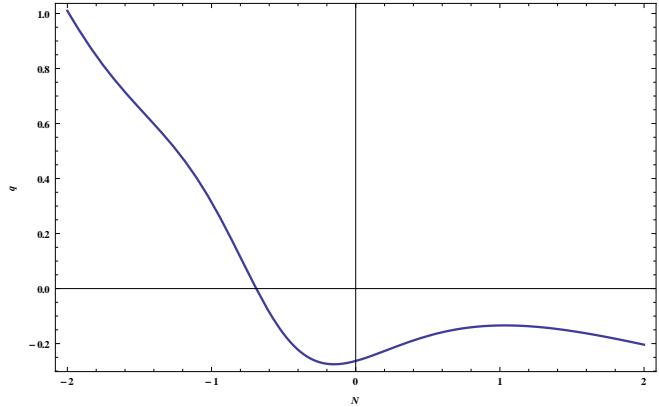}
 % sol1.eps: 0x0 pixel, 300dpi, 0.00x0.00 cm, bb=0 0 407 263
 \caption{Plot of deceleration parameter ``q'' against $N$ for $\lambda_0 = -1.5$ and $\delta_0 = 0.1$.}
 \label{c13}
 \end{figure}

The example is clearly in the Class I of the present work where both of $V$ and $f$ are non-exponential functions. In this case $\Gamma = 1 + \frac{1}{n^2}$ and $\tau = 0$, both are independent of $\xi$. As an illustration, we take up the case for $n=4$. With the boundary values of $x$, $z$ chosen from observational constraints as in section 3.4.4, one can plot the variables $x, z, \lambda$ and $\delta$ (figure \ref{c12}). The plot of the deceleration parameter $q$ against $N$ (figure \ref{c13}) shows that the universe has a smooth transition from a decelerated to an accelerated expansion approximately at $N = -0.55$ which corresponds to a red-shift of 0.74, which perfectly matches the observation\cite{ob}. So we find one example where this transition happens independent of the coupling of the scalar field with matter and hence posing no threat to the allowed band between $\xi$ and the mass of the scalar field.  It also deserves mention that the system has evolved from an unstable fixed point $(1,0,0,0)$ and approaches to a stable fixed point $(0,0,0,\delta)$, which also agrees very well with the present dynamical systems analysis of Class I models. 

\vspace{-2em}
\section{Discussion}
\label{s6}
The stability of the various chameleon scalar field models are investigated in the context of the present accelerated expansion of the universe in the present chapter. The aim is not to suggest any new model of dark energy, but rather to look at the various possibilities where a chameleon field can indeed serve as driver of the acceleration, starting from a decelerated situation. An unstable fixed point might describe the initial stage of the universe from where a small perturbation could trigger the start of the evolution whereas a stable fixed point is apt to describe the final stage of the universe. Unlike many of the dark energy models, a chameleon field has the distinct possibilities of being detected and hence that of being nullified as well. Thus it warrants attention regarding the choice of the favoured combination of the dark energy potential and the coupling with matter. 

For the sake of convenience, the functions $V$ and $f$ are classified as either exponential or not. So in all there are four such combinations, which are quite extensively studied in the present work. In fact any old (or new) chameleon model with given $V=V(\phi)$ and $f=f(\phi)$, the stability criteria need not be checked afresh. The present investigation provides a complete set of choices for $V(\phi)$ and $f(\phi)$ and can serve as the diagnostics. 

It has been mentioned in Section 2.2 that there are actually four unknown quantities, namely $a, \phi, V$ and $f$. In class I unknown and the autonomous system is 4 dimensional. Class II and III have one of $v$ and $f$ exponential and the system reduces to a 3-dimensional one. Class IV has both of $V$ and $f$ exponential and the system reduces to a 2-dimensional problem. The autonomous system is thus verified to be consistent with the number of unknowns. 

 It is found that if both $V$ and $f$ are exponential functions,  $\delta$ and $\lambda$ are not dynamical variables but are rather some parameters. The stability of the fixed points depends on the values of these parameters and it opens up a possibility for a transient acceleration for the universe around the present epoch. The other observation is that when $V$ is exponential and $f$ is not, $V$ effectively resembles a cosmological constant and when $f$ is exponential but $V$ is not, the matter-chameleon coupling is actually broken. 

It is interesting to note that the fixed point $n_3$ with $\lambda = 0$ and $p_1$, $q_1$ and $m_1$ are in fact the same, where there is no fluid and the scalar field sits at a non-zero minimum. So all the different combinations of $V$ and $f$ coincide at this fixed point which formally resembles a de Sitter model. 

We can also find at least one example where the cosmological observations of the transition from the decelerated phase to the accelerated one can happen without any contradiction with the allowed band between the mass of the scalar field and its coupling with matter. 

As already mentioned, the basic aim was not to propose a new chameleon model but rather to provide an exhaustive study of the stability characteristics of all possible combination of $V$ and $f$. This purpose is achieved quite comprehensively and as a bonus some interesting physical features are also noted. For example, it is noted that a chameleon field can give rise to a transient acceleration for the universe and certain combination of the potential $V$ and the coupling $f$ leads to a breakdown of the chameleon mechanism itself. 

\chapter{Stability analysis of a holographic dark energy model}

% **************************** Define Graphics Path **************************
\ifpdf
    \graphicspath{{Chapter4/Figs/Raster/}{Chapter4/Figs/PDF/}{Chapter4/Figs/}}
\else
    \graphicspath{{Chapter4/Figs/Vector/}{Chapter4/Figs/}}
\fi

\section{Introduction:}
Holographic principle was proposed by Gerard 't Hooft and Leonard Susskind \cite{'thooft, susskind}. According to the holographic principle, the degrees of freedom of any system is determined by the area of the boundary and not really by the volume. It was originally inspired by black hole thermodynamics. The principle states that the surface fluctuations of the event horizon of a black hole contain description of all the objects ever fall in. Mathematically, the energy inside a region of length $L$, must not  exceed the mass of a black hole of the same size. From effective field theory $L^3 \rho \lesssim L M_{p} ^2$, where $L$ is the infra-red cut-off and this has to be a cosmological length scale if we consider cosmological holography. In cosmological holography the universe can be thought of as a two dimensional information structure painting on the cosmological horizon. There are three different choices of $L$, the Hubble radius, the particle and the event horizon and Ricci scalar curvature. The dark energy models in which the cut-off length is identified with the Hubble radius ($H^{-1}$) \cite{Cohen}, calculated dark energy density is close to the observed  effective cosmological constant. Li in \cite{li} discussed that considering the particle horizon and the event horizon as the cut-off length scale, it is possible to construct viable dark energy models. The cosmological dynamics of the Ricci dark energy models, in which the length scale $L$ is identified with the Ricci scalar is discussed in \cite{Gao, Camco}.

Pavon  in \cite{diego}, gave a brief but comprehensive review of this infra-red cut-off and its application in cosmology as a ``holographic'' dark energy. Cosmological implications of this holographic dark energy, particularly its role in driving the accelerated expansion of the universe have been quite thoroughly discussed\cite{li, diego1, diego2}. Holographic dark energy has been discussed in non-minimally couple theories as well, such in Brans-Dicke theory by Banerjee and Pavon\cite{diego3}, and in a chameleon scalar field model by Setare and Jamil\cite{setare}.

The holographic dark energy attracted attention as it can alleviate, if not resolve, the issue of cosmic coincidence, i.e., why the energy densities due the dark matter and the dark energy should have a constant ratio of order unity for the present universe\cite{diego2}.      

This chapter deals with the stability of a holographic dark energy model in standard Einstein's gravity. The model is quite general as it allows interaction between the dark energy and the dark matter, so that one can grow at the expense of the other. The dynamical systems analysis has been used to study the system . The field equations are written as an autonomous system and the fixed points are found out. A stable fixed point, an ``attractor'', is likely to describe the final state of the universe. As already  mentioned, this kind of dynamical systems study is not new in cosmology. There are excellent reviews on this topic\cite{r13}. However, such analysis is more frequently used where a scalar field is involved. A few dynamical systems study on holographic dark energy models, such as one where the infra-red cut-off given by the Ricci length\cite{nairi1} and by the future event horizon\cite{nairi2} are there in the literature. Setare and Vagenas\cite{setare1} discussed the bounds on the effective equation of state parameters on the basis of dynamical systems study, with an infra-red cut off given by the future event horizon. The present deals with a holographic dark energy model where the cut-off is determined by the Hubble length\cite{diego, diego2}.

\section{Phase space analysis of the model:}                                          
We consider an interacting holographic dark energy model in which the universe is filled with a pressureless matter component with energy density $\rho_{m}$ and a holographic dark energy of density $\rho_h$. There is an interaction between these two components. The total energy density of the universe is $\rho = \rho_m + \rho_h$.

In a spatially flat FRW universe with the line element
\begin{equation} \label{metric}
ds^2 = - dt^2 + a^2(t) (dr^2 + r^2 d \omega^2),
\end{equation}
Einstein's field equations are
\begin{equation} \label{field1}
3 H^2 = 8 \pi G (\rho_{m} + \rho_{h}),
\end{equation}
\begin{equation} \label{field2}
\dot{H} = - \frac{3}{2} H^2 (1 + \frac{w}{1+r}),
\end{equation}
where $w = \frac{p_{h}}{\rho_{h}}$, is the equation of state parameter of holographic dark energy, $p_{h}$ is the contribution to the pressure by the holographic dark energy and $r = \frac{\rho_{m}}{\rho_{h}}$, is the ratio of the two energy densities\cite{li, diego}.

The conservation equations for $\rho_{m}$ and $\rho_{h}$ are
\begin{equation} \label{matter}
\dot{\rho_{m}} + 3 H \rho_{m} = Q,
\end{equation}
and
\begin{equation} \label{holo}
\dot{\rho_{h}} + 3 H (1+ w) \rho_{h} = -Q,
\end{equation}
respectively. The dark matter and the dark energy are assumed to interact amongst themselves and hence do not conserve separately. They conserve together and $Q$ is the rate of loss of one form  and hence the gain of the other and is assumed to be proportional to the dark energy density given by $Q= \Gamma \rho_{h}$, where $\Gamma$ is the decay rate\cite{diego1}. 

Using (\ref{matter}) and (\ref{holo}) , one can write time evolution of $r$ as 
\begin{equation} \label{r}
\dot{r} = 3 H r (1+r) [\frac{w}{1+r} + \frac{Q}{3 H \rho_{m}}].
\end{equation}

If the holographic bound is saturated, one has\cite{li}
\begin{equation} \label{bound}
\rho_{h} = 3 M_{p}^2 C^2 / L^2,
\end{equation}  
where $L$ is the infra-red cut-off that sets the holographic bound. The Plank mass $M_p$ is given by $M_p ^2 = \frac{1}{8 \pi G}$. If the cut-off is chosen to be the Hubble length, which has a possible contribution towards the resolution of the coincidence problem\cite{diego1}, one has

\begin{equation} \label{holodensity}
\rho_{h} = 3 C^2 M_{p}^2 H^2.
\end{equation}

By differentiating equation (\ref{holodensity}) and using (\ref{field2}) in the result, one can write

\begin{equation} \label{rhohdot}
\dot{\rho_{h}} = -3 H (1 + \frac{w}{1+r}) \rho_{h}.
\end{equation}

%It is interesting to note that for this choice of the infra-red cut-off, the dimensionless density parameter for the dark energy, ${\Omega}_{h} = \frac{\rho_{h}}{3M^{2}_{p} H^{2}}$ becomes equal to $C^{2}$. 

Equations (\ref{holo}) and (\ref{rhohdot}) yield the expression for the equation of state parameter $w$ for the  holographic dark energy as,
\begin{equation} \label{w}
w = - (\frac{1+r}{r}) \frac{\Gamma}{3 H}.
\end{equation}

From equations (\ref{bound}) and (\ref{field1}), one has $ \rho_{m} = 3 M_p ^2 H^2 (1-C^2)$. Considering a saturation of the holographic dark energy, equations (\ref{field2}) and (\ref{holo}) now yield
\begin{equation} \label{dot1}
\dot{H} = - \frac{3}{2} H^2 (1 - \frac{C^2}{3 (1- C^2)} \frac{\Gamma}{H}),
\end{equation}
and,
\begin{equation} \label{dot2}
\dot{\rho_{m}} + 3 H \rho_{m} = 3 C^2 M_p ^2 H^2 \Gamma.
\end{equation}
In the subsequent discussion, equations (\ref{dot1}) and (\ref{dot2}) will replace the field equations (\ref{field1}) and (\ref{field2}). For the study of the phase space behaviour of the system,  we introduce a new set of variables $x = \rho_{m}$ , $y = \frac{\Gamma}{H}$ and $N = \ln a$.  The system of equations can now be written in the form of an autonomous system in terms of new dynamical variables as 

\begin{equation} \label{x}
x^{\prime} = -3 x + \frac{C^2}{1-C^2} x y,
\end{equation}
\begin{equation} \label{y}
y^{\prime} = - \frac{3}{2} (\lambda - 1) y (1- \frac{C^2}{3 (1- C^2)} y).
\end{equation}

Here $\lambda = (\dfrac{d\Gamma}{dH}) / (\frac{\Gamma}{H})$ and a `prime' indicates a differentiation with respect to $N$, where $N= \ln(\frac{a}{a_0})$. In an FRW cosmology, $H$, the fractional rate of change of the length scale of the universe, is the naturally available rate. We assume the decay rate $\Gamma$ to be a function of $H$, $\Gamma = \Gamma(H)$. Depending on the value of $\lambda$, we have classified our system into two classes. \\

I) When $\lambda \neq 1$, $\Gamma$ is any function of $H$, except a linear function.

II) When $\lambda =1 $, $\Gamma$ is linear functions of H. 

\subsection{Class I: $\lambda \neq 1$}
In this case the problem is a two-dimensional one and equations (\ref{x}) and (\ref{y}) form our system. Fixed points of the system are the simultaneous solutions of the equation $x^{\prime} = 0$ and $y^{\prime} = 0$. So it is easy to check that it admits two fixed points, namely, $p_1 : (x=0, y=0)$ and $p_2 : (x$ is arbitrary, $y = \frac{3 (1-C^2)}{C^2})$. The second fixed point is a set of non isolated fixed points, a straight line parallel to $x$ - axis. If Jacobian matrix at any point of a set of non isolated fixed points has at least one zero eigenvalue the set of fixed points is called normally hyperbolic\cite{r19}. Stability of a normally hyperbolic\cite{reza} set of fixed points can be analysed from sign of remaining eigenvalues. If remaining eigenvalues are negative then the fixed point is stable. Eigenvalues and stability condition of the fixed points are given in the following table.

\begin{table}[H]
\centering
\caption{Fixed points and eigenvalues of Class I models}
\label{ht1}
\begin{tabular}{|c|c|c|c|}
\hline 
Fixed Points & Co-ordinate & eigenvalues &  Condition of stability \\ 
\hline 
$p_1$ & $x=0,y=0$ & $-3, - \frac{3}{2} (\lambda -1)$  & $\lambda > 1$ \\ 
\hline 
$p_2$ & $x(N), y = \frac{3 (1-C^2)}{C^2} $ & $0,  \frac{3}{2} (\lambda -1)$  & $\lambda < 1$ \\ 
\hline 
\end{tabular}   
\end{table}
\vspace{2em}

\begin{figure}[H]
\centering
\includegraphics[scale=.6]{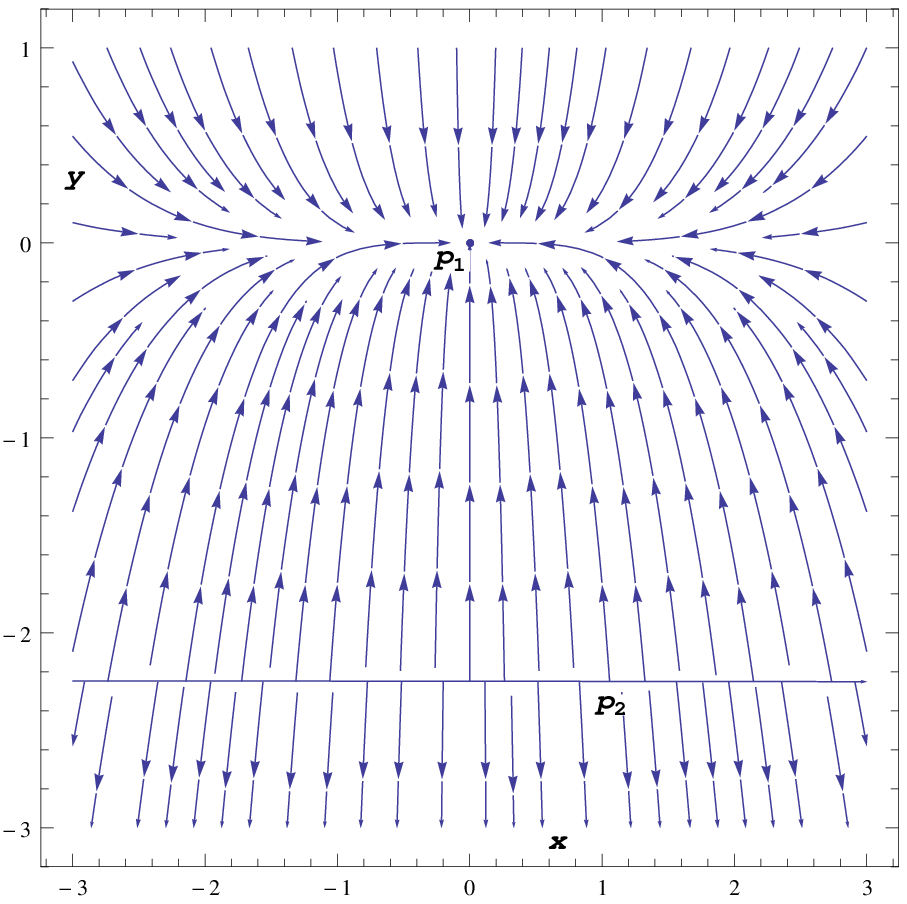}
\caption{Phase plot of the system when $\lambda = 10$ and $C=2$.}
\label{h1}
\end{figure}

\begin{figure}[H]
\centering
\includegraphics[scale=.6]{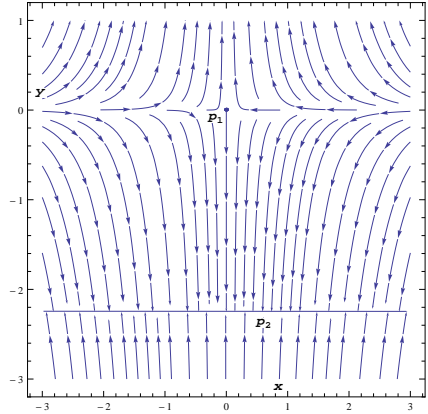}
\caption{Phase plot of the system when $\lambda = -10$ and $C=2$.}
\label{h2}
\end{figure}

 Phase plots of this system has been shown in figure \ref{h1} and figure \ref{h2} for $\lambda>1$ and $\lambda < 1$ respectively. The plots show that fixed points $p_1$ and $p_2$ are indeed stable for $\lambda>1$ and $\lambda<1$ respectively. This is consistent with the fact that negative eigenvalues indicate stable fixed points (see table \ref{ht1}).

Form the field equations and the definition of $x$ and $y$, the deceleration parameter can be written as
\begin{equation}\label{q}
 q = -1 + \frac{3}{2} (1- \frac{C^2}{3(1-C^2)} y).
\end{equation}

As we have two fixed points, the system may be treated as a heteroclinic one where solutions join two fixed points.

For $\lambda>1$, $p_1$ is a stable fixed point and $p_2$ is an unstable one, thus indicating a sink and source respectively. So the universe can originate from  $p_2$ with an acceleration ($q=-1$) and an arbitrary $\rho_{m}$ and can settle down to $p_1$, the stable fixed point where the expansion is decelerated ($q=\frac{1}{2}$) with $\rho_{m} \longrightarrow 0$. This squarely contradicts the observation which indicates an exactly opposite situation!

For $\lambda<1$, however, the fixed points actually reverse their roles as the source and the sink. In this case $p_1$ is unstable, thus, for a small perturbation,  the universe starts evolving with a deceleration ($q=\frac{1}{2}$) and settles down to the final configuration of an accelerated expansion ($q=-1$). The final  $\rho_{m}$ is an arbitrary function of $N$. This situation is indeed realistic. 

 It deserves mention that $\lambda$ is not actually a constant. So the figures \ref{h1} and \ref{h2} represents a section of the 3-dimensional figure at particular values of $\lambda$, i.e., snapshots at those values. If one changes the values of $\lambda$, the nature of the figures remain the same. For a particular value of $\lambda$, equations (\ref{x}) and (\ref{y}) lead to an expression for the deceleration parameter $q$ in terms of scale factor $a$ as, $q = -1 + \frac{3}{2} (1 - \frac{C^2}{3(1-C^2)} \frac{b a^{-\frac{3}{2} (\lambda - 1)}}{1 + \frac{b C^2}{3(1-C^2)} a^{-\frac{3}{2} (\lambda - 1)} })$ where $b$ is integration constant. 
 
\begin{figure}[H]
\centering
\includegraphics[scale=.7]{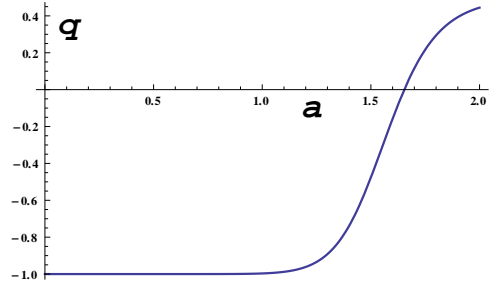}
\caption{$q$ vs $a$, when $\lambda = 10, C=2$ and $ b = -1000$.}
\label{h3}
\end{figure}

\begin{figure}[H]
\centering
\includegraphics[scale=.6]{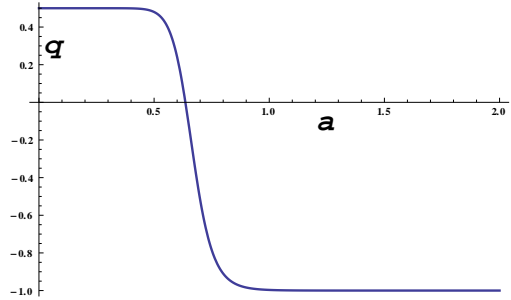}
\caption{$q$ vs $a$, when $\lambda = -10, C=2$ and $ b = -1000$.}
\label{h4}
\end{figure}

 In figure \ref{h3} and \ref{h4}, the evolution of the deceleration parameter $q$ against $\frac{a}{a_{0}}$ is shown where $a_{0}$ is the present value of the scale factor. It is clearly seen that $\lambda <1$ is favoured in order to describe the observed dynamics of the universe. It should be noted that, for, $\lambda <1$, the interaction rate $\Gamma$ decays at a slower rate than the decay of $H$ and for a negative $\lambda$, the rate actually grows with some power of the decay of $H$.
 
 %For $\lambda >1$, the universe starts from a negative $q$ ($q=-1$) with $x=0$ i.e., $\rho_{m}=0$ and settles into the stable configuration of a decelerated expansion. For $\lambda<1$, however, one obtains the desired behaviour, the universe starts with a deceleration ($q=\frac{1}{2}$) and settles into an accelerated phase with $q=-1$ and an arbitrary $\rho_{m}$.

\subsection{Class II : $\lambda =1$}
In this case $\Gamma$ is a linear function of $H$, given by $\Gamma = \alpha H$. Our system of equations reduces to 
\begin{equation} \label{x1}
x^{\prime} = - 3 x + \frac{c^2}{1-c^2} x y, 
\end{equation}
\begin{equation} \label{y1}
y^{\prime} = 0.
\end{equation}

\begin{figure}[H]
\centering
\includegraphics[scale=.8]{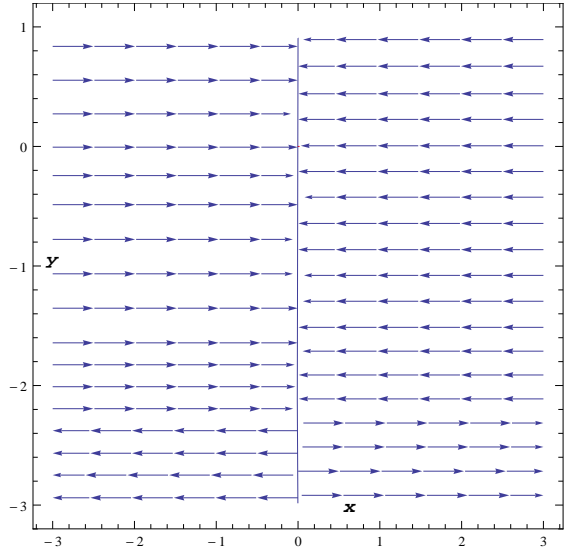}
\caption{Phase plot of the system when $\lambda = 1$ and $C=2$.}
\label{h5}
\end{figure}

As $y$ is a constant, this is essentially a one dimensional system with $x=0$ and $y=\alpha$, a constant. Integration of the equation (\ref{x1}) yields the solution for $x$ as $x = A e^{-3 k N}$ where A is a constant of integration and $k= (1-\frac{C^2}{3(1-C^2)}\alpha)$. If $k>0$, the solution is indeed stable, as for $N \longrightarrow \infty$, one has $x \longrightarrow 0$. For $k<0$, the fixed point is unstable. The phase plot is shown in figure \ref{h5}, which indicates that the $y$-axis is the attractor of all solutions for $k>0$. Since $y$ is a constant, equation (\ref{q}) indicates that there is no transition from a decelerated to an accelerated expansion for the universe.

\section{Bifurcation in the system:}
It is interesting, from the point of view of the dynamical systems, to note that there is a clear bifurcation in the system, where $\lambda$ is the bifurcation parameter and $\lambda =1$ is the bifurcation point. When $\lambda<1$, there are two fixed points, $p_1$ and $p_2$ where $p_1$ is unstable but $p_2$ is stable. As $\lambda$ approaches unity these two merge into a single fixed point where the stability depends on the choice of the values of the constants $C$ and $\alpha$. With $\lambda>1$, i.e., when it attains values on the other side of the bifurcation point, one has the same two fixed points with their roles interchanged so far as the stability is concerned. Figures \ref{h1},\ref{h2} and figure \ref{h5} show the change of the behaviour of the phase space with variation of $\lambda$, which clearly indicate the occurrence  of the bifurcation in the system. 
At the bifurcation point, $\lambda =1$, $y=\frac{\Gamma}{H}$ is a constant. As already mentioned,  there is no transition from a decelerated to an accelerated expansion for the universe for a spatially flat FRW metric at bifurcation. This result is completely consistent with that obtained by Pavon and Zimdahl\cite{diego2}. 

\section{Discussion}
The holographic dark energy, tipped by many as the possible saviour from the coincidence problem, is analyzed as an autonomous system. It is found that the system indeed has unstable and stable fixed points. It is found that for $\lambda <1$, where $\lambda = (\dfrac{d\Gamma}{dH}) / (\frac{\Gamma}{H})$, the system indeed has at least one natural description of the universe  which starts with a decelerated expansion ($q=\frac{1}{2}$) and settles down to an accelerated expansion for the universe with dark matter completely giving way to the dark energy. So the interacting holographic dark energy warrants more attention, particularly when $\dfrac{d\Gamma}{dH} < \frac{\Gamma}{H}$ leading to $\lambda <1$. We see that the  interaction rate $\Gamma$ actually plays a crucial role in this.

\chapter{Conclusion}

In this thesis, we used a dynamical systems analysis to find the qualitative behaviour of some dark energy models. Specifically, quintessence scalar field models, chameleon scalar field models and holographic models of dark energy are discussed in the present thesis.

Einstein's field equations form a set of nonlinear differential equations. So finding out an exact solution is not always easy. But, using a dynamical systems analysis the qualitative behaviour of the non-linear system can be investigated. Usually a set of suitable new variables and a dimensionless time variable are introduced so as to write the field equations as an autonomous system of equations. The fixed points of the system lead to the stability analysis of the system. Stable fixed points are the late time attractors and unstable fixed points are the repellers. Any heteroclinic solution of the system starts from an unstable fixed point and approaches to a stable fixed point. So one can study the beginning and possible ultimate fate of the universe by finding out the behaviour of the physical parameters near the fixed points. 

In the second chapter, Quintessence scalar field models, with and without a tracking condition, are analyzed. In the first part, the qualitative behaviour of the system in the neighborhood of the fixed points is analyzed with a tracking condition, where the evolution of the scalar field almost mimics that of the matter field and picks up the dominant role very slowly in the later stage. This alleviates the so-called coincidence problem. It is found that, only one fixed point is of physical relevance in such a situation. Two specific forms of the potential are discussed, one is an exponential potential and the other is in a cosine hyperbolic form. It is also shown that each of them can drive the accelerated expansion of the universe near the fixed point. 

In the second part, the tracking condition is relaxed, and with the boundary values consistent with the present observations,  the system has been numerically evolved for two type of specific potentials to find the qualitative behaviour of the universe. It is found for both the potentials, one an exponential and the other being a power-law, the solution starts from an unstable fixed point and evolve to a stable fixed point. From this heteroclinic behaviour of the solution, a qualitative study of the beginning, present phase of accelerated expansion and possible ultimate fate of the universe has been carried out. It is intriguing to note that in both the examples, the scalar field energy density dominates over the dark matter density not only at present, but also at the beginning. With a favourable equation of state parameter (${\gamma}_{\phi} > 1$) in the beginning, the effective pressure is positive leading to a decelerated expansion. The parameter ${\gamma}_{\phi}$ evolves to a negative value only at a later epoch. So the model successfully drives a late time acceleration. However, this may encounter problems with other physical requirements such as Big Bang Neucleosynthesis.

The third chapter deals with the dynamical systems study of the Chameleon dark energy models. In Chameleon cosmology models the scalar field is non-minimally coupled to the  matter sector and it's effective mass depends on ambient matter density. Actually, this chapter provides an exhaustive diagnostic in terms of stability of the model for the combinations of all types of chameleon potential and its coupling with the matter sector. Phase space behaviour of each class has been investigated along with the nature and asymptotic behaviour of the fixed points.

In the fourth chapter, the field equations of a holographic dark energy model are  written as an autonomous system and the standard dynamical systems analysis is done. Assuming the decay rate to be function of $H$, the system is classified into two classes depending on the functional form of the decay rate. It is interesting to note that the system undergoes a bifurcation at a parametric value of $\lambda =1$. It is also found that when decay rate is a simple linear function of $H$, there is no transition from decelerated expansion phase to the accelerated expansion phase of the universe. 

Amongst the models discussed, there is hardly any pointer to pick up one as the most favoured candidate as a dark energy. But certainly the chameleon fields have the widest variety of possibilities.

%\include{Chapter6/chapter6}
%\include{Chapter7/chapter7}

% ********************************** Back Matter *******************************
% Backmatter should be commented out, if you are using appendices after References
%\backmatter

% ********************************** Bibliography ******************************
\begin{spacing}{0.9}

%\bibliographystyle{myapsrev}
%\bibliography{dissertation}
%\addcontentsline{toc}{chapter}{Bibliography}
% To use the conventional natbib style referencing
% Bibliography style previews: http://nodonn.tipido.net/bibstyle.php
% Reference styles: http://sites.stat.psu.edu/~surajit/present/bib.htm

\bibliographystyle{apalike}
\bibliographystyle{plainnat} % use this to have URLs listed in References
\cleardoublepage
\bibliography{References/thesis} % Path to your References.bib file

\begin{thebibliography}{200}

\bibitem{r1} S. Perlmutter et al, Bull. Am. Astron.Soc., {\bf 29}, 1351 (1997).\\
             S. Perlmutter et al, Astrophys. J., {\bf 517}, 565 (1999).\\
             J. L. Tonry et al, Astrophys. J., {\bf 594}, 1 (2003).\\
             S. Bridle, O. Lahav, J.P. Ostriker and P.J. Steihardt, Science, {\bf 299}. 1532 (2003).\\
             G. Hinshaw et al, Astrophys. J. Suppl., {\bf 148}, 135 (2003).\\
             A. Kogut et al, AstroAstrophys. J. Suppl., {\bf 148}, 161 (2003).\\
             D.N. Spergel et al, Astrophys. J. Suppl., {\bf 148}, 175 (2003).\\
             C.L. Bennet at al, Astrophys. J. Suppl., {\bf 148}, 1 (2003).

\bibitem{r20} A.G. Riess et al, Astrophys. J., {\bf 560}, 49 (2001).

\bibitem{r2} T. Padmanabhan and T. Roy Choudury, Mon. Not. R. Astron. Soc., {\bf 344}, 823 (2003).\\
             T. Roy Choudury and T. Padmanabhan, Astron. Astrophys., {\bf 823}, 807 (2005).

\bibitem{r3} V. Sahni and A. Starobinsky, Int. J. Mod. Phys. D, {\bf 9}, 373 (1000).\\
             T. Padmanabhan, Phys. Rep., {\bf 380}, 235 (2003).

\bibitem{r4} J. Martin, astro-ph/0803.4076.

\bibitem{r5} I. Zlatev and P.J. Steinhardt. Phys.Lett.B, {\bf 459}, 570 (1999).

\bibitem{r} N. Banerjee and S. Das, Mod. Phys. Lett. A, {\bf 21}, 2663 (2006).

\bibitem{r6} I. Zlatev, L. Wang and P.J. Steinhardt, Phys. Rev. Lett. {\bf 82}, 896 (1999).\\
             P.J. Steinhardt, L. Wang and I. Zlatev, Phys. Rev.D, {\bf 59}, 123504(1999).\\
             L. Wang, R.R. Caldwell, J.P. Ostriker and P.J. Steinhardt, Astrophys. J., {\bf 530}, 17 (2000).

\bibitem{r7} V.B.Johri Class.Quant.Grav {\bf 19}, 5959 (2002).

\bibitem{r8} L.A.Urena-Lopez and T.Matos Phys.Rev.D, {\bf 62}, 081302 (2000).

\bibitem{r9} M.Sahlen, AR.Liddle, D. Parkinson Phys.Rev.D, {\bf 75}, 023502 (2007).

\bibitem{r10} S.Dodelson, M.Kaplinghat and E. Stewart Phys.Rev.Lett, {\bf 85}, 5276(2000).

\bibitem{r11} P-Y. Wang, C.W Chen and P.Chen. JCAP 2012.


\bibitem{r12} S.C.C.Ng, N.J. Nunes, F.Rosati, Phys.Rev.D, {\bf64} ,083510(2001).

\bibitem{r13} \textit{Dynamical Systems in Cosmology}, J. Wainwright and G.F.R.Ellis (eds); Cambridge University Press, (1997).\\   \textit{Dynamical System and Cosmology}, A.A.Coley, Kluwer Academic Publishers (2003).

\bibitem{r14} L.Lara and M.Castagnins, Int.J.Theor. Phys. {\bf 44}, 1839(2005).

\bibitem{r15} E.Gunzig, V.Faraoni, A.Figeredo, T.M.Rocha Filho and L.Brenig, Class.Quant.Grav., {\bf 17}, 1783(2000).

\bibitem{r16} J.Carot and M.M.collinge, Class.Quanta.Grav., {\bf 20}, 707(2003).

\bibitem{r17} L.Arturo, Urena-Lopez, JCAP, {\bf 0509}, 013(2005).

\bibitem{r18} S.J.Kolitch and B.Hall, arxiv:[gr-qc /9410039].\\
              S.J. Kolitch and D.M. Eardley, Ann. Phys., {\bf 241}, 128, 1995.

\bibitem{r19} \textit{Nonlinear dynamics and chaos: With applications to Physics, Biology, Chemistry and Engineering}, S.H. Strogatz, Westview Press (2001).

\bibitem{r21} V. Sahni, arxiv:{astro-ph/0403324}.
%---------------------------------------------------------------------------------------CHAMELEON--------------------------------------
\bibitem{varun} V. Sahni and A. Starobinsky, Int. J Mod. Phys. D, {\bf 9}, 373 (2000).
\bibitem{paddy} T. Padmanabhan, Phys. Rep., {\bf 380}, 235 (2003).
\bibitem{sami} E.J. Copeland, M. Sami and S. Tsujikawa, Int. J Mod. Phys. D, {\bf 15}, 1753 (2006).
\bibitem{martin} J. Martin, Mod. Phys. Lett. A, {\bf 23}, 1252 (2008).
\bibitem{review} T.P. Sotiriou and V. Faraoni; arxiv:gr-qc/0805.1726;\\
                 A. De Felice  and S. Tsujikawa; arxiv:gr-qc/1002.4928.\\
                 S. Nojiri and S.D. Odintsov, Phys. Rep. {\bf 505}, 59 (2011).
\bibitem{nbdp}O. Bertolami and P.J. Martins, Phys. Rev. D, {\bf 61}, 064007 (2000).\\
              N. Banerjee and D. Pavon, Phys. Rev. D, {\bf 63}, 043504 (2001).\\
              S. Sen and A.A. Sen, Phys. Rev. D, {\bf 63}, 124006 (2001).\\
              E. Elizalde, S. Nojiri and S.D. Odintsov, Phys. Rev. D, {\bf 70}, 043539 (2004).\\
            
\bibitem{justin1} J. Khoury and A. Weltman, Phys. Rev D, {\bf 69}, 044026 (2004).
\bibitem{justin2} J. Khoury and A. Weltman, Phys. Rev. Lett., {\bf 93}, 171104 (2004).
\bibitem{mota1} D.F. Mota and D.J. Shaw; arxiv:0805.3430. 
\bibitem{mota2} D.F. Mota and D.J. Shaw Phys. Rev. Lett., {\bf 97}, 151102 (2006).
\bibitem{shaw1} C. Burrage, A.C. Davis and D.J. Shaw, Phys. Rev. D, {\bf 79}, 044028 (2009).
\bibitem{shaw2} A.C. Davis, C.A.O. Schelpe and D.J. Shaw, Phys.Rev. D, {\bf 80}, 064016 (2009).
\bibitem{buri} P. Burikham and S. Panpanich, Int. J. Mod. Phys. D, {\bf 21}, 1250041 (2012).
\bibitem{brax1} P. Brax, C. Burrage, A.C. Davis, D. Seery and A. Weltman, Phys. Lett. B, {\bf 699}, 5 (2011).
\bibitem{mota3} D.F. Mota and C.A.O. Schelpe, Phys. Rev. D, {\bf 86}, 123002 (2012).
\bibitem{khoury} J. Khoury, Class. Quant. Grav., {\bf 30}, 214004 (2013).
\bibitem{brax} P. Brax, C. van de Bruck, A.C. Davis, J. Khoury and A. Weltman, Phys. Rev. D, {\bf 70}, 123518 (2004).
\bibitem{das} S. Das, P.S. Corasaniti and J. Khoury, Phys. Rev. D, {\bf 73}, 083509 (2006).
\bibitem{nbsdkg} N. Banerjee, S. Das and K. Ganguly, Pramana, {\bf 74}, L481 (2010).
\bibitem{sdnb} S. Das and N. Banerjee, Phys. Rev.D, {\bf 78}, 043512 (2008).
\bibitem{jamil} M. Setare and M. Jamil. Phys. Lett. B, {\bf 690}, 1 (2014).
\bibitem{tavakol} S. Tsujikawa, T. Tamaki and R. Tavakol, JCAP, {\bf 05}, 020 (2009).
\bibitem{gunzig} E. Gunzig, V. Faraoni, A. Figeredo and L. Brenig, Class. Quantum Grav., {\bf 17}, 1783 (2000).\\
                 J. Carot and M.M. Collinge, Class. Quantum Grav., {\bf 20}, 707 (2003).
                 L.A. Urena-Lopez, JCAP, {\bf 0509}, 013 (2005).
                 S. Sen, A.A. Sen and M. Sami, Phys. Lett. B, {\bf 686}, 1 (2010).
                 S. Kumar, S. Panda and A.A. Sen, Class. Quantum Grav., {\bf 30}, 155011 (2013).
                 N. Roy and N. Banerjee, Gen. Rel. Grav., {\bf 46}, 1651 (2014).
\bibitem{coley}  A.A. Coley, {\it Dynamical Systems and Cosmology}, Cambridge University Press (2003).
\bibitem{ellis}  J. Wainwright and G.F.R. Ellis, {\it Dynamical Systems in Cosmology}, Springer (2005).
\bibitem{harko1} O. Minazzoli and T.Harko, Phys. Rev. D, {\bf 86}, 087502 (2012).
\bibitem{harko2} T.Harko, Phys. Rev. D, {\bf 81}, 044021 (2010).
\bibitem{strogatz}S.H. Strogatz, {\it Nonlinear dynamics and chaos: With Applications to Physics, Biology Chemistry and  		             Engineering}, Westview Press, Boulder (2001).
\bibitem{nrnb} N. Roy and N. Banerjee, Euro. Phys. J Plus., {\bf 129}, 162 (2014).
\bibitem{giostri}R. Giostri, M.V. dos Santos, I. Waga, R.R.R. Reis, M.O. Calvao and B.L. Lago, J. Cos. Astrophys., {\bf 027}, 
                 1203 (2012).
%------------------------------------Holo-----------------------------------------------------------------------------------
\bibitem{'thooft} G. 't Hooft, arxiv: gr-qc/9310026.
%\bibitem{susskind} L. Susskind, J. Math. Phys., {\bf 36}, 6377 (1995).
%\bibitem{diego} D. Pavon, J. Phys. A:Math. Theor., {\bf 40}, 6865 (2007).
%\bibitem{li} M.Li, Phys. Lett. B {\bf 603}, 1 (2004).
%\bibitem{diego1} D.Pavon and W. Zimdahl, Phys. Lett. B {\bf 628}, 206 (2005).
%\bibitem{diego2} D.Pavon and W. Zimdahl, Class. Quantum Grav. {\bf 24}, 5461 (2007).
%\bibitem{diego3} N. Banerjee and D. Pavon, Phy. Lett. B {\bf 647}, 477 (2007).
%\bibitem{setare} M.R. Setare and M. Jamil, Phys. Lett B {\bf 690}, 1 (2010).
%\bibitem{ellis} J. Wainwright and G. F. R. Ellis, \textit{Dynamical System in Cosmology} (Cambridge University Press, 2005).
%\bibitem{coley} A.A.Coley, \textit{Dynamical System and Cosmology} (Springer, 2003).
%\bibitem{gunzig} E. Gunzig, V. Faraoni, A. Figeredo and L. Brenig, Class. Quantum Grav. {\bf 17}, 1783 (2000).
%\bibitem{carot} J. Carot and M.M. Collinge, Class. Quantum Grav. {\bf 20}, 707 (2003).
%\bibitem{urena} L.A. Urena-Lopez, JCAP {\bf 0509}, 013 (2005).
%\bibitem{nandan2} N. Roy and N. Banerjee, Eur. Phys. J. Plus, {\bf 129}, 162 (2014).
%\bibitem{anjan} S. Kumar, S. Panda and A.A. Sen, Quantum Grav. {\bf 30}, 155011 (2013).
%\bibitem{soma} S. Sen, A.A. Sen and M. Sami, Phys. Lett B. {\bf 686}, 1 (2010).
%\bibitem{nandan1} N. Roy and N. Banerjee, Gen. Rel. Grav. {\bf 46}, 1651 (2014).
%\bibitem{fang} W. Fang, H. Tu, J. Huang and C. Shu, arxiv:[1402.4005].
%\bibitem{nairi1} N. Mazumder, R. Biswas and S. Chakraborty, arxiv:[1106.4627].
%\bibitem{nairi2} N. Mazumder, R. Biswas and S. Chakraborty, arxiv:[1106.4626].
%\bibitem{setare1} M.R. Setare and E.C. Vagenas, Int. J. Mod. Phys. D, {\bf 18}, 147 (2009).
%\bibitem{strog} S.H. Strogatz, \textit{Nonlinear Dynamics and Chaos: With Applications to Physics, Biology, Chemistry and     Engineering};                             Westview Press, Boulder (2001). 

%\bibitem{reza} R. Tavakol, `` Introduction to dynamical systems '' in ref \cite{ellis} .
                
\end{thebibliography}


\begin{thebibliography}{300}


\bibitem{c1} A. R. Liddle and D. H. Lyth, Cosmological inflation and large-scale structure, Cambridge University Press (2000).

\bibitem{c2} S. Weinberg, Gravitation and Cosmology, John Wiley and Sons, Inc. (1972).

\bibitem{c3} E. W. Kolb and M. Turner, The Early Universe, Addison-Wesley (1990)

\bibitem{c4} T. Padmanabhan, Theoretical Astrophysics, Cambridge Univ. Press (2000).

\bibitem{c5} A. G. Riess et al., Astron. J. 116, 1009 (1998); Astron. J. {\bf 117}, 707 (1999)

\bibitem{c6} S. Perlmutter et al., Astrophys. J. {\bf 517}, 565 (1999).


\bibitem{c7} V. Sahni and A. A. Starobinsky, Int. J. Mod. Phys. D {\bf 9,} 373 (2000).\\
             V. Sahni, Lect. Notes Phys. {\bf 653}, 141 (2004) [arXiv:astro-ph/0403324].


\bibitem{c8} T. Padmanabhan, Phys. Rept. {\bf 380}, 235 (2003).\\
             T. Padmanabhan, Current Science, {\bf 88}, 1057 (2005) [arXiv:astro-ph/0510492].

\bibitem{c9} A. G. Riess et al. [Supernova Search Team Collabora-tion], Astrophys. J., {\bf 607}, 665 (2004).

\bibitem{c10} P. Astier et al., Astron.Astrophys., {\bf 447}, 31 (2006).

\bibitem{c11} H. K. Jassal, J. S. Bagla and T. Padmanabhan, Mon.Not.Roy.Astron.Soc., { \bf 405}, 2639 (2010).


\bibitem{c12} S. Podariu, R. A. Daly, M. P. Mory and B. Ratra, Astrophys. J. {\bf 584}, 577 (2003).


\bibitem{c13} R. A. Daly and E. J. Guerra, Astron. J. {\bf 124}, 1831 (2002).


\bibitem{c14} R. A. Daly and S. G. Djorgovski, Astrophys. J. {\bf 597}, 9 (2003).

\bibitem{c15} D. Hooper and S. Dodelson, Astropart.Phys, {\bf 27}, 113 (2007).


\bibitem{c16} M. Tegmark et al. [SDSS Collaboration], Phys. Rev. D {\bf 69}, 103501 (2004).

\bibitem{c17} U. Seljak et al., Phys. Rev. D {\bf 71}, 103515 (2005).

\bibitem{c18} D. N. Spergel et al. [WMAP Collaboration], Astrophys.J.Suppl., {\bf 170}, 377(2007).





\bibitem{m1} Tsujikawa, Shinji. "Modified gravity models of dark energy." In Lectures on Cosmology, pp. 99-145. Springer Berlin Heidelberg, 2010.

\bibitem{m2} S. Capozziello, Int. J. Mod. Phys. D { \bf 11}, 483, (2002).

\bibitem{m3} S. Capozziello, V. F. Cardone, S. Carloni and A. Troisi, Int. J. Mod. Phys. D, { \bf 12}, 1969 (2003).

\bibitem{m4} S. M. Carroll, V. Duvvuri, M. Trodden and M. S. Turner, Phys. Rev. D {\bf 70}, 043528 (2004).

\bibitem{m5} S. Nojiri and S. D. Odintsov, Phys. Rev. D {\bf 68}, 123512 (2003).

\bibitem{m6} L. Amendola, Phys. Rev. D {\bf 60}, 043501 (1999).

\bibitem{m7} J. P. Uzan, Phys. Rev. D {\bf 59}, 123510 (1999).

\bibitem{m8} T. Chiba, Phys. Rev. D {\bf 60}, 083508 (1999).

\bibitem{m9} N. Bartolo and M. Pietroni, Phys. Rev. D {\bf61} 023518 (2000).

\bibitem{m10} F. Perrotta, C. Baccigalupi and S. Matarrese, Phys. Rev. D {\bf 61}, 023507 (2000).

\bibitem{m11} A. Riazuelo and J. P. Uzan, Phys. Rev. D {\bf 66}, 023525 (2002).

\bibitem{m12} G. R. Dvali, G. Gabadadze and M. Porrati, Phys. Lett. B {\bf 485}, 208 (2000).

\bibitem{m13} A. Nicolis, R. Rattazzi and E. Trincherini, Phys. Rev. D {\bf 79}, 064036 (2009).

\bibitem{m14} S. Nojiri, S. D. Odintsov and M. Sasaki, Phys. Rev. D {\bf 71}, 123509 (2005).

\bibitem{m15} S. Nojiri and S. D. Odintsov, Phys. Lett. B {\bf 631}, 1 (2005).

\bibitem{m24} G. Felder, A. Frolov, L. Kofman, and A. Linde, Phys. Rev. D {\bf 66}, 023507 (2002).

\bibitem{m25} R. Bean, D. Bernat, L. Pogosian, A. Silvestri, and M. Trodden, Phys. Rev. D {\bf 75}, 064020 (2007).

\bibitem{m26} L. Amendola, R. Gannouji, D. Polarski, and S. Tsujikawa, Phys. Rev. D {\bf 75}, 083504 (2007).

\bibitem{m27} L. Pogosian and A. Silvestri, Phys. Rev. D {\bf77}, 023503(2008).

\bibitem{m28} J. D. Evans, L. M. H. Hall, and P. Caillol, Phys. Rev. D { \bf 77} , 083514 (2008)

\bibitem{m29} J. Q. Guo, A. V. Frolov, Phys. Rev. D {\bf 88} 124036(2013).

\bibitem{m17} De Felice, Antonio, Pia Mukherjee, and Yun Wang, Phys. Rev. D {\bf 77}, 2 (2008).

\bibitem{m16} Amendola, Luca, David Polarski, and Shinji Tsujikawa., Phys. Rev. Lett., {\bf 98} 13 (2007).



\bibitem{m18} S. Capozziello and V. Salzano, Advances in Astronomy 2009 (2010).

\bibitem{m19} C. Brans and R. H. Dicke, Phys. Rev. {\bf 124}, 925 (1961).

\bibitem{m30} N. Banerjee ans S. Sen, Phys. Rev. D, {\bf 56} 1334 (1997).

\bibitem{m20} J. O’Hanlon, Phys. Rev. Lett. {\bf 29}, 137 (1972).

\bibitem{m21} T. Chiba, Phys. Lett. B {\bf 575}, 1 (2003).

\bibitem{m22} L. Amendola, Phys. Rev. D {\bf 62}, 043511 (2000).

\bibitem{m23} N. Banerjee and D. Pavon, Phys. Rev. D, {\bf 63}, 043504 (2001).







\bibitem{c19} S. Weinberg, Rev. Mod. Phys. {\bf 61}, 1 (1989).

\bibitem{c20} A. R. Liddle and D. H. Lyth, Cosmological inflation and large-scale structure, Cambridge University Press (2000).

\bibitem{c}  E.J. Copeland, M. Sami, and Shinji Tsujikawa,  Int.J.Mod.Phys.D {\bf 15}, 1753-1936 (2006)

\bibitem{c21} P. Steinhardt, in {\it Critical Problems in Physics}, edited by V.L. Fitch and D.R. Marlow(Princeton University Press, Princeton, NJ, 1997) 

\bibitem{c22} I. Zlatev, L. Wang, and P.J. Steinhardt, Phys. Rev. Lett. {\bf 82} 896 (1999).

\bibitem{c23} P.J. Steinhardt,  L. Wang, and  I. Zlatev, Phys. Rev. D {\bf 59} 123504 (1999).

\bibitem{c24} C. Armendariz-Picon, T. Damour, and V. Mukhanov, Phys. Lett. B {\bf 458}, (1999) 209.
              J. Garriga and V. Mukhanov, Phys. Lett. B {\bf 458}, 219 (1999).

\bibitem{c25} T. Chiba, T. Okabe and M. Yamaguchi, Phys. Rev. D {\bf 62}, 023511 (2000).

\bibitem{c26} C. Armendariz-Picon, V. Mukhanov, and P. J. Steinhardt, Phys. Rev. Lett. {\bf 85}, 4438 (2000).

\bibitem{c27} C. Armendariz-Picon, V. Mukhanov, and P. J. Steinhardt, Phys. Rev. D {\bf 63}, 103510 (2001).

\bibitem{c28} N. Arkani-Hamed, P. Creminelli, S. Mukohyama and M.Zaldarriaga, JCAP {\bf 0404}, 001 (2004)

\bibitem{c29} V. D. Barger and D. Marfatia, Phys. Lett. B 498, {\bf 67} (2001).\\
              M. Li and X. Zhang, Phys. Lett. B {\bf 573}, 20 (2003).\\
              M. Malquarti, E. J. Copeland, A. R. Liddle and M. Trodden, Phys. Rev. D {\bf 67}, 123503 (2003).\\
              M. Malquarti, E. J. Copeland and A. R. Liddle, Phys. Rev. D {\bf 68}, 023512 (2003).\\

\bibitem{c30}A. Sen, JHEP 0204, 048 (2002); JHEP {\bf 0207}, 065 (2002).


\bibitem{c31} A. Sen, JHEP {\bf 9910}, 008 (1999).\\
              M. R. Garousi, Nucl. Phys. {\bf B584}, 284 (2000).\\
              Nucl. Phys. B {\bf 647}, 117 (2002); JHEP {\bf 0305}, 058 (2003).\\
              E. A. Bergshoeff, M. de Roo, T. C. de Wit, E. Eyras, S. Panda, JHEP {\bf 0005}, 009 (2000).\\
              J. Kluson, Phys. Rev. D {\bf 62}, 126003 (2000).\\

\bibitem{c32} G. W. Gibbons, Phys. Lett. B {\bf 537}, 1 (2002).

\bibitem{c33} T. Padmanabhan, Phys. Rev. D {\bf 66}, 021301 (2002).\\
              J. S. Bagla, H. K. Jassal and T. Padmanabhan, Phys. Rev. D {\bf 67}, 063504 (2003).\\
              L. R. W. Abramo and F. Finelli, Phys. Lett. B {\bf 575} 165(2003).\\
              J. M. Aguirregabiria and R. Lazkoz, Phys. Rev. D {\bf 69}, 123502 (2004).\\
              Z. K. Guo and Y. Z. Zhang, JCAP {\bf 0408}, 010 (2004).\\
              E. J. Copeland, M. R. Garousi, M. Sami and S. Tsujikawa, Phys. Rev. D {\bf 71}, 043003 (2005).\\

\bibitem{c34} F. Hoyle, Mon. Not. R. Astr. Soc. {\bf 108}, 372 (1948); {\bf 109}, 365 (1949).

\bibitem{c35} F. Hoyle and J. V. Narlikar, Proc. Roy. Soc. {\bf A282}, 191 (1964); Mon. Not. R. Astr. Soc. {\bf 155}, 305 (1972).


\bibitem{c36}  R.R. Caldwell, Phys.Lett. B {\bf 54523} 29(2002)

\bibitem{c37} A. Y. Kamenshchik, U. Moschella and V. Pasquier, Phys. Lett. B {\bf 511}, 265 (2001).

\bibitem{c50} M. C. Bento, O. Bertolami, A. A. Sen, Phys. Rev. D, {\bf 66}, 043507 (2002).

\bibitem{c38} L. Amendola, F. Finelli, C. Burigana and D. Carturan, JCAP {\bf 0307}, 005 (2003).

\bibitem{c39} M. d. C. Bento, O. Bertolami and A. A. Sen, Phys. Rev. D {\bf 67}, 063003 (2003).


\bibitem{c40} L. Perko, {\it Differential equations and dynamical systems}, Springer Science \& Business Media , 2001. 

\bibitem{c41} J. D. Meiss, {\it Differential dynamical systems}, Siam, 2007.

\bibitem{c42} V. Faraoni, Physical Review D {\bf 70}, 044037 (2004).

\bibitem{c43} J.Q. Guo and A.V. Frolov, Physical Review D {\bf 88} 124036 (2013).

\bibitem{c44} H. Sahabani and M. Farhoudi, Physical Review D {\bf 88} 044048 (2013).

\bibitem{c45} A. Batista, J. Fabris, S. Goncalves, and J. Tossa, Int. J. Mod. Phys. A {\bf 16}, 4527 (2001).

\bibitem{c46} S. J. Kolitch and D. M. Eardley, Annala of Physics {\bf 241}, 128 (1995).

\bibitem{c47} L. Jarv, P. Kuusk, and M. Saal,  Physical Review D {\bf 78},083530 (2008).

\bibitem{c48} E. J. Copeland, A. R. Liddle and D. Wands, Phys. Rev. D {\bf 57}, 4686 (1998).

\bibitem{c49} A. de la Macorra and G. Piccinelli, Phys. Rev. D {\bf 61}, 123503 (2000).










%...............................................................................................
\bibitem{r1} S. Perlmutter et al, Bull. Am. Astron.Soc., {\bf 29}, 1351 (1997).\\
             S. Perlmutter et al, Astrophys. J., {\bf 517}, 565 (1999).\\
             J. L. Tonry et al, Astrophys. J., {\bf 594}, 1 (2003).\\
             S. Bridle, O. Lahav, J.P. Ostriker and P.J. Steihardt, Science, {\bf 299}. 1532 (2003).\\
             G. Hinshaw et al, Astrophys. J. Suppl., {\bf 148}, 135 (2003).\\
             A. Kogut et al, AstroAstrophys. J. Suppl., {\bf 148}, 161 (2003).\\
             D.N. Spergel et al, Astrophys. J. Suppl., {\bf 148}, 175 (2003).\\
             C.L. Bennet at al, Astrophys. J. Suppl., {\bf 148}, 1 (2003).

\bibitem{r20} A.G. Riess et al, Astrophys. J., {\bf 560}, 49 (2001).

\bibitem{r2} T. Padmanabhan and T. Roy Choudury, Mon. Not. R. Astron. Soc., {\bf 344}, 823 (2003).\\
             T. Roy Choudury and T. Padmanabhan, Astron. Astrophys., {\bf 823}, 807 (2005).

\bibitem{r3} V. Sahni and A. Starobinsky, Int. J. Mod. Phys. D, {\bf 9}, 373 (1000).\\
             T. Padmanabhan, Phys. Rep., {\bf 380}, 235 (2003).

\bibitem{r4} J. Martin, Mod.Phys.Lett.A, {\bf 23}, 1252 (2008).

\bibitem{r5} I. Zlatev and P.J. Steinhardt. Phys.Lett.B, {\bf 459}, 570 (1999).

%\bibitem{r} N. Banerjee and S. Das, Mod. Phys. Lett. A, {\bf 21}, 2663 (2006).

\bibitem{r6} I. Zlatev, L. Wang and P.J. Steinhardt, Phys. Rev. Lett. {\bf 82}, 896 (1999).\\
             P.J. Steinhardt, L. Wang and I. Zlatev, Phys. Rev.D, {\bf 59}, 123504(1999).\\
             L. Wang, R.R. Caldwell, J.P. Ostriker and P.J. Steinhardt, Astrophys. J., {\bf 530}, 17 (2000).

\bibitem{r7} V.B.Johri Class.Quant.Grav {\bf 19}, 5959 (2002).

\bibitem{r8} L.A.Urena-Lopez and T.Matos Phys.Rev.D, {\bf 62}, 081302 (2000).

\bibitem{r9} M.Sahlen, AR.Liddle, D. Parkinson Phys.Rev.D, {\bf 75}, 023502 (2007).

\bibitem{r10} S.Dodelson, M.Kaplinghat and E. Stewart Phys.Rev.Lett, {\bf 85}, 5276(2000).

\bibitem{r11} P-Y. Wang, C.W Chen and P.Chen. JCAP, {\bf 02}, 016, 2012.

\bibitem{r} N. Banerjee and S. Das, Mod. Phys. Lett. A, {\bf 21}, 2663 (2006).

\bibitem{r13} \textit{Dynamical Systems in Cosmology}, J. Wainwright and G.F.R.Ellis (eds); Cambridge University Press, (1997).\\   \textit{Dynamical System and Cosmology}, A.A.Coley, Kluwer Academic Publishers (2003).

\bibitem{r12} S.C.C.Ng, N.J. Nunes, F.Rosati, Phys.Rev.D, {\bf64} ,083510(2001).

\bibitem{r14} L.Lara and M.Castagnins, Int.J.Theor. Phys. {\bf 44}, 1839(2005).

\bibitem{r15} E.Gunzig, V.Faraoni, A.Figeredo, T.M.Rocha Filho and L.Brenig, Class.Quant.Grav., {\bf 17}, 1783(2000).

\bibitem{r16} J.Carot and M.M.collinge, Class.Quanta.Grav., {\bf 20}, 707(2003).

\bibitem{r17} L.Arturo. Urena-Lopez, JCAP, {\bf 0509}, 013(2005).

\bibitem{r31} W. Fang, H. Tu, J. Huang and C. Shu, arxiv:[1402.4005]

\bibitem{r22} S. Kumar, S. Panda and A.A. Sen, Quantum Grav. {\bf 30}, 155011 (2013).

\bibitem{r23} S. Sen, A.A. Sen and M. Sami, Phys. Lett B. {\bf 686}, 1 (2010).

\bibitem{r18} S.J.Kolitch and B.Hall, arxiv:[gr-qc /9410039].\\
              S.J. Kolitch and D.M. Eardley, Ann. Phys., {\bf 241}, 128, 1995.

\bibitem{r19} \textit{Nonlinear dynamics and chaos: With applications to Physics, Biology, Chemistry and Engineering}, S.H. Strogatz, Westview Press (2001).

\bibitem{r21} V. Sahni, Lect.NotesPhys., {\bf 653}, 141 (2004).

\bibitem{r29} R. Giostri, M.V. dos Santos, I. Waga, R.R.R. Reis, M.O. Calvalo and B.L. Lago, J. Cos. Astrophysics, {\bf 027}, 1203 (2012).
%---------------------------------------------------------------------------------------CHAMELEON--------------------------------------
%\bibitem{varun} V. Sahni and A. Starobinsky, Int. J Mod. Phys. D, {\bf 9}, 373 (2000).
%\bibitem{paddy} T. Padmanabhan, Phys. Rep., {\bf 380}, 235 (2003).
%\bibitem{sami} E.J. Copeland, M. Sami and S. Tsujikawa, Int. J Mod. Phys. D, {\bf 15}, 1753 (2006).
%\bibitem{martin} J. Martin, Mod. Phys. Lett. A, {\bf 23}, 1252 (2008).
%\bibitem{review} T.P. Sotiriou and V. Faraoni; arxiv:gr-qc/0805.1726;\\
                % A. De Felice  and S. Tsujikawa; arxiv:gr-qc/1002.4928.\\
                 %S. Nojiri and S.D. Odintsov, Phys. Rep. {\bf 505}, 59 (2011).
%\bibitem{nbdp}O. Bertolami and P.J. Martins, Phys. Rev. D, {\bf 61}, 064007 (2000).\\
             % N. Banerjee and D. Pavon, Phys. Rev. D, {\bf 63}, 043504 (2001).\\
            %  S. Sen and A.A. Sen, Phys. Rev. D, {\bf 63}, 124006 (2001).\\
            %  E. Elizalde, S. Nojiri and S.D. Odintsov, Phys. Rev. D, {\bf 70}, 043539 (2004).\\
\bibitem{webb} J.K. Webb \textit{et al}., Astrophysics. Space Sci. {\bf283}, 565 (2003).
\bibitem{justin1} J. Khoury and A. Weltman, Phys. Rev D, {\bf 69}, 044026 (2004).
\bibitem{justin2} J. Khoury and A. Weltman, Phys. Rev. Lett., {\bf 93}, 171104 (2004).
\bibitem{khoury} J. Khoury, Class. Quant. Grav., {\bf 30}, 214004 (2013).
\bibitem{mota1} D.F. Mota and D.J. Shaw; arxiv:0805.3430. 
\bibitem{mota2} D.F. Mota and D.J. Shaw Phys. Rev. Lett., {\bf 97}, 151102 (2006).
\bibitem{shaw1} C. Burrage, A.C. Davis and D.J. Shaw, Phys. Rev. D, {\bf 79}, 044028 (2009).
\bibitem{shaw2} A.C. Davis, C.A.O. Schelpe and D.J. Shaw, Phys.Rev. D, {\bf 80}, 064016 (2009).
\bibitem{buri} P. Burikham and S. Panpanich, Int. J. Mod. Phys. D, {\bf 21}, 1250041 (2012).
\bibitem{brax1} P. Brax, C. Burrage, A.C. Davis, D. Seery and A. Weltman, Phys. Lett. B, {\bf 699}, 5 (2011).
\bibitem{mota3} D.F. Mota and C.A.O. Schelpe, Phys. Rev. D, {\bf 86}, 123002 (2012).
\bibitem{brax} P. Brax, C. van de Bruck, A.C. Davis, J. Khoury and A. Weltman, Phys. Rev. D, {\bf 70}, 123518 (2004).
\bibitem{das} S. Das, P.S. Corasaniti and J. Khoury, Phys. Rev. D, {\bf 73}, 083509 (2006).
\bibitem{nbsdkg} N. Banerjee, S. Das and K. Ganguly, Pramana, {\bf 74}, L481 (2010).
\bibitem{sdnb} S. Das and N. Banerjee, Phys. Rev.D, {\bf 78}, 043512 (2008).
\bibitem{jamil} M. Setare and M. Jamil. Phys. Lett. B, {\bf 690}, 1 (2014).
\bibitem{tavakol} S. Tsujikawa, T. Tamaki and R. Tavakol, JCAP, {\bf 05}, 020 (2009).
\bibitem{gunzig} E. Gunzig, V. Faraoni, A. Figeredo and L. Brenig, Class. Quantum Grav., {\bf 17}, 1783 (2000).\\
                 J. Carot and M.M. Collinge, Class. Quantum Grav., {\bf 20}, 707 (2003).\\
                 L.A. Urena-Lopez, JCAP, {\bf 0509}, 013 (2005).\\
                 S. Sen, A.A. Sen and M. Sami, Phys. Lett. B, {\bf 686}, 1 (2010).\\
                 S. Kumar, S. Panda and A.A. Sen, Class. Quantum Grav., {\bf 30}, 155011 (2013).\\
                 N. Roy and N. Banerjee, Gen. Rel. Grav., {\bf 46}, 1651 (2014).\\
%\bibitem{coley}  A.A. Coley, {\it Dynamical Systems and Cosmology}, Cambridge University Press (2003).
%\bibitem{ellis}  J. Wainwright and G.F.R. Ellis, {\it Dynamical Systems in Cosmology}, Springer (2005).
\bibitem{harko1} O. Minazzoli and T.Harko, Phys. Rev. D, {\bf 86}, 087502 (2012).
\bibitem{harko2} T.Harko, Phys. Rev. D, {\bf 81}, 044021 (2010).
%\bibitem{strogatz}S.H. Strogatz, {\it Nonlinear dynamics and chaos: With Applications to Physics, Biology Chemistry and  		             Engineering}, Westview Press, Boulder (2001).
\bibitem{nrnb} N. Roy and N. Banerjee, Euro. Phys. J Plus., {\bf 129}, 162 (2014).
%\bibitem{giostri}R. Giostri, M.V. dos Santos, I. Waga, R.R.R. Reis, M.O. Calvao and B.L. Lago, J. Cos. Astrophys., {\bf 027}, 1203 (2012).
\bibitem{ob} O. Farooq and B. Ratra, Astrophysics. J., {\bf 766} L7(2013)
%------------------------------------Holo-----------------------------------------------------------------------------------
\bibitem{'thooft} G. 't Hooft, arxiv: gr-qc/9310026.
\bibitem{susskind} L. Susskind, J. Math. Phys., {\bf 36}, 6377 (1995).
\bibitem{Cohen} A. G. Cohen, D. B. Kaplan and A. E. Nelson, Phys. Rev. Lett. {\bf 82}, 4971 (1999).
%\bibitem{li} M. Li, Phys. Lett. B {\bf 603}, 1 (2004)
\bibitem{Gao} C. Gao, F. Q. Wu, X. Chen and Y. G. Shen, Phys. Rev. D {\bf 79}, 043511 (2009).
\bibitem{Camco} Del Campo, Sergio, et al. Phys. Rev. D {\bf 83} 123006 (2011).
\bibitem{diego} D. Pavon, J. Phys. A:Math. Theor., {\bf 40}, 6865 (2007).
\bibitem{li} M.Li, Phys. Lett. B {\bf 603}, 1 (2004).
\bibitem{diego1} D.Pavon and W. Zimdahl, Phys. Lett. B {\bf 628}, 206 (2005).
\bibitem{diego2} D.Pavon and W. Zimdahl, Class. Quantum Grav. {\bf 24}, 5461 (2007).
\bibitem{diego3} N. Banerjee and D. Pavon, Phy. Lett. B {\bf 647}, 477 (2007).
\bibitem{setare} M.R. Setare and M. Jamil, Phys. Lett B {\bf 690}, 1 (2010).
%\bibitem{ellis} J. Wainwright and G. F. R. Ellis, \textit{Dynamical System in Cosmology} (Cambridge University Press, 2005).
%\bibitem{coley} A.A.Coley, \textit{Dynamical System and Cosmology} (Springer, 2003).
%\bibitem{gunzig} E. Gunzig, V. Faraoni, A. Figeredo and L. Brenig, Class. Quantum Grav. {\bf 17}, 1783 (2000).
%\bibitem{carot} J. Carot and M.M. Collinge, Class. Quantum Grav. {\bf 20}, 707 (2003).
%\bibitem{urena} L.A. Urena-Lopez, JCAP {\bf 0509}, 013 (2005).
%\bibitem{nandan2} N. Roy and N. Banerjee, Eur. Phys. J. Plus, {\bf 129}, 162 (2014).
%\bibitem{anjan} S. Kumar, S. Panda and A.A. Sen, Quantum Grav. {\bf 30}, 155011 (2013).
%\bibitem{soma} S. Sen, A.A. Sen and M. Sami, Phys. Lett B. {\bf 686}, 1 (2010).
%\bibitem{nandan1} N. Roy and N. Banerjee, Gen. Rel. Grav. {\bf 46}, 1651 (2014).
%\bibitem{fang} W. Fang, H. Tu, J. Huang and C. Shu, arxiv:[1402.4005].
\bibitem{nairi1} N. Mazumder, R. Biswas and S. Chakraborty, arxiv:[1106.4627].
\bibitem{nairi2} N. Mazumder, R. Biswas and S. Chakraborty, arxiv:[1106.4626].
\bibitem{setare1} M.R. Setare and E.C. Vagenas, Int. J. Mod. Phys. D, {\bf 18}, 147 (2009).
%\bibitem{strog} S.H. Strogatz, \textit{Nonlinear Dynamics and Chaos: With Applications to Physics, Biology, Chemistry and     Engineering};                             Westview Press, Boulder (2001). 

\bibitem{reza} R. Tavakol, `` Introduction to dynamical systems '' in \textit{Dynamical Systems in Cosmology}, J. Wainwright and G.F.R.Ellis (eds); Cambridge University Press, (1997). .
                
\end{thebibliography}

% If you would like to use BibLaTeX for your references, pass `custombib' as
% an option in the document class. The location of 'reference.bib' should be
% specified in the preamble.tex file in the custombib section.
% Comment out the lines related to natbib above and uncomment the following line.

%\printbibliography[heading=bibintoc, title={References}]

\end{spacing}

% ********************************** Appendices ********************************

%\begin{appendices} % Using appendices environment for more functunality

%\include{Appendix1/appendix1}
%\include{Appendix2/appendix2} 

%\end{appendices}

% *************************************** Index ********************************
\printthesisindex % If index is present

\end{document}